\documentclass[nonacm]{acmart}

\AtBeginDocument{%
  \providecommand\BibTeX{{%
    \normalfont B\kern-0.5em{\scshape i\kern-0.25em b}\kern-0.8em\TeX}}}

\setcopyright{acmcopyright}
\copyrightyear{2020}
\acmYear{2020}
\acmDOI{10.1145/1122445.1122456}

\acmJournal{TOCHI}
\acmVolume{27}
\acmNumber{2}
\acmArticle{111}
\acmMonth{8}

\usepackage{balance}     
\usepackage{graphics}    
\usepackage[T1]{fontenc}   

\makeatletter
\g@addto@macro{\UrlBreaks}{\UrlOrds}
\makeatother

\usepackage{color}
\usepackage{booktabs}

\usepackage{microtype}    
\usepackage{ccicons}      

\usepackage{todonotes}
\usepackage[T1]{fontenc}
\usepackage[scaled]{beramono}
\usepackage{enumitem}
\usepackage{cleveref}
\usepackage{multirow}

\setlist{itemsep=0pt}

\usepackage{listings}

\lstdefinelanguage{obsidian}{
	keywords=[1]{contract,state,transaction,disown,new,asset,this,in,return,returns,remote,main,private},
	keywordstyle=[1]\color{blue}\bfseries,
	keywords=[2]{int},
	keywordstyle=[2]\color{teal}\bfseries,
	keywords=[3]{->},
	keywordstyle=[3]\color{blue}\bfseries,
	keywords=[4]{Owned,Unowned,Shared},
	keywordstyle=[4]\color{black}\bfseries,
	comment=[l]{//},
	morecomment=[s]{/*}{*/},
	commentstyle=\color{gray}\ttfamily,
	morestring=[b]",
	alsoother={@},
}

\newcommand{\code}[1]{\texttt{#1}}
\newcommand*\LSTfont{\Small\ttfamily\SetTracking{encoding=*}{-60}\lsstyle}

\makeatletter
\def\url@leostyle{%
  \@ifundefined{selectfont}{
  \def\UrlFont{\sf}
  }{
  \def\UrlFont{\small\bf\ttfamily}
  }}
\makeatother
\makeatletter
\newcommand*{\compress}{\@minipagetrue}
\makeatother

\crefname{section}{Section}{Sections}
\crefname{table}{Table}{Tables}
\crefname{figure}{Figure}{Figures}

\begin{document}

\author{Michael Coblenz}
\orcid{0000-0002-9369-4069}             
\affiliation{
  \position{Doctoral Candidate}
  \department{Computer Science Department}              
  \institution{Carnegie Mellon University}            
  \streetaddress{5000 Forbes Ave.}
  \city{Pittsburgh}
  \state{PA}
  \postcode{15213}
  \country{USA}                    
}
\email{mcoblenz@cs.cmu.edu}          

\author{Gauri Kambhatla}
\authornote{Work conducted while at Carnegie Mellon University.}          
\affiliation{
  \department{Computer Science Department}              
  \institution{University of Michigan}            
  \streetaddress{500 S. State St.}
  \city{Ann Arbor}
  \state{MI}
  \postcode{48109}
  \country{USA}                    
}
\email{gkambhat@umich.edu}          

\author{Paulette Koronkevich}
\affiliation{
  \department{Computer Science Department}              
  \institution{University of British Columbia}            
  \streetaddress{2329 West Mall}
  \city{Vancouver}
  \state{BC}
  \postcode{V6T 1Z4}
  \country{Canada}                    
}
\email{pletrec@cs.ubc.ca}          
\authornotemark[1]

\author{Jenna L. Wise}
\affiliation{
  \position{Doctoral Student}
  \department{Institute for Software Research}              
  \institution{Carnegie Mellon University}            
  \streetaddress{5000 Forbes Ave.}
  \city{Pittsburgh}
  \state{PA}
  \postcode{15213}
  \country{USA}                    
}
\email{jlwise@cs.cmu.edu}          

\author{Celeste Barnaby}
\affiliation{
  \institution{Facebook, Inc.}            
  \streetaddress{1 Hacker Way}
  \city{Menlo Park}
  \state{CA}
  \postcode{94025}
  \country{USA}                    
}
\email{celestebarnaby@gmail.com}          
\authornotemark[1]

\author{Joshua Sunshine}
\orcid{0000-0002-9672-5297}             
\affiliation{
  \position{Systems Scientist}
  \department{Institute for Software Research}              
  \institution{Carnegie Mellon University}            
  \streetaddress{5000 Forbes Ave.}
  \city{Pittsburgh}
  \state{PA}
  \postcode{15213}
  \country{USA}                    
}
\email{joshua.sunshine@cs.cmu.edu}          

\author{Jonathan Aldrich}
\orcid{0000-0003-0631-5591}             
\affiliation{
  \position{Professor}
  \department{Institute for Software Research}              
  \institution{Carnegie Mellon University}            
  \streetaddress{5000 Forbes Ave.}
  \city{Pittsburgh}
  \state{PA}
  \postcode{15213}
  \country{USA}                    
}
\email{jonathan.aldrich@cs.cmu.edu}          

\author{Brad A. Myers}
\orcid{0000-0002-4769-0219}             
\affiliation{
  \position{Professor}
  \department{Human-Computer Interaction Institute}              
  \institution{Carnegie Mellon University}            
  \streetaddress{5000 Forbes Ave.}
  \city{Pittsburgh}
  \state{PA}
  \postcode{15213}
  \country{USA}                    
}
\email{bam@cs.cmu.edu}          

\renewcommand{\shortauthors}{Coblenz et al.}

\title[PLIERS: A User-Centered Process for Programming Language Design]{PLIERS: A Process that Integrates User-Centered
Methods into Programming Language Design}


\begin{abstract}
Programming language design requires making many usability-related design decisions. However, existing HCI methods can be impractical to apply to programming languages: they have high iteration costs, programmers require significant learning time, and user performance has high variance. To address these problems, we adapted both formative and summative HCI methods to make them more suitable for programming language design. We integrated these methods into a new process, PLIERS, for designing programming languages in a user-centered way. We evaluated PLIERS by using it to design two new programming languages. Glacier extends Java to enable programmers to express immutability properties effectively and easily. Obsidian is a language for blockchains that includes verification of critical safety properties. Summative usability studies showed that programmers were able to program effectively in both languages after short training periods.

	 
  \end{abstract}


\maketitle

\section{Introduction}
\label{introduction}


Programming languages serve as interfaces through which programmers and software engineers can create software. The ability of these users to achieve their goals, as with other kinds of interfaces, depends on the usability of the languages in which they do their work. For example, the presence of \code{null} in Java results in a particular kind of error-proneness, since programmers can easily accidentally write code that dereferences \code{null}~\cite{Hoare2009:Null}. These kinds of mistakes persist despite the training and experience that professional software engineers have. 

There is a long history of research on understanding how programmers' cognitive processes relate to the tools they use~\cite{Miller1974:Programming, Shneiderman1986:Empirical, Sime1977:Scope, Soloway1984:Empirical}. This work was described in part in the proceedings of the Empirical Studies of Programmers (ESP) and Psychology of Programming Interest Group (PPIG) workshops. More recently, this line of work has continued at conferences such as CHI, ICSE, and VL/HCC. Our focus is a practical one: \textit{How can programming language designers leverage data from users to improve language usability?} 

Our high-level focus can be refined in terms of three research questions:
\begin{description}
	\item[Naturalness:] How can we obtain insights as to what language designs will be \textit{natural} for programmers, given that we are trying to obtain particular static safety guarantees in the language?
	\item[Iteration:] How can we iterate on a particular design so it continually gets to be more effective for users?
	\item[Comparison:] How can we compare multiple language designs to see which are more effective for users?
\end{description}

Some authors, such as Stefik and Hanenberg, have focused on using quantitative approaches~\cite{Stefik14:Programming,stefik2017methodological}. Others have focused on in-depth case studies to evaluate their languages~\cite{Ahmad2011:Jabberwocky}. Our approach is to integrate a wide variety of both \textit{qualitative} and \textit{quantitative} user-focused methods with \textit{formal theory-based methods}~\cite{Pierce:TypeSystems} to design programming languages~\cite{Coblenz18:Interdisciplinary, Myers2016:Programmers}. This approach allows us to integrate user research into many different stages of the design process. 

In order to address our three research questions, we adapted traditional HCI methods to the context of the design of programming languages that target professional software engineers. Then, we applied those adaptations to the design process of two new languages, Glacier and Obsidian, which we used as testbeds for language design methods. Finally, we synthesized the methodology into a process we call \textit{PLIERS}: \textbf{P}rogramming \textbf{L}anguage \textbf{I}terative \textbf{E}valuation and \textbf{R}efinement \textbf{S}ystem. This paper describes the methods and process we developed, and motivates them by showing the insights that we obtained by using the process on Glacier and Obsidian. The Obsidian work is new in this paper, and the Glacier work was partially discussed previously~\cite{Coblenz2017:Glacier}. 

It is not enough to only use methods from HCI to design a programming language because, if a designer wants programs to have well-defined meanings (semantics) and well-understood safety properties (soundness), the programming language must also be subject to a collection of semantic constraints from the theory of programming languages. We have integrated strategic application of our adapted programming language design and evaluation methods into a process that also incorporates formal methods so that the resulting languages can be both usable and sound. Although the individual methods have been applied in the HCI literature in a variety of contexts, this paper shows how we have adapted the methods and combined them to obtain insights regarding programming languages that target professional software engineers. PLIERS is not a recipe for language design, just as \textit{agile} is not a recipe for software engineering. Instead, PLIERS provides a process for organizing language design work around human-centered methods. For each step in the design process, PLIERS suggests ways of integrating human-centered methods to inform the designers.

Designing programming languages requires expertise in type systems, compilers or interpreters, and language runtimes. As such, PLIERS is aimed at showing people with those technical tools that it is feasible and effective to include user-centered methods. By doing so, the goal is that the languages will be more effective for programmers than they might be otherwise. 

This paper makes three main contributions:

\begin{enumerate}
\item We define the PLIERS programming language design process, which shows how user-centered methods can contribute to many different phases of programming language creation. We evaluated PLIERS by using it to develop two programming languages; we describe the benefits and areas for improvement that we observed in the process (\cref{sec:evaluating-PLIERS}).

\item We show how we have adapted several \textit{formative} study techniques, such as natural programming, Wizard of Oz, rapid prototyping, cognitive dimensions of notations analysis, and interview studies to inform the design of Glacier and Obsidian. We found that our adapted methods were effective when used in the context of PLIERS. 

\item We show how we conducted \textit{summative} usability studies on new programming languages. By developing ways of teaching the languages efficiently, effectively, and consistently, we were able to conduct usability studies of programmers using novel programming languages.

\end{enumerate}

We designed PLIERS for use with language designs that require developers to learn challenging programming concepts or think in a new way. The safety properties that motivated the two languages we discuss in this paper provided opportunities for language constructs that could potentially be confusing, which made them ideal testbeds for PLIERS. However, other contexts provide other kinds of conceptual challenges for programmers, and PLIERS may be similarly useful in those contexts. For example, multicore programming results in concurrency challenges; distributed programming requires managing asynchronous requests that may fail; and domain-specific languages (DSLs) may require mastery of domain concepts. We expect that PLIERS will be useful for helping language designers with any language that requires programmers to master difficult concepts.

In developing PLIERS, we initially intended to apply known HCI methods, such as \textit{natural programming}~\cite{naturalprogramming}, \textit{Wizard of Oz}~\cite{dahlback1993wizard}, \textit{interviews}, and \textit{rapid prototyping}. However, we found that the study design process was very challenging due to the nature of programming and the complexity of the design space. For example, some of the challenges we faced included: 
\begin{description}
    \item[Recruiting:] how could we recruit participants who have sufficient programming skill and whose results would generalize beyond the population of students, despite having limited access to professional software engineers? 
    \item[Training:] how could we train participants in a new programming language in a short enough amount of time to make studies practical? 
    \item[High prototyping cost:] how could we conduct user studies on programming languages that have only informal designs and no implementations, since the cost of building working prototypes is high? 
    \item[Variance and external validity:] how would we mitigate high variance, which is typical in programming tasks, without constraining the tasks so much that they were no longer representative of real-world programming tasks? 
\end{description}

The problems of variance and external validity were particularly relevant for quantitative studies, which needed to be practical in the context of our university setting. Programming tasks that are \textit{not} extremely constrained tend to produce results with high variance, making statistical significance hard to obtain. On the other hand, tasks that \textit{are} highly constrained suffer from low external validity, since real-world programming tasks are typically long and complex. 

Therefore, we had to modify existing methods to address these challenges. Our study design contributions are summarized in \cref{approach-summary-table}. For example:
\begin{itemize}
	\item By adapting the \textit{natural programming} technique to allow \textit{progressive prompting}, we were able to obtain both unbiased responses as well as data that were relevant to the particular designs we were considering.
	
	\item By \textit{back-porting} language design questions to languages with which participants were familiar and by using the \textit{Wizard of Oz} evaluation technique, we were able to obtain usability insights on incomplete designs, and \textit{isolate} the design questions of interest from confounding variables.
	
	\item By dividing large tasks into multiple, smaller tasks, and by using pilot studies to set task time limits effectively in quantitative studies, we were able to reduce variance sufficiently to obtain meaningful results in complex programming tasks.
	
	\item By recruiting participants who were \textit{representative of at least some junior-level professional developers}, we were able to maximize external validity in our studies while still conducting them practically at a university setting. We were also able to show usability impacts of the language designs under consideration.
		
	\item By developing incremental tutorials with integrated practice opportunities, we were able to \textit{teach} the languages to participants in a short time (for Obsidian, about 90 minutes was typical). 
\end{itemize}

In order to contextualize the methods we describe in this paper, \cref{glacier-language,obsidian-language} explain the two languages that we used to develop the methods. \cref{sec:related-work} discusses related work, and \cref{sec:PLIERS} introduces PLIERS. The rest of the paper proceeds by showing how we used PLIERS in Glacier (\cref{sec:glacier}) and Obsidian (\cref{sec:obsidian-formative,Obsidian-summative,sec:Obsidian-RCT}). Then, we discuss the challenges to effective study design that we observed while creating those two languages and how we addressed those challenges (\cref{challenges}). We  propose future work and conclude in \cref{sec:future-work,sec:conclusion}.

\subsection{Glacier}
\label{glacier-language}

Glacier~\cite{Coblenz2017:Glacier} is an extension to Java that supports \textit{transitive class immutability}. Although security experts had recommended expressing state in an immutable way whenever possible, it was unclear how programming languages should support immutability. For example, Java includes the \code{final} keyword, but because \code{final} only restricts assignment to variables and not mutation of referenced state, actually enforcing immutability in Java is very difficult. The Java code below compiles without error despite the assignment and the use of \code{final}:

\begin{lstlisting}[language=Java]
final int a[] = {0};
a[0] = 42;
\end{lstlisting}

In designing Glacier, we sought to show how a language design might support the use of immutability in practical programming languages. \textit{Immutability} means that objects cannot be changed after they are created. Several organizations recommend the use of immutability to prevent security vulnerabilities in software. For example, Oracle's \textit{Secure Coding Guidelines for Java}~\cite{securecoding} and Microsoft's \textit{Framework Design Guidelines}~\cite{microsoft-struct} both recommend using immutability for security reasons. However, we found that there were hundreds of different possible designs for immutability protection in programming languages, and it was unclear which approaches might be usable by programmers and which might actually support programmers' needs~\cite{Coblenz2016:Exploring}.

To determine a point in the design space that might be useful and effective, we conducted semi-structured interviews with eight software engineers. In those interviews, we asked questions about bugs, such as ``Can you think of a bug you investigated or fixed that was caused by a data structure changing when it should not have?''~\cite{Coblenz2016:Exploring-extended}. All the participants who worked on software with significant amounts of state said that incorrect state change was a major source of bugs.

As a result, we hypothesized that \textit{transitive} immutability might be a useful point in the design space to pursue. \textit{Transitive} means that the restriction applies not just to a class, but recursively to all of its fields. \textit{Immutability} means that objects to which the restriction applies cannot have any of their data modified through \textit{any} reference, as opposed to the restriction only applying to certain references to a given object. This kind of immutability would provide strong guarantees, which developers could rely on to protect against improper changes to state.

We initially adapted an existing system, IGJ~\cite{IGJ}, to enforce transitivity. We found in a usability study, however, that there were significant usability challenges with this approach, which related to the flexibility provided by IGJ to apply restrictions to individual objects. This led us to create Glacier, which supports transitive \textit{class} immutability, in which classes that are declared immutable have all instances immutable. We were able to show that Glacier (a) could be used effectively by our study participants to specify immutability; and (b) detected improper-mutation bugs that participants frequently inserted in the codebase when they were using regular Java.

The final version of Glacier provides a new annotation, \code{@Immutable}, which can be applied to class definitions. \code{@Immutable} indicates that every instance of that class must be transitively immutable. The alternative to \code{@Immutable} is \code{@MaybeMutable}, which applies by default. The compiler checks classes that are annotated \code{@Immutable} and gives an error if any of the fields are references to classes that are themselves not \code{@Immutable}. Also, assignment is disallowed to fields of \code{@Immutable} classes except in their constructors. For example, the comments indicate that the compiler would emit errors for lines 3 and 6:

\begin{lstlisting}[language=Java,numbers=left,xleftmargin=2em]
@Immutable class Person {
    String name; // OK; String is @Immutable
    Date birthdate; // Error; Date is @MaybeMutable
  
    void setName(String n) {
        name = n; // Error; cannot assign to fields of @Immutable classes
    }
}
\end{lstlisting}

In \cref{sec:glacier}, we describe how we used qualitative methods to ground our initial design in user data and how we conducted an RCT to compare Glacier to Java's \code{final} feature. In that study, we recruited 20 participants and found that although none of those assigned to use \code{final} were able to specify immutability in an existing program correctly, almost all of the Glacier participants were able to do so. We also gave the participants two maintenance tasks. Among 14 tasks that Glacier participants said they completed, all were done correctly. However, of 18 tasks that \code{final} participants said they completed, only seven were completed correctly.

\subsection{Obsidian}
\label{obsidian-language}

Obsidian is targeted at programming \textit{blockchains}~\cite{herlihy2019blockchains}, in which a decentralized network of computers maintains system state and executes transactions. Blockchains support deploying \textit{smart contracts}, which are programs that maintain state. Typically, each deployment is an instance of a class, though in a blockchain context, the keyword \code{contract} is used instead of \code{class}. In contrast to most of the existing user-centered programming language work, which often focuses on novice or end-user programmers~\cite{Kelleher2005:Lowering}, Obsidian is intended for use by professional programmers and software engineers. This presents additional challenges, since we are interested in evaluating how the language will be used in the long term despite being limited in our ability to recruit software engineers to work for extended periods of time in our studies. After our design was complete, we found in a summative study that most of the participants were able to complete programming tasks successfully in Obsidian.

Blockchains, which have been proposed for high-stakes applications such as financial transactions, health care~\cite{HealthCare}, supply chain management~\cite{SupplyChain}, and others~\cite{Elsden18:Making}, are an ideal testbed for a new language design process. The need for a safer language is motivated by the history of security vulnerabilities, through which over \$80 million worth of virtual currency has been stolen~\cite{DAO, CNBC}. However, it is not realistic to assume, as some other projects do~\cite{Bhargavan2016,Harz2018:Towards,Alt2018:SMT-based}, that the developers will be experts in formal verification or that companies will invest the resources required to formally verify that their programs are correct. Instead, we seek a more \textit{lightweight} approach that provides additional safety guarantees at low cost to developers.

We established several objectives in our design of Obsidian:
\begin{enumerate}
	\item Improve safety by detecting more bugs than current smart contract languages do, preferably at compile time, to prevent deployment of buggy programs.
	\item Maximize usability by ensuring that programmers can complete domain-appropriate programming tasks, ideally with little training in the language.
	\item Advance the science of programming language design by developing user-centered methods that can contribute to a more usable language.
\end{enumerate}

Detecting bugs was our initial objective, so we considered bugs, such as the DAO hack~\cite{DAO-details}, which resulted from a reentrant invocation in which a contract allowed itself to be invoked while in an inconsistent state. We also analyzed characteristics of proposed blockchain applications. In general, we observed that proposed blockchain applications typically maintain high-level state, which governs which operations are safe. 

For example, a \code{Casino} can accept \code{bet} invocations only before the \code{Game} has been played. More generally, the authors of Solidity~\cite{Solidity}, a commonly-used smart contract language, observed that many contracts implement state machines~\cite{Solidity-Patterns}. Unfortunately, in Solidity, users must define states via enumerated types and then manually ensure that methods are only invoked when the target object is in an appropriate state. Although methods that can only be invoked in particular states are common~\cite{Beckman:2011:ESO:2032497.2032501}, writing programs that only invoke methods when appropriate has been shown to be hard for users~\cite{Sunshine:2015:SSS:2820282.2820295}, and Solidity includes no mechanism to ensure safety.

Smart contracts commonly manipulate \textit{assets}, which are objects that have value (such as cryptocurrencies). In Solidity, it is possible to lose track of money and other assets~\cite{Delmolino2016:Step}, resulting in their value being permanently irretrievable. We were interested in designing a language in which many kinds of asset loss could be detected by the compiler.

In order to leverage those observations, we became interested in a \textit{typestate}-oriented approach~\cite{Aldrich09:Typestate}, in which \textit{states} of objects are incorporated into types. For example, rather than merely having a \code{LightSwitch} type, we can have \code{LightSwitch@On} be the type of a reference to an object that is in the \code{On} state. Then, if the user attempts an invalid operation, such as turning on a switch that is already on, the compiler can issue an error.

Typestate-based types are in a class called \textit{linear types}. Unlike traditional types, linear types can change as operations are performed. For example, invoking \code{turnOff()} on a reference of type \code{LightSwitch@On} \textit{changes the type} of the reference to \code{LightSwitch@Off}. Conveniently, linear types are also what are needed to ensure that assets are never lost. Obsidian includes \textit{owned} objects: for each owned object, there is an object that owns it via an owning reference. If a local variable that owns an asset goes out of scope, the compiler emits an error message. Fields that own assets can only exist in contracts that are themselves assets. This way, each asset always has an owner.

We selected an object-oriented approach, since object-oriented approaches are well-suited for representing state and corresponding updates. We avoided inheritance, since we wanted to avoid the fragility that results~\cite{Mikhajlov:1998:SFB:646155.679700}.  For a full description of the language, please refer to~\citet{Coblenz19:Obsidian-arxiv}. However, \cref{tiny-vending-machine} shows some of the key features of the \textit{final} version of Obsidian using the example of a \textit{tiny vending machine} (TVM). \code{TVM} is a \code{main} contract, so it can be deployed independently to a blockchain. A \code{TVM} has a very small inventory: just one candy bar. It is either \code{Full}, with one candy bar in inventory, or \code{Empty}. Clients may invoke \code{buy} on a vending machine that is in \code{Full} state, passing a \code{Coin} as payment. When \code{buy} is invoked, the caller must initially \textit{own} the \code{Coin}, but after \code{buy} returns, the caller no longer owns it. \code{buy} returns a \code{Candy} to the caller, which the caller then owns. After \code{buy} returns, the vending machine is in state \code{Empty}.

\begin{figure}[ht]
\begin{lstlisting}[numbers=left, framexleftmargin=0em, xleftmargin=1em, basicstyle=\LSTfont, language=obsidian]
// TVM is a Tiny Vending Machine.
main asset contract TVM {
  Coins @ Owned coinBin;

  state Full {
    Candy @ Owned inventory;
  }
  
  // No candy if the machine is empty.
  state Empty;

  TVM() {
    // Start with no coins, and go to the Empty state.
    coinBin = new Coins(); 
    ->Empty;
  }

  // restock transitions from Empty to Full by taking ownership of candy.
  transaction restock(TVM @ Empty >> Full this,
                      Candy @ Owned >> Unowned candy) 
  {
    ->Full(inventory = candy);
  }

  // buy transitions from Full to Empty by taking ownership of a coin.
  // buy returns the purchased candy.
  transaction buy(TVM @ Full >> Empty this,
                  Coin @ Owned >> Unowned coin) 
                  returns Candy @ Owned 
  {
    coinBin.deposit(coin);
    Candy result = inventory;
    ->Empty;
    return result;
  }

  // withdraw removes any accumulated coins and returns them to the caller.
  transaction withdraw() returns Coins @ Owned 
  {
    Coins result = coinBin;
    coinBin = new Coins();
    return result;
  }
}
\end{lstlisting}
\caption{A tiny vending machine that shows key features of Obsidian.}
\label{tiny-vending-machine}
\end{figure}

In \cref{sec:obsidian-formative}, we describe techniques we used to obtain data about individual language design choices. In the summative portion of the evaluation process, which we describe in \cref{Obsidian-summative}, we recruited six participants. We trained them in Obsidian and then asked them to complete three programming tasks. Two of the participants inserted bugs that the Obsidian compiler detected, but which the Solidity compiler would not have been able to detect. Although we observed high variance in performance, two of the participants were able to complete all of the tasks. 

We also conducted an experiment comparing Obsidian and Solidity, which we summarize in \cref{sec:Obsidian-RCT}. In one task, Obsidian programmers were able to avoid losing an asset that the Solidity programmers frequently lost track of. In another, six of ten Obsidian participants were able to use ownership to fix a security problem. In a final, more open-ended task, we observed that surprisingly, the Obsidian programmers abused a feature of the language, resulting in asset loss, and that Solidity programmers (who were not constrained by Obsidian's strong type system) were generally able to complete the task faster.

\section{Related Work}
\label{sec:related-work}

Newell and Card argued for the use of HCI methods in programming language design in 1985: ``Millions for compilers but hardly a penny for understanding human programming language use~\cite{Newell1985:Prospects}.'' Morrisett reiterated this problem in 2009, arguing that a programming language is a medium for communication among humans, but we lack principles for evaluating this aspect of languages~\cite{Morrisett2009:Grand}. Our earlier essay argued for using many different methods in language design~\cite{Coblenz18:Interdisciplinary}. Although that article promoted the use of formative methods (among others), this paper describes the methods in much more detail, giving recommendations for how other designers might use them in their own work. This paper also includes our experiences with Obsidian, including techniques we developed during that work. Finally, it describes PLIERS, which is our overall language design process.

The Empirical Studies of Programmers workshops focused in large part on a cognitive science approach to studying programmers: can we build models of cognition that explain programmer behavior? Key results include describing techniques used by programmers when working with code, such as identifying key lines (beacons), relating program details to the problem domain, and using both top-down and bottom-up understanding techniques. The ESP literature provided insights into the problems that people have when using existing languages. 

Some of the work in ESP workshops studied professional programmers. For example, Pennington used theories of understanding of natural language text to model expert programmers' comprehension of programs~\cite{Pennington1987:Stimulus}, finding that procedural knowledge (rather than knowledge of functional units) dominated their understanding. The study was conducted on COBOL programs, which were likely structured substantially differently from modern software. Furthermore, no libraries or frameworks were used, so the fact that the programmers could see and consider all relevant code may have resulted in a substantially different kind of programming task than the ones that we consider today. However, the approach suggests that a cognitive modeling approach may help derive a theory of programmers that is relevant for designing programming languages. In this work, we rely on more recent, heuristic-based approaches, such as the Cognitive Dimensions of Notations~\cite{green1996usability}, which are applicable in more general domains, and which came out of the cognitive science-based approach that was central in the ESP work.

Visser described a four-week observational study of one professional programmer working in a declarative, domain-specific language~\cite{Visser1987:Strategies}. Visser noted that obtaining mental models was challenging because the programmer found think-aloud very difficult while working on problems, but made observations about the structure of the programmer's work. For example, the programmer used analogical reasoning and examples to help reason about the software; used both top-down and bottom-up work styles; and sought consistency in the program. This work provides an empirical foundation for some requirements of programming environments, such as allowing creation of interfaces separately from implementations, and providing tools to standardize notation (e.g. style linting tools).

Vans et al. conducted study of the comprehension process and information needs of programmers in industry doing maintenance tasks~\cite{Vans1999:Program}. Some of the understanding techniques that the programmers used were similar to methods that were observed in novices as well, including top-down, bottom-up, and code-tracing methods, but the professionals used a much wider variety of techniques than had been observed in novices, such as generating and abandoning large numbers of hypotheses regarding the programs. This suggests that programming language design studies conducted with students can give some guidance regarding languages intended for professionals, but such studies may be limited in the kinds of techniques that the participants use. In many of the studies presented in this paper, we recruited experienced students, who in many cases had several years of professional programming experience. This approach allowed us to broaden the set of techniques our participants would use to accomplish their tasks and make our results more generalizable relative to using only novice programmers.

Guindon et al. used protocol analysis~\cite{Ericsson1984:Protocol} to analyze think-aloud protocols from three experienced programmers who were asked to design software to solve an elevator-scheduling problem~\cite{Guindon1987:Breakdowns}. Guindon et al. observed breakdowns in process that arose from lack of knowledge (e.g., of the problem domain) and from cognitive limitations (e.g., capacity of short-term memory). Because this work consisted of a think-aloud study of programmers, it shares several threats to validity with the qualitative work we describe here: a task that may not reflect real-world tasks; short duration of the task, which was concentrated in a lab-based two-hour session; and a sample of programmers that may not be representative of programmers in general. Our approaches to mitigating external validity were similar to theirs. We recruited from students who were likely to have some professional experience and do not claim that all programmers will encounter the same difficulties that they did. We do not claim that we observed all possible problems that users might have when using the tools we gave them. Instead, we argue that addressing the problems we observed is likely to help \textit{some} relevant users be more effective in achieving their goals. In our approach to think-aloud studies, we analyzed notes taken by the experimenter and screen recordings of the participants doing the tasks.

Direct observations of work and interviews have both been used to understand how teams work together to develop software. Walz et al. analyzed videos of teams conducting a requirements analysis to study conflict patterns~\cite{Walz1987:Methodology}; Krasner et al. interviewed members of 19 software development teams to understand team communication~\cite{Krasner1987:Communication}. Although our work focused on individual developers, we used multiple methods where appropriate. This work shares threats to external validity with other small-sample studies, including the ones we conducted. Krasner et al. mitigated these risks by choosing diverse teams to study. Although Walz et al. only studied one team, they studied the team over 43 meetings over four months.


A substantial amount of prior work on the usability of programming languages focuses on novices. For example, HANDS~\cite{Pane2002:Using}, Helena~\cite{Chasins2017:Helena}, and Scratch~\cite{resnick2009scratch} aimed to make it easier for novices to write programs. HANDS, in particular, introduced the \textit{Natural Programming} technique, which we leveraged and adapted in this work. Stefik et al. also focused on novices, collecting quantitative data on their error rates~\cite{Stefik:2011:ECA:2089155.2089159}. Designing languages for novices is substantially different from designing languages for experienced programmers. For example, languages for novices typically focus on learnability. In contrast, languages for professionals commonly include additional complexity, in part resulting from the kinds of safety properties that are beneficial when building real systems. 

Other work has focused on programming tools for end-user programmers, whose primary goal is not to write software but rather to accomplish goals in some particular domain~\cite{Ko2011:State}. For example, Blackwell and Burnett developed Attention Investment and applied it to a research spreadsheet tool, Forms/3~\cite{Blackwell2002:Applying}. Peyton Jones et al. used Cognitive Dimensions~\cite{green1996usability} and Attention Investment to provide a new kind of user-defined functions in Excel~\cite{Jones2003:User}. Our work is focused on methods that address the unique challenges of complexity that result from targeting professional programmers and software engineers.

RCTs (randomized controlled trials) have been used to compare different programming language designs. For example, Uesbeck et al. investigated the impact of lambdas in C++~\cite{Uesbeck:2016:ESI:2884781.2884849}, and Endrikat et al. looked at static typing~\cite{Endrikat:2014:ADS:2568225.2568299}. That work is a useful complement to this work, but the focus here is on using low-cost, practical \textit{qualitative} methods to inform the entire language design process. In contrast, quantitative summative studies require high-fidelity prototypes in order to obtain  measurements that can be expected to generalize to the final system. These prototypes can be very expensive to build for complex programming languages.

Another approach to programming language evaluation is to compare designs via crowdsourcing methods. \citet{Chamberlain2017:Assessing} compared functional-style to literal-style approaches for specifying topology of streaming applications (i.e., pipes-and-filters style applications) using Mechanical Turk, finding that users were more likely to prefer literal-style specifications, and experienced programmers were more likely to understand the literal-style specifications than the functional-style ones. \citet{Wilson2017:Crowdsource} investigated crowdsourcing more esoteric language design decisions, finding low consistency (people did not give consistent answers when asked similar questions repeatedly) and low consensus (people did not agree with each other on which design choice was best). Crowdsourcing approaches can scale well, but typically require that the studies be of relatively short duration. This paper focuses on higher-bandwidth qualitative methods and on evaluation approaches for languages for professionals, and serves to complement crowdsourcing approaches.

The HCI literature includes many different language designs as well as other kinds of tools for programmers. For example, Dog/Jabberwocky~\cite{Ahmad2011:Jabberwocky}, Protovis~\cite{Bostock2009:Protovis}, Reactive Vega~\cite{Satyanarayan2015:Reactive}, and InterState~\cite{Oney2014:Interstate} are all languages or APIs that make it easier for programmers to accomplish their goals. Those papers describe only the final designs of those systems and summative usability studies. This paper focuses on methods that can be used \textit{during} the design process and gives recommendations that are useful in preparing a summative evaluation.

Finally, there is a variety of methodological guidance in SE and HCI that is applicable to studies of programming languages. Ko et al. discussed techniques for doing empirical studies of tools for software engineers~\cite{Ko2015:practical}.  Buse et al. conducted a systematic literature review, observing increasing use of user evaluations in software engineering research~\cite{Buse2011:benefits}. 
Verner et al. gave guidelines for industrial case studies in software engineering research~\cite{Verner2009:Guidelines}. Perry et al. gave a tutorial on case study methodology for software engineers~\cite{Perry2004:Case}. Likewise, Shneiderman and Plaisant gave recommendations for using case studies for information visualization tools~\cite{Shneiderman2006:Strategies}. 

The technical details of the Obsidian language are described in a separate paper~\cite{Coblenz19:Obsidian-arxiv}; Glacier is also described in more detail separately~\cite{Coblenz2017:Glacier}. This paper focuses on the user-centered process that we used to design and evaluate the languages.

\section{PLIERS}
\label{sec:PLIERS}

\begin{figure}[b]
\centering
\includegraphics[width=.8\columnwidth]{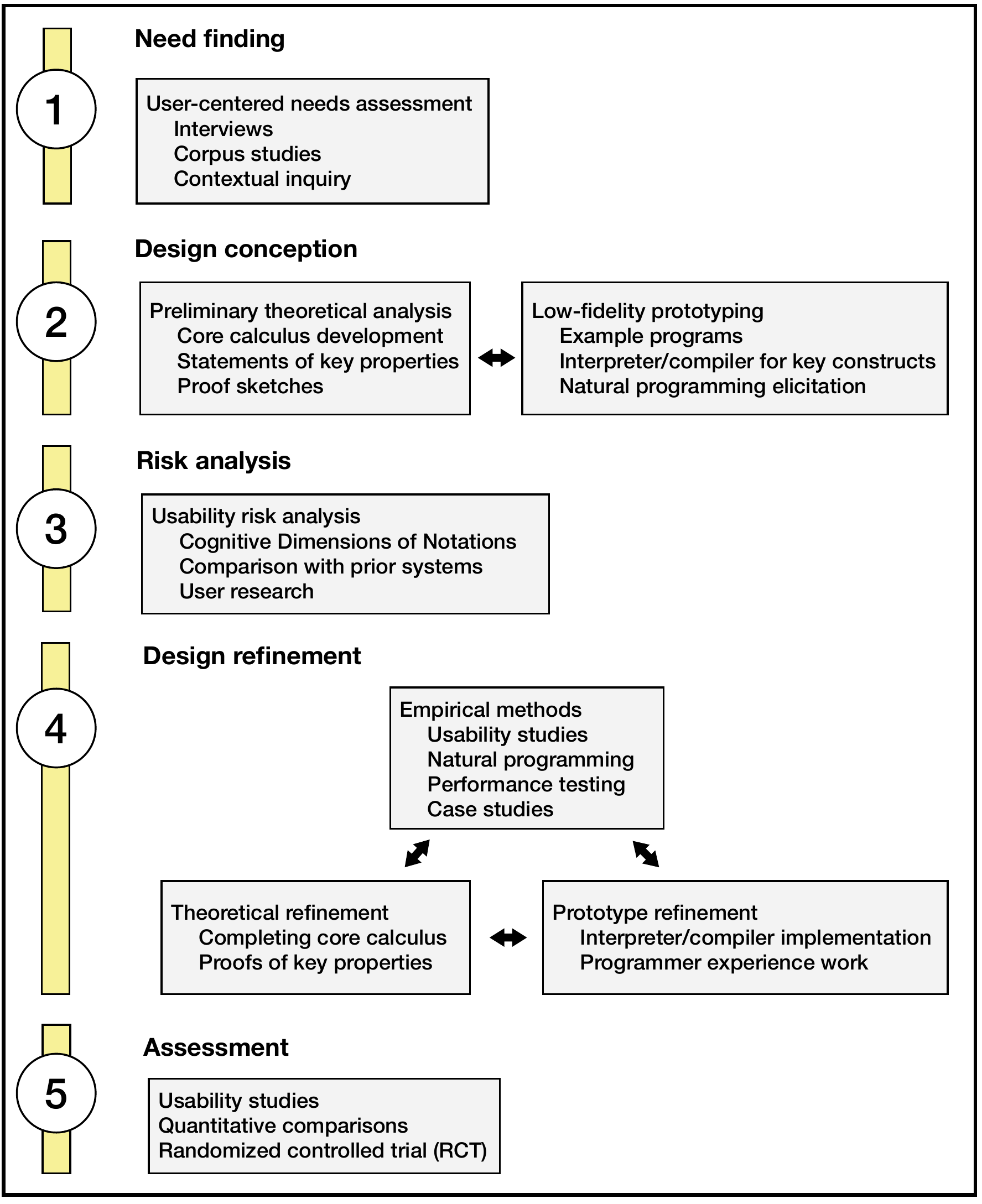}
\caption{The phases of the PLIERS process, showing activities conducted in each phase. Designers can return to previous phases if evaluation identifies opportunities for improvement.}
\label{PLIERS-method}
\end{figure}

PLIERS is summarized in \cref{PLIERS-method}. User-centered design methodology~\cite{Gulliksen2003:Key} seeks to leverage data from users to improve the design of systems. PLIERS is a specialization of user-centered design for programming languages to enable designers to incorporate ideas from user studies as well as from the theory of programming languages. PLIERS consists of five phases: need finding, design conception, risk analysis, design refinement, and assessment. In each phase, the designer seeks and leverages input from or about the users that the designer is hoping will use the programming language. If work in any phase calls into question the work done in a previous phase, the designer may return to the previous phase and conduct more work according to the difficulties that were identified.

\begin{description}[listparindent=\parindent]
	\item[Need finding:] The process begins by assessing the user's needs. What programming problems does the user have, and how might a new programming language help the user achieve their programming goals? Some have advocated that language designers should design languages for their own use~\cite{Graham}. In contrast, PLIERS uses user-centered methods, such as corpus study, interview, or contextual inquiry to understand the target audience and what their needs are. The designer chooses which particular user-centered methods to use according to the available resources and the goals of the design project. These user needs may be stated as hypotheses regarding what kinds of languages might benefit users, what \textit{benefit} means to those users, and how those benefits might be assessed.
	\item[Design conception:] After assessing users' needs, the designer must conceive of initial language ideas that might satisfy those needs. As in other design situations, this process often requires significant creativity. The designer iterates between two kinds of work: theoretical work, in which the designer develops a theoretical foundation for the programming language (a \textit{core calculus}), and prototyping work, in which the programmer directly works on the language that programmers will see (the \textit{surface language}). The result of this work is expressed as a low-fidelity prototype, such as a corpus of code samples (to demonstrate the surface syntax, by which users will write and edit programs) and a core calculus.
	\item[Risk analysis:] In the risk analysis phase, the designer assesses and prioritizes usability risks in the proposed design. This step leverages theoretical models of cognition and usability inspection techniques to identify the aspects of the language design that are most likely to present difficulties to users. For example, cognitive dimensions of notations~\cite{green1996usability} can help identify usability risks that are worthy of further study. 
	
	User research might be needed in this phase to better understand the target audience. For example, if the designer is considering an approach that requires particular skills, then risk analysis might include assessing to what extent the target audience has those skills or whether those skills can be taught in an acceptable amount of time. If the language targets professionals, substantial training may be acceptable. If the language targets end-user programmers, the designer may want to limit the training that would be required.
	
	The prototype design likely leverages some elements of existing languages, while creating new features that may be unfamiliar to users and have unknown usability characteristics. These new features may be particularly worth evaluating. Each risk corresponds to a design or research question, and provides an opportunity for learning more about how to make the programming language as usable as possible.
	\item[Design refinement:] Using empirical methods, the designer assesses the usability risks identified in the prior phase. Then, the designer refines the prototype, successively increasing prototype fidelity as usability risks are addressed. The designer also considers other language requirements, such as \textit{expressiveness} (can the language be used to write a particular kind of program?) and \textit{performance} (does the program, when run, meet the designer's performance goals?). Then, the results are used to revise the theoretical model and the prototype. 
	By using theory, the designer can ensure that any changes retain any formal guarantees that the language promises. Alternatively, the designer may choose to guarantee different properties in order to allow the desired modifications.
	
	Because the theoretical model and prototype are related, changes in one frequently lead to changes in the other. Eventually, the theoretical analysis will include proofs of key properties, and the prototype will be high-fidelity, typically including IDE support and a compiler or interpreter.
	\item[Assessment:] A summative usability study can assess whether the final design has achieved the designers' usability objectives. In contrast to the usability studies in the previous phase, which assess specific aspects of the language design, a summative study is intended to assess programmers' abilities to complete realistic programming tasks.

	In this paper, we focus on usability-related objectives, but the designer may want to conduct performance testing as well. For performance evaluation, we refer readers to the SIGPLAN empirical evaluation checklist~\cite{sigplan-checklist}. 
\end{description}

In developing the PLIERS process, we identified a collection of adaptations to traditional HCI methods, which helped us obtain useful information, primarily in the \textit{design refinement} and \textit{assessment} phases. Here, we describe some of the key ways in which we modified existing methods, which are further discussed in \cref{challenges}. The methods we have found useful are summarized in \cref{PLIERS-methods}.

\begin{description}[listparindent=\parindent]
	\item[Back-porting design questions to existing languages:] To study the usability of a design decision in isolation, we start from an existing language with which participants would already be familiar. For example, rather than asking participants in our early formative studies to learn a whole programming language, we told them that we were adding certain features to Java, and then asked them to complete programming tasks in the Java variant. This substantially reduced the training time and allowed us to reason that any confusion was likely related to the new features, since our participants were already familiar with Java.
	
	Selecting a target language for back-porting depends on several factors:
	\begin{enumerate}[leftmargin=2\parindent]
	    \item Availability of participants skilled in the target language
	    \item High-level similarity between the novel language and the target language (e.g., both object-oriented, both functional, etc.)
	    \item For high-fidelity prototypes: development cost of modifications to target language
	\end{enumerate}
	
	We also favored high-level design decisions that allowed us to attract participants who had relevant background. If we had tried to teach participants a completely novel language paradigm even though the basic assumptions of the language paradigm were not the targets of our research, we would have needed to try to distinguish the \textit{relevant} mistakes from all the novice-level mistakes that the new programmers would be likely to make.
	\item[Wizard of Oz:] Implementing a programming language is expensive. Rather than implementing a full compiler for each language variant we wanted to test, we adapted the Wizard of Oz technique~\cite{dahlback1993wizard}. In a classic Wizard of Oz study, an experimenter pretends a system is working by remote-controlling it in order to obtain insights about potential designs without having to actually build the system.
	
	In early Obsidian studies, we gave participants a text editor, documentation, and programming tasks to do. Then, an experimenter verbally simulated compiler errors. Like a modern IDE, the experimenter could interject with errors, and could provide error messages when participants asked whether the compiler would emit any errors on their current code. For example, the experimenter might say ``Suppose your compiler indicated that there was asset loss that occurred on line 42.'' If the error messages were unclear, the experimenter could revise them with more detail, helping us understand how to write clear error messages for the compiler. The technique we developed allowed efficient iteration on our design ideas, since design changes only required updating the documentation, not a potentially complex implementation. Unlike in a traditional Wizard of Oz study, participants were aware that the feedback was being provided manually, but we observed that this did not present an obstacle to the effectiveness of the technique.
	\item[Multi-part tutorials:] In our studies, we needed to teach participants a new programming language in a consistent and efficient way. A traditional course would not be effective, since we could not recruit our participants into a semester-long course. Instead, we developed multi-part language tutorials. By breaking the tutorial into sections (each about ten minutes long), including practice problems for participants to do, and having an experimenter available to answer questions, we were able to convey the knowledge we needed in a relatively short period of time. Our longest tutorial, for example, took participants an average of 1 hour, 35 minutes.

\end{description}

Next, we show how we applied PLIERS when designing Glacier (\cref{sec:glacier}) and Obsidian (\cref{sec:obsidian-formative,Obsidian-summative}). Then we discuss how our adaptations to traditional HCI methods helped address some of the challenges of using HCI methods on programming languages for software engineers (\cref{challenges}).

\begin{table}
\begin{tabular}{p{2.5cm}p{3.1cm}p{7cm}}
\toprule
\textbf{Design Task} & \parbox[t]{3cm}{\textbf{Method or \\Component}} & \textbf{Details} \\
\midrule
\multirow{2}{*}{\parbox[t]{2.5cm}{\raggedright Need finding and hypothesis generation}} & Interviews & 
	\begin{itemize}[leftmargin=8pt, nosep, before=\compress]
		\item Interview practitioners (e.g., software engineers) to identify problems and formulate hypotheses 
	\end{itemize}\\
	& Corpus analysis & 
		\begin{itemize}[leftmargin=8pt, nosep, before=\compress]
			\item Analyze corpora of applications and bugs to identify common goals and obstacles 
	\end{itemize}\\
\parbox[t]{2.5cm}{Formative design evaluation} & Natural programming elicitation &
	\begin{itemize}[leftmargin=8pt, nosep, before=\compress]
		\item Ask participants to do programming problems without giving them syntax or identifiers in order to help design a syntax and vocabulary that matches their expectations
	\end{itemize}\\
	& Programming tasks & 
		\begin{itemize}[leftmargin=8pt, nosep, before=\compress]
			\item Back-port design components to an existing language to isolate variables of interest
			\item Break larger tasks into subtasks to constrain unproductive exploration
			\item Include a range of task difficulties to obtain data from both more- and less- successful participants
			\item Use low-fidelity prototypes to obtain early feedback on designs
			\item Use Wizard of Oz to enable studies of incomplete prototypes
		\end{itemize}\\
    Summative design evaluation & Usability studies & 
	\begin{itemize}[leftmargin=8pt, nosep, before=\compress]
		\item Assess what barriers users face when attempting to complete relevant programming tasks
	\end{itemize}\\
	& Randomized controlled trials (RCTs) & 
	\begin{itemize}[leftmargin=8pt, nosep, before=\compress]
		\item Compare task times and success rates between different languages
	\end{itemize}\\
\bottomrule
\end{tabular}
\caption{A summary of user-centered methods that we have found useful for studies in PLIERS.}
\label{PLIERS-methods}
\end{table}

\subsection{Evaluating PLIERS}
\label{sec:evaluating-PLIERS}
To evaluate PLIERS, one might like to teach PLIERS to a collection of programming language designers and conduct a qualitative study regarding the insights the designers obtained. Better yet, one might like to recruit language designers and assign them to either use or not use PLIERS to design a language in a domain, and then conduct usability studies of the resulting languages. Unfortunately, these approaches are impractical: language design and implementation is typically a long process, taking months or years, and language designers cannot be recruited for such studies.

Another approach might be to teach PLIERS to designers who have recently completed their designs and observe what changes the designers make as a result. However, a main benefit of PLIERS is that it provides a framework for the entire language creation process; many of the methods can be applied to incomplete prototypes or design concepts. Such an analysis, though useful, would only obtain insight on some of the components of PLIERS.

Because of these practical considerations, we evaluated PLIERS by using it ourselves to create Glacier and Obsidian. In the process, we observed how the approach helped us create and iterate on the language designs. This approach has significant limitations. Our evaluation does not show that other designers can use the process effectively, that it works on a wide variety of different programming languages, that languages produced with the method are necessarily superior to languages produced without the method, or even that the process does not make languages \textit{worse}. However, using PLIERS ourselves was a necessary part of developing the process; in this paper, we leverage our experience creating PLIERS in the hope that others may benefit from it and iterate on its component methods. As this is the first work of which we are aware that integrates HCI methods into the design of languages for professional software engineers, we view this is the first step toward creating a design process that will aid designers in making professional-level languages more usable.

Venable et al. provide a framework for evaluating design methodologies~\cite{Venable2012:Comprehensive}. In that framework, our approach to evaluating PLIERS would be considered \textit{ex ante} (formative, evaluating the design method before it is complete). The ex ante approach was appropriate for PLIERS since the work was conducted in large part to create and iterate on PLIERS. The evaluation included some aspects of \textit{naturalistic} evaluations (it was evaluated in the context of real language design projects) and some aspects of \textit{artificial} evaluations (the designers of the method used the method instead of third parties). This choice was driven by the impracticality of recruiting programming language designers other than ourselves to use our process over the long period of time required to conduct a language design and implementation project.

Blandford and Green have acknowledged the lack of an established path to acceptance for new methods and the difficulty of conducting rigorous evaluations of design methods~\cite{Blandford2008:Methodological}. We regard our work as the first step of many in evaluating PLIERS.

\section{PLIERS for Glacier}
\label{sec:glacier}

\textit{Formative} studies are conducted before a system is built, or in the process of building the system, to learn about the target users or problem space, or to refine a design. \textit{Summative} studies are conducted after an artifact is created to evaluate whether the designers have achieved their design goals. In Glacier, we conducted studies of both types. Before we developed Glacier, we conducted formative studies (\cref{sec:glacier-formative}) to develop a hypothesis regarding which particular kind of immutability might have direct benefits to programmers. We developed IGJ-T based on this hypothesis, but our risk analysis suggested that the flexibility of IGJ-T might result in usability barriers. We conducted a usability study of IGJ-T, which confirmed our hypothesis. Then, we refined our design to create Glacier, and conducted a summative study to answer two usability questions: first, can people easily specify immutability with Glacier, and if so, is it easier with Glacier than it is with Java? Second, does providing compile-time enforcement of immutability prevent bugs that would likely be inserted otherwise? \Cref{sec:glacier-summative} shows that the answers are in the affirmative.

\subsection{Formative studies}
\label{sec:glacier-formative}
We used the \textit{Cognitive Dimensions of Notations} framework~\cite{green1996usability} to reason about some of the design choices. For example, including features that provided weaker guarantees than programmers actually needed could be \textit{error-prone} if those features could be easily confused with stronger ones. Likewise, the inverse is error-prone too: if a programmer applied a weaker specification than could actually be applied, this could lead to undesirable tradeoffs. For example, if an interface is annotated to return a read-only object (indicating that the object potentially could be mutated through \textit{other} references), the programmer might add locks to ensure safety in a concurrent context. But if the object is actually \textit{immutable} (that is, \textit{no} reference could be used to mutate the object), then the locks would be unnecessary and reduce performance. More details about our Cognitive Dimensions analysis appear in our earlier paper~\cite{Coblenz2016:Exploring}.

Although the Cognitive Dimensions analysis was lightweight, it did not answer some of our higher-level design questions. Cognitive Dimensions provides a vocabulary for discussing and analyzing tradeoffs, but it does not provide ground truth regarding how usable particular approaches will be for people. In order to narrow the space of possible language designs, we conducted semi-structured interviews with eight software engineers who were working on large software projects at several organizations. Our participants had an average of fifteen years of experience, with a minimum of seven years, and had worked on projects with millions of lines of code and hundreds of people.

In order to both obtain unbiased data on problems with mutability in general as well as to obtain feedback on concrete language designs, we carefully ordered the interview questions. First we asked general questions, such as ``How do you make sure that state in running programs remains valid?'' We got wide-ranging answers, including ones such as ``We've essentially done away with mutability to avoid security and concurrency problems'' as well as recommendations for regular use of testing and assertions. Afterward, we asked about existing language features, such as \code{const} and \code{final} and their use. Then we asked about specific related areas, including concurrency and security. Finally, we asked about our own language design ideas, including immutable classes. The full set of interview questions is included in a previous paper~\cite{Coblenz2016:Exploring-extended}.

Our interview participants said that bugs in which state changes when it is not supposed to are frequent. They also described how the language features they had available did not provide guarantees that were sufficient for their purposes. For example, when reusing existing code, participants could not typically tell whether the code was thread-safe, so they had to assume that it was not. If a component came with an appropriate compiler-checked immutability specification, then they could be confident of safety, but languages did not provide such a feature.  We concluded that \textit{transitive} immutability provided the strong safety properties that our interview participants requested: a transitively immutable object can be shared safely among threads without locks.

An interesting observation that came out of the interview studies is that typically, for a given class, either all instances are mutable or all instances are immutable. In contrast, some prior systems, such as IGJ~\cite{IGJ}, supported immutability at the object level of granularity (\textit{object immutability}). We evaluated an initial prototype, IGJ-T, that extended IGJ with transitivity. We found that participants had great difficulty managing the complexity, which was in part because IGJ's syntax focused on object immutability, not class immutability. We reasoned that if we designed our system to support \textit{class immutability} only, our system would be simpler and therefore likely easier to use without sacrificing much expressiveness. This motivated our new tool, Glacier, which was centered around \textit{transitive class immutability}. 

\Cref{fig:glacier-design-alternatives} shows several design alternatives. \#1 shows how in IGJ, immutability is a property of references, not of classes. Compared with IGJ, IGJ-T (\#2) also enforces that fields of classes for which there are immutable instances must have all-immutable fields (enforcing transitivity). In Glacier (\#3), immutability is a property of classes, not of references. Variant \#4 explores a possible extension of Glacier, in which the compiler can automatically derive immutable versions of mutable classes. We elected not to pursue that approach, since our interview had indicated that most classes are used either in an immutable or a mutable way, but not both.

\begin{figure}
\raggedright
\textbf{1: IGJ. Immutability is a property of references. Immutability is not transitive.}
\begin{lstlisting}[numbers=left, framexleftmargin=0em, xleftmargin=2em, language=Java]
public class Game  {
  // No need for 'outcome' to be immutable, since immutability is not transitive.
  private Outcome outcome; 

  // Constructor returns a reference to an immutable object.
  @Immutable Game (Outcome outcome) {
    this.outcome = outcome;
  }
}
	
\end{lstlisting}

\textbf{2: IGJ-T. Immutability is a property of references. Immutability is transitive.}
\begin{lstlisting}[numbers=left, framexleftmargin=0em, xleftmargin=2em, language=Java]
public class Game {
  // @Immutable is required when declaring 'outcome' because there is  
  // at least one constructor that returns an @Immutable reference.
  private @Immutable Outcome outcome; 

  // Constructor returns a reference to an immutable object.
  @Immutable Game (@Immutable Outcome outcome) {
    this.outcome = outcome;
  }
}
\end{lstlisting}

\textbf{3: Glacier. Immutability is a property of classes. Immutability is transitive.}
\begin{lstlisting}[numbers=left, framexleftmargin=0em, xleftmargin=2em, language=Java]
@Immutable public class Game {
  private Outcome outcome; // OK because Outcome class was declared @Immutable

  // Every instance of Game is immutable, so no need to specify immutability.
  Game (Outcome outcome) {
    this.outcome = outcome;
  }
}
\end{lstlisting}

\textbf{4: An variant of Glacier, in which the compiler can synthesize immutable subsets of mutable classes.}
\begin{lstlisting}[numbers=left, framexleftmargin=0em, xleftmargin=2em, language=Java]
// Assume Outcome is mutable, but the compiler can synthesize an
// immutable version by leaving out the mutating methods
@Immutable public class Game {
  Outcome outcome; // error: Outcome is mutable
  @Immutable private Outcome immutOutcome; // OK: use immutable subset
  
  Game (Outcome outcome) {
    // Assignment is always permitted in the constructor
    this.outcome = outcome;
  }
  
  void test () {
    this.immutOutcome.setOutcome(WON); // compile error: no such method
    this.outcome = ... // compile error because Game is @Immutable
  }
}
\end{lstlisting}
\caption{Alternatives we considered for immutability systems.}
\label{fig:glacier-design-alternatives}
\end{figure}

\textit{Triangulation}~\cite{Given2008:Sage}, in which a designer combines results of multiple qualitative studies, was a key aspect of the design process. We sought designs that resembled the approaches participants proposed via natural programming and which also enabled participants to complete programming tasks as effectively as possible in our task-based studies. We also leveraged real-world evidence of security vulnerabilities to motivate our safety objectives. At the same time, we were guided by the theory of programming languages, which we used to ensure that our language would provide the guarantees that our design intended to achieve.

\subsection{Summative studies}
\label{sec:glacier-summative}
In addition to doing two case studies to evaluate \textit{expressiveness}~\cite{Coblenz19:Obsidian-arxiv}, we conducted a lab study to answer two research questions relating to our \textit{comparison} question in \S \ref{introduction} (``How can we compare multiple language designs to see which are more effective for users?''):

\begin{enumerate}
	\item Can participants express immutability more successfully in Glacier than with Java's \code{final} keyword?
	\item Without Glacier (using only standard Java), are programmers likely to accidentally insert the kinds of bugs that Glacier detects?
\end{enumerate}

We recruited 20 Java programmers. We randomly assigned participants to use either Glacier or \code{final}, and we gave participants a tutorial in their given tool (two pages for Glacier, three pages for \code{final}). In addition, we gave the \code{final} participants a page from \textit{Effective Java}~\cite{bloch2008effective} explaining how to safely enforce immutability with \code{final}. Then, to address the first question, we asked them to change one class in each of two small projects (\code{Person} and \code{Accounts}) so that those classes were immutable.  None of the participants in the \code{final} condition were able to do their task successfully because it was too easy to forget to do one of the changes required, such as copying mutable inputs to constructors. Of the 20 Glacier tasks attempted, participants completed 19 correctly. The difference between success rates across conditions is significant in each task (Person: $p \approx 1.08 \times 10^{-5}$; Accounts: $p \approx 1.2 \times 10^{-4}$, Fisher's exact test). Among users who said they were done with both tasks, \code{final} users completed both annotation tasks in an average of 14 minutes. The average Glacier user completed the task in 11 minutes. This difference is significant with $p \approx 0.1$ (Wilcoxon rank sum test).

To address the second question, we asked our participants to do two programming tasks (\code{FileRequest.execute} and \code{HashBucket.put}) on two small immutable classes. Although we did not verbally tell them that the classes were immutable, the classes were adapted from real-world code, and the participants had just completed the tasks above pertaining to immutability. In the Glacier condition, each task was completed successfully by seven participants; of course, no one accidentally mutated immutable state because Glacier disallowed it. In the \code{final} condition, however, four of eight participants who finished the first task completed it successfully, and only three of ten participants who finished the second task completed it successfully. Fisher's exact test indicates these differences are significant at $p \approx .077$ for \code{FileRequest.execute} and $p \approx .0098$ for \code{HashBucket.put}. These results are summarized in Table \ref{Glacier-results}.

\begin{table}
\begin{tabular}{p{8.5 cm} l l}
\toprule
&	\textbf{final} &	\textbf{Glacier}\\
\midrule
Correctly enforced immutability in \code{Person} & 0/10 & 10/10 \\
Correctly enforced immutability in \code{Accounts} & 0/10 & 9/10 \\
Average time for enforcing immutability across both tasks & 14 min. & 11 min. \\
\code{FileRequest.execute()} \\
\hspace{.5 cm} Tasks without security vulnerabilities & 4/8 	& 7/7 \\
\hspace{.5 cm} Average time spent (among participants who finished) & 14 min. & 14 min. \\
\code{HashBucket.put()} \\
\hspace{.5 cm} Tasks without bugs	& 3/10 	& 7/7  \\
\hspace{.5 cm} Average time spent (among participants who finished) & 18 min. & 14 min. \\
\bottomrule
\end{tabular}
\caption{Summary of summative study results for Glacier}
\label{Glacier-results}
\end{table}

These results seems surprising: although we tried to design the experiment to be as unbiased as possible, the programming tasks were actually biased toward the control condition (\code{final}) in that participants had just been trained to consider immutability. One would expect, then, that in a real-world scenario, programmers might perform even more poorly. The success of this study teaches us some lessons about study design:

\begin{description}[listparindent=\parindent]
	\item[Errors are frequent:] Programming is so difficult that participants are likely to make errors very frequently, consistent with the \textit{variance} challenge. Some of these errors will be ones that the experiment designer was hoping to observe, but many of them will be irrelevant. To mitigate this, ensure that participants are given enough time to correct their mistakes and actually finish tasks. Any task can be made difficult enough that participants will not finish it within a given amount of time, so it is imperative to pilot studies to identify an appropriate amount of time to allocate. A corollary, however, is that it is not difficult to run a study in which at least some participants make a particular error of interest.
	\item[Training may have limited effectiveness:] In the Glacier study, participants in the \code{final} condition were unable to correctly follow the advice we gave them on using \code{final}, despite having both documentation and a relevant page from a textbook. This is an example of the \textit{training} challenge. Likewise, in the second part of the study, they frequently failed to identify that the class they were working on was immutable, despite having just spent time studying immutability. This leads to two lessons. First, attempts to change programmer behavior with only training materials, without actually modifying the tools programmers use, may have limited effectiveness. Second, in retrospect, considering our observations teaching Obsidian (\cref{Obsidian-summative}), the training might have been more effective if we had required participants to do exercises with the new knowledge rather than assuming that they could read documentation and follow directions.
	\item[Bias toward control condition may be acceptable:] The effect of tool-based interventions can be so dramatic that it is much better to potentially bias the study toward the control condition than it is to introduce threats to validity or make the study harder to execute. For example, we might have seen even more errors in the second part of the experiment if we had not previously trained the participants in immutability, but that would have required either getting a second set of participants or changing the task order. If we had conducted the programming tasks first, then those tasks could have served to bias the \textit{other} set of tasks. Because recruiting participants is challenging (\textit{recruitment} challenge), we opted to do both the immutability-specification and the immutable-class-programming tasks with one set of participants.
\end{description}

Replication materials for the study can be found online~\cite{Coblenz2020:Glacier}.

\section{Formative studies for Obsidian}
\label{sec:obsidian-formative}

Obsidian was a much larger language design project than Glacier was, so there were many more design questions to address. We started with an analysis of proposals for blockchain applications and by studying bugs that had significant implications on existing blockchain platforms. We identified hypotheses for technical approaches that would provide safety properties to address the key problems we identified. Then, we conducted formative studies to explore whether we could design a language to be as usable as possible while still achieving our safety goals. This section focuses on how those formative studies informed the language design. In the next section (\cref{Obsidian-summative}), we show how we used summative studies to assess to what extent we had achieved our usability objectives.

We conducted formative studies, which consisted of traditional programming tasks as well as natural programming tasks. In various studies, we used our adapted Wizard of Oz method, back-ported our design to the context of Java to isolate our research questions, provided participants with tasks spanning a range of difficulties, and divided larger tasks into subtasks. The formative studies spanned a variety of central design questions, which corresponded to usability risks we identified in our early designs:

\begin{itemize}
 \item How should lexical scoping should work for states? (\cref{basic-design})
 \item How should the language help programmers manage fields that enter and exit scope when state changes? (\cref{fields-in-states})
 \item How could programmers use \textit{permissions} to express different kinds of references to objects? (\cref{permissions-usability})
 \item How should the language represent the relationship between typestate and permissions? (\cref{comparing-typestate-ownership})
\end{itemize}

Due to the complexity of each design problem, we refer readers to the individual sections for details of the findings.

In this section, we describe studies that helped us identify a suitable design and iterate on our initial design ideas for Obsidian. For each study, we identify our research questions, methodology, and results. We started by assuming that we would use \textit{typestate} to achieve the desired safety guarantees but that expressing typestate in a usable way would require substantial iteration with users. The latter assumption was based on past work on typestate systems, such as Plural~\cite{Bierhoff:2008:PCP:1370175.1370213} and Plaid~\cite{sunshine2011first}, which researchers had found were difficult for users to use. All of the studies were approved by our IRB. Because we needed skilled programmers, we recruited from appropriate academic programs, by posting flyers, and by contacting our acquaintances. Except where noted below, we paid participants \$10/hour for participating. Materials used in the studies can be found in the replication package~\cite{Coblenz2020:ObsidianReplication}.

Although \cref{tiny-vending-machine} uses the final version of the language, because the formative studies were done earlier, they use code from earlier versions of the language. In this way, the reader can see how we changed the language as a result of the user studies. For example, \cref{decl-approaches} shows different approaches that we considered using for declaring local variables.

\Cref{table:all-participants} summarizes all the Obsidian user studies that are described in this paper.

\begin{table}
\begin{tabular}{p{7cm} l p{3.5cm}}
\toprule
\textbf{Topic} & \textbf{Participants} & \textbf{Methods} \\
\midrule
Basic design of typestate (\cref{basic-design}) & P1--P12 & Natural programming \\
Fields in states (\cref{fields-in-states}) & P20, P66--P68 & Natural programming; Usability study \\ 
Permissions (\cref{permissions-usability}) & P14--P19 & Natural programming; Usability study \\
Typestate and ownership approaches (\cref{comparing-typestate-ownership}) & P21--P25 & Usability study \\
Summative usability study pilots & P26--P34 & Usability study \\ 
Summative usability study (\cref{Obsidian-summative}) & P35--P40 & Usability study \\
RCT pilots & P41--P44 & RCT \\
RCT (\cref{sec:Obsidian-RCT}) & P45--P65 & RCT \\
\bottomrule
\end{tabular}
\caption{Summary of all Obsidian user studies described in this paper and their participants. Participant P13 was in pilot studies. Participant numbering is consistent with prior distributed drafts, e.g.,~\cite{Coblenz2019:Usability}.}
\label{table:all-participants}
\end{table}

\subsection{Basic design of typestate}
\label{basic-design}
In order to minimize assumptions regarding how Obsidian should best represent typestate, we conducted a \textit{natural programming} study, which we described in an earlier paper~\cite{barnaby}. We focus here on two of the research questions we had:

\begin{itemize}
    \item Are states a natural way of approaching the challenges that arise in blockchain programming?
    \item  Which (if any) of our proposed ways of presenting states and state transitions is most understandable and usable by programmers?
\end{itemize}

These are examples of the \textit{naturalness} research question in \S \ref{introduction} (``How can we obtain insights as to what language designs will be natural for programmers?''). We gave participants a description of a voter registration system, in which we would investigate to what extent state machines were a natural way to write smart contracts. The first task used a natural programming methodology: we asked participants to implement the system using pseudocode using any language features they wanted to solve the problem. Next, we gave participants a state diagram that modeled the system, and asked them to modify their pseudocode to include states and state transitions. In the third task, we gave participants a two-page Obsidian tutorial that described state blocks. However, the tutorial omitted any description of how state transitions should be written; we gave participants an Obsidian program that implemented the voter registration system but which omitted state transitions. We asked participants to fill in the missing transitions by inventing their own syntax to do so. In the fourth task, we gave participants three options for the syntax and semantics of state transitions and asked them to use each option once in an example program we provided. The first and third options are explained in of \cref{fig:nesting-options}. An additional option involved a constructor in each state that would be invoked on transitions to that state and a rule that no code could follow a state transition.

\begin{figure}
    \centering
    \begin{minipage}[t]{.45\textwidth}
\begin{lstlisting}[numbers=left, xleftmargin=2em]
contract C { 
  state Start {
    transaction t(int x) {
      ->S1{x1 = x};
      toS2(); 
    }
  }
  
  state S1 { 
    int x1;
    transaction toS2() {
      ->S2{x2 = x1};
    } 
  }
  
  state S2 { 
    int x2;
  } 
}
\end{lstlisting}
\vspace{3em}
(a) Option 1. The dynamic state (not the lexical structure) governs which transactions may be called. For example, line 5 calls \code{toS2()} even though \code{t} is lexically in the \code{Start} state and \code{toS2()} is defined in \code{S1}.
    \end{minipage}
    \vspace{5ex}
    \hspace{2em}
    \begin{minipage}[t]{.45\textwidth}
\begin{lstlisting}[numbers=left, xleftmargin=2em]
contract C { 
  state Start {
    transaction t(int x) {
      ->S1({x1 = x})
      if in S1 {
        ->S2({x2 = x1})
      } 
      if in S2 {
        ...
      }
    }
  }
  
  state S1 {
    int x1;
  }
  
  state S2 {
    int x2;
  }
}
\end{lstlisting}
\vspace{1em}
(b) Option 3. Fields can only be referenced by code that \textit{lexically} is in the state in which those fields are defined. An \code{if in} block can be used to enclose code that must reference fields of other states, as in line 6.
    \end{minipage}
    \caption{Two of the options given to participants in the \textit{basic design} study.}
    \label{fig:nesting-options}
\end{figure}

Finally, in a fifth task, we asked participants to select one of the three options and use it to complete the voter registration program they started earlier.

We recruited a convenience sample of seven participants, most of whom were computer science undergraduates. Each participant was given a description of a program to implement and one hour to complete the implementation. We paid participants \$10/hour for their time.

Only two participants invented syntax denoting states and state transitions; the rest used a conventional approach, such as an enumerated type. However, many of the approaches the remaining five participants used were unsafe, helping to justify using typestate to improve safety. For example, creating separate lists for unregistered and unregistered citizens results in the possibility of citizens appearing on both lists. 

We asked six of the participants to modify their pseudocode to use states. Two created explicit state blocks with states and variables nested inside. The remaining four either maintained global state for each citizen, or gave each citizen a state field, or created empty, immutable states at the top of the program. Although the instructions forbid allowing duplicate registrations, several participants did not check for existing registrations before processing applications.

Regarding the syntactic choices we offered in the third task, three participants preferred state constructors (part (a) in \cref{fig:nesting-options}), one preferred nested state blocks (part (b) in \cref{fig:nesting-options}), and the remaining three either did not indicate a preference or did not complete this task.

Although most of this study focused on participant \textit{behavior}, we took the opportunity to also ask participants for their syntactic \textit{preferences}. Five participants preferred a syntax where all the actions of a state must be lexically encapsulated in that state, as in the first alternative in \cref{lexical-encapsulation}. Likewise, P4 felt it should not be permitted to call transactions from one state while lexically in another state: ``I'm calling \code{S1}'s transaction from code for \code{Start}.''

\begin{figure}
\raggedright
\textbf{Transaction lexically \textit{nested inside} state declaration:}
\begin{lstlisting}[numbers=left, framexleftmargin=0em, xleftmargin=2em]
contract Wallet {
   state Empty;
   state Full {
      int balance;
      
      // spend() is nested inside the declaration of the state it belongs to.
      transaction spend(Wallet@Full this) {
        // use 'balance'...
      } 
   }
}
\end{lstlisting}
\textbf{Transaction lexically \textit{outside} state declaration:}
\begin{lstlisting}[numbers=left, framexleftmargin=0em, xleftmargin=2em]
contract Wallet {
   state Empty;
   state Full {
      int balance;
   }
   // spend() is not nested in Full, even though it can only be called in Full state
   transaction spend(Wallet@Full this) {
      // use 'balance'...
   } 
}
\end{lstlisting}
\textbf{Transitions when inside a state could be confusing:}
\begin{lstlisting}[numbers=left, framexleftmargin=0em, xleftmargin=2em]
contract Wallet {
   state Empty {
      transaction fill(int amount) {
        ->Full(balance = amount);
        // Now balance should be in scope, since new state is Full
      }
   }
   
   state Full {
      int balance;
      // ...
   }
}
\end{lstlisting}

\caption{Although participants preferred to have transactions nested inside state declarations (the first alternative), this desire conflicted with the need for transactions only reference fields that were in lexical scope.}
\label{lexical-encapsulation}
\end{figure}

This preference led to a conflict in the design. We found through work with example programs after the study that Obsidian needed to support transactions that could be executed in several different states. For example, in the third example of \cref{lexical-encapsulation}, line 5 may reference \code{balance}, even though line 5 is lexically enclosed in the \code{Empty} state, in which \code{balance} is not in scope. This represents a conflict between a syntactic preference and an expressivity concern.

Another difficulty with the constructor-based approach is that states might be entered for a variety of different reasons, requiring different code to run after the transitions. This makes the constructor-based approach likely too inflexible. These challenges underscore the importance of sufficient example-based work before conducting user studies; it is easy to design a study that provides plausible options that turn out to not support critical use cases.

This problem led us to run the study described in \cref{fields-in-states}. As a result of that study, we addressed the conflict by requiring that transactions are lexically \textit{outside} of state declarations, like the second example in \cref{lexical-encapsulation}. Future IDE tools could show all transactions that are possible for an object in a given state, even though their declarations are lexically outside that state's declaration.

\subsection{Fields in states}
\label{fields-in-states}
States in contracts can have different sets of fields, so transitioning can cause some fields to exit scope and others to enter scope. For example, in \cref{tiny-vending-machine}, the \code{Full} state has the \code{inventory} field, but the \code{Empty} state has no fields. This study used \textit{natural programming} and \textit{code understanding} methods to investigate how users specify cleanup of old fields and initialization of new fields when invoking state transitions. 


We recruited participants until the data we were obtaining from recent participants duplicated data we had obtained earlier. This led to recruiting four participants. All were Ph.D. students 
studying software engineering.
They had an average of seven years of programming experience (ranging from three to fifteen years) and an average of 1.5 years of Java experience. Two identified as male, and two identified as female. We did not limit their time in completing the tasks. The mean time to completion was of 1 hour, 27 minutes. The times per user are shown in \cref{table:fields-in-states-times}. Due to a miscommunication between co-authors, the participant identifiers are not contiguous, but the experiments occurred sequentially over about a month.

\begin{table}[t]
\begin{center}
\begin{tabular}{ll}
\toprule
\textbf{Participant} & \textbf{Time} \\
\midrule
P20	& 54m \\
P66	& 1h 39m \\
P67	& 1h 35m \\
P68	& 1h 41m \\
\bottomrule
\end{tabular}
\end{center}
\caption{Times spent by the four participants in the Fields in States study.}
\label{table:fields-in-states-times}
\end{table}

In \textbf{Part 1}, we gave participants a state transition diagram for a \code{Wallet} object, which could hold a license and money, and which had four states corresponding to the possible combinations of contents. Participants were also given code partially implementing the \code{Wallet}, with several TODO comments asking participants to invent code to add money to the \code{Wallet}, remove money from the \code{Wallet}, etc. Participants were told that the money and license should be thought of as assets, so they could not be duplicated, used more than once, or lost. The code they were given was in a language similar to Obsidian but which used some keywords that would be more familiar to a Java programmer, such as \code{class} instead of \code{contract}. As such, this was a staged natural programming study, since we progressively gave participants more detail about the language we were designing.

All four participants prepared assets for a state transition before making the state transition (corresponding to option (2) in Part 2 below, \code{S::x = a1; ->S}). Two participants felt they needed to write code to handle failures during the asset preparation stage, which might lead to an improperly initialized state upon transition. One of them suggested a try-catch type wrapper for the asset preparation and transition phases. 

In \textbf{Parts 2} through \textbf{4} of the study, participants were given several options. Then they were asked to implement each of the options within a given partially-implemented transaction. Finally, they were asked for their preferences.

\textbf{Part 2} compared approaches for initializing fields in states during transitions. Options were:
\begin{enumerate}
\item Assets are assigned to fields in the transition, e.g. \code{->S(x = a1)} assigns the value of \code{a1} to field \code{x} of state \code{S}.
\item Assets are assigned to fields \textit{before} the transition, e.g. \code{S::x = a1; ->S}.
\item Assets are assigned to fields \textit{before} the transition, but the fields are in local scope even though the state has not changed yet, e.g. \code{x = a1; ->S}.
\item Assets are assigned to fields \textit{after} the transition, e.g. \code{->S; x = a1}.
\end{enumerate}

The participants successfully used all the approaches, but most of the participants preferred assigning assets to fields \textit{before} the transition with destination state scoping (option 2). Before the study, Obsidian supported only atomic assignment (option 1, shown in \cref{tiny-vending-machine} on line 22). The results of these two parts motivated a language change: Obsidian now also supports option 2.

\textbf{Part 3} presented two options for handling assets when transitioning from a state with an asset to a state without it:
\begin{enumerate}
\item The transition evaluates to a collection containing the old assets, e.g. \code{x = ->S} indicates that \code{x} is assigned the leftover assets after the transition to state \code{S}. If the current state is unknown statically, the contents of the collection are determined dynamically.
\item The transition evaluates to a tuple, e.g. \code{(x = a1) = ->S} indicates that \code{x} will be assigned the asset \code{a1} which is not present in state \code{S}.
\end{enumerate}

There was consistent confusion about which leftover assets are assigned to option 1's collection after a transition. All participants understood the need for both options in certain cases, but would choose the tuple-like collection for more control and explicitness when the use of either approach is acceptable. We would like to implement this approach in the future but so far have not prioritized it, since the existing approach (described in Part 4, option 1), which requires that ownership of assets be surrendered before transitioning, has been effective for participants.

\textbf{Part 4} focused on releasing assets owned by state fields when transitioning to states in which those fields do not exist. In contrast to \textit{part 3}, this approach added the option of releasing assets before the transition. The choices were:
\begin{enumerate}
\item Assets must be released before the transition, e.g. \code{release(a1); ->S}.
\item The transition evaluates to a tuple of assets that are no longer owned, e.g. \code{a1 = ->S}. The tuple is necessary since there may be several asset-owning fields going out of scope, so there would be one element per field.
\end{enumerate}

All the participants understood the options and implemented them without mistakes. Implementing using option 2 (evaluating to a tuple) enables both approaches, so participants were asked to indicate scenarios where one option would be preferred over the other. The participants consistently indicated that assets should be released before a transition if they are no longer needed; otherwise, they should evaluate to a tuple. This helped us prioritize our features, since releasing assets before the transition seemed to suffice.

\subsection{Permissions: a qualitative study}
\label{permissions-usability}
Soundly enforcing typestate requires knowledge about all references to an object, which is afforded by a \textit{permission system}.~\cite{Bierhoff:2007:MTC:1297027.1297050}. Permission systems allow the programmer to express what a particular reference can be used for (and therefore also what it \textit{cannot} be used for). Is there a permission system that users can understand and use effectively (a question of \textit{naturalness})? If so, what can we learn from users about how to design it (a question of how to iterate on designs)? In this work, we conducted the first studies (of which we are aware) in which people other than the designers of the system were asked to use a permission system to restrict references in a programming language. We found that our initial system design was surprisingly difficult to use, and iterated the design until it was more successful.

In order to study permissions while mitigating the \textit{interdependency of features}, \textit{training} and \textit{recruiting} challenges, we extracted the permission system from Obsidian and re-cast it in Java as a set of annotations. We conducted a Wizard of Oz study where participants received documentation on a Java extension and the experimenter gave simulated compiler error messages. This approach minimized training time for participants, minimized implementation cost for ourselves, and allowed us to isolate this design decision from many others that would have otherwise distinguished the language from Java. At this point in the development of Obsidian, we assumed that it would be best to separate the notions of permissions and typestate; this approach was reflected in the study materials but may surprise a reader who has studied Fig. \ref{tiny-vending-machine}, which reflects the final Obsidian version, which combines the two. The training materials explained the annotations: \texttt{@Asset}, which applied to classes; and \texttt{@Owned}, \texttt{@Shared} (\texttt{@Shared} means there are no restrictions because the object has no owner), and \texttt{@ReadOnlyState} (restricting state modification), all of which applied to references.

Our objective was not to obtain as much data as possible about the current design, but rather to identify a forward path through the design process. Relevance of data depends on the fidelity of the prototypes used; because this study was with a relatively low-fidelity prototype, we conducted a relatively small number of trials to help us make key design decisions. As with the prior study, we continued recruiting participants until most of the new data duplicated earlier data and we had identified a concrete plan for continued language revision and evaluation. For this study, this resulted in recruiting six participants (P14--P19).
They had a mean of six years of programming experience (ranging from three to nine years), a mean of one year of professional experience, and a mean of two years of Java experience. All identified as male.

The study included five parts. Since our goal was to identify as many usability problems as possible in each trial, we revised the design and instructions after each participant. This approach (of changing the tasks between participants) is an accepted practice in usability studies in order to obtain the most useful data from the study~\cite{Lazar2010:Research,Dumas1999:Practical}. The first three participants were given 1.5 hours to do the first four parts; the last three were given two hours to fit in a fifth part of the study. An experimenter was available to answer questions.

\textbf{Part 1}. To motivate the need for language features to prevent bugs, we gave participants a 163-line Java medical records system and asked the first two participants to find a bug in which a patient could refill the prescription more times than specified. The first participant did not find the bug within 30 minutes; the second did so just as time expired. To conserve time, we gave the other participants five minutes to inspect the code and then we explained the problem to them.

We conclude that at least some programmers who use traditional languages would have difficulty detecting the kind of bug that Obsidian prevents. This provides further evidence that if users use Obsidian, the compiler will help them detect bugs that otherwise might go undetected. 

\textbf{Part 2}. We told participants we would prevent the previous bug by distinguishing between two kinds of references. ``Considering an object \textit{o}:
\textit{Kind \#1}: There is only one reference of kind \#1 to \textit{o} at a time. \textit{Kind \#2}: There may be many references of kind \#2 to \textit{o} at a time.''
We asked participants to propose names for the two kinds of references. Note the careful language avoiding bias toward specific vocabulary. Participants' name suggestions included:

\begin{description}
\item[Kind \#1:] KeyReference, UniqueReference, Owned, Singleton reference, Resource handle, @default
\item[Kind \#2:] DuplicateReference, ForeignKeyReference, Borrowed, Flyweight pattern reference, const pointer
\end{description}

The results were too inconsistent to justify an particular choice in the language; all the suggestions were distinct, and some of them were not appropriate in context (\textit{unsound proposals} challenge). Obsidian uses \code{Owned}, which is at least consistent with one suggestion, and \code{Unowned}. 

\textbf{Part 3}. To evaluate the usability of ownership, we gave participants an ownership tutorial and told them we had chosen [\textit{no annotation}, \texttt{@ReadOnly}] (first participant) or [\texttt{@Owned}, \textit{no annotation}] (later participants) as keywords. We asked them to modify the code from Part 1 to fix the bug. We hoped participants would require that \code{Prescriptions} deposited in a \code{Pharmacy} be owned and that the \code{Pharmacy} take ownership; thus, a deposited \code{Prescription} could not be deposited in a second \code{Pharmacy}. Completion times ranged from 3 minutes to 40 minutes (\textit{variance} challenge). Two participants did not finish, one of whom we stopped after 38 minutes to prioritize other tasks.

We were surprised that many of the participants found this task very difficult. We expanded the tutorial to include a practice section for later participants. In general, participants were not prepared to use a type system to address a bug that they thought of in a dynamic way. For example, P16 wrote \texttt{if (@Owned prescription)}, attempting to indicate a dynamic check of ownership. We asked participants who wanted to use dynamic approaches for enforcement to use the language feature instead. P14 commented ``I haven't seen\ldots types that complex in an actual language \dots enforced at compile time.'' 

%


P17 had trouble guessing what the compiler could know, expecting an interprocedural analysis (which would be non-modular). For example, in a case where an owned object was being consumed twice, P17 expected the compiler to give an error on the second \code{spend} invocation. Instead, because the second invocation was inside a helper method, the compiler reported the error on the invocation to the helper method, which took an owned argument and invoked the second \code{spend}. 

P17, P18, and P19 had difficulty determining which variables should be annotated \texttt{@Owned}. In one case, a lookup method took an object to search for, but P17 specified that it should take an owned reference. Then he was stuck after invoking it: ``How can I get the annotation back?'' But  this was impossible except via adding another method, since he had already given away ownership. Likewise P17 was confused by whether accessors should return owned references. Mistakes could be costly. For example, P19 unnecessarily annotated as \texttt{@Owned} a class that was contained in a collection, which caused a problem iterating through the collection. He made the reference to the current list element \texttt{@Owned}, which would require removing each item from the collection when iterating over it in code that was not supposed to modify the container at all.

Parameter-passing and assignment were common points of confusion. P18 asked what happens when passing an \texttt{@Owned} object to a method with an unowned formal parameter (ownership was not passed in this case).  P19 said, ``when I [annotate this constructor type \texttt{@Owned}], I'm not sure if I'm making a variable owned or I'm transferring ownership.'' P17 was surprised that assignment from an owned reference to an unowned-type variable did not transfer ownership. We later addressed this confusion by making assignment always transfer ownership; participants in later studies were generally not confused about which assignments transfer ownership.

From this portion of the study, we came to two general conclusions. First, the semantics of ownership needed to be as explicit and as simple as possible. This likely generalizes to many different kinds of complex language constructs: implicit behavior, although sometimes convenient for experts, can be baffling to novices. When the behavior can be made explicit without making the language inconvenient for experts, that should be done. Second, language design decisions that have structural implications (as is the case for ownership) require substantial high-level training; we refined the training materials in future studies to give more explanation and examples.

\textbf{Part 4} introduced the notion of \textit{assets}. After a tutorial explaining the properties of assets, participants were asked to invent code that could indicate a particular owned reference was \textit{intentionally} going out of scope. Two participants suggested \texttt{@Disown} and  \texttt{free} to abandon owned references; the rest did not have time to answer or had no suggestions. We chose \code{disown} for Obsidian, since \code{free} has additional memory management connotations that are not relevant here.

\textbf{Part 5} introduced typestate, starting with the fourth participant. Participants read 2.5 pages on typestate in Obsidian (as it existed then), including \texttt{@ReadOnlyState}, \texttt{@Shared}, and \texttt{@Borrowed} (which was for temporary ownership transfer in invocations). Ownership was the default, so no \code{@Owned} was needed. The tutorial also explained \texttt{available in} and \texttt{ends in}, which at the time specified state assumptions and guarantees for methods (before we changed to using \code{this} parameters instead, e.g., as on lines 19 and 27 of Figure \ref{tiny-vending-machine}). Then, they were asked to annotate uses of \texttt{Bond} in a 212-line Java program implementing a financial market. They were told to use ownership and state specifications whenever possible.



Consistent with Part 3, some participants were more comfortable with a dynamic perspective on ownership rather than a static one. 
P18 felt that \texttt{ends in} declarations were redundant with the transition code already in the method implementations, but these declarations allow separation of interface and implementation and modular checking. P19 wanted to use borrowing to represent the notion that the \texttt{BondMarket} owns a \texttt{Bond}, but an \texttt{Investor} borrows it for a while. In fact, borrowing was only appropriate for the duration of a method invocation. We later changed the design of the formal parameter syntax to remove the need for \code{@Borrowed}; now, if no ownership change is specified (via the \code{>>} operator), ownership remains unchanged.

P19 required significant prompting by the experimenter to make maximum use of typestate. First, P19 added annotations on methods but not on any variables. After prompting, he added dynamic checks in one place but required prompting to add static typestate specifications. This suggests that tools may be needed to help users obtain the most benefits from the language. On the other hand, P18 specified \texttt{@Asset} on \texttt{Bond} without being asked to do so, explaining ``because it's something important and I don't want to get it out of scope\ldots'' 

Overall, understanding the limitations of the type system and compiler may be an obstacle for some people. Users will need training to reason about what typestate can do, but the observations above motivated language changes that simplified the design without lowering the expressivity or safety. Tools could mitigate the limitations of traditional type systems by providing sophisticated static analyses rather than taking a traditional type checking approach (as Obsidian does), and by providing detailed, explanatory errors.

\subsection{Comparing typestate and ownership approaches}
\label{comparing-typestate-ownership}
We were interested in evaluating a new approach we invented, which was motivated by the confusion we observed in the prior study (in part a question of \textit{naturalness} and in part a challenge of \textit{training}). We invented a new approach: fuse the notions of ownership and typestate in order to simplify the type system, and the next study refined this design. This design has the benefit of eliminating \code{Shared} references that also specify typestate, which would then have to be disallowed to preserve soundness. Thus, the type \texttt{Bond@S} is always implicitly an owned reference for any state \texttt{S}, and users can write any permission instead of \texttt{S}, as in \texttt{Bond@Unowned}.

We were also interested in another usability concern. Consider \textit{Approach 1} in Fig. \ref{decl-approaches}. A reader of line 1 might expect that the type of \texttt{bond} would always be \texttt{Bond@Offered}. In fact, after line 2, the type is \texttt{Bond@Sold} due to the call to \texttt{buy}. A fundamental aspect of Obsidian is that ownership can change, so if a variable declaration includes any ownership information, the variable's ownership status may later be inconsistent with its declaration.

We initially invented two possible approaches to address this problem. One idea involved incorporating types into variable names, shown in \textit{Approach 2}. The annotations pertain to the \textit{current} type rather than the \textit{new} type. The reader would have to look at only the most recent operation to infer the new type of a variable rather than having to potentially read the whole sequence since the declaration. 

\textit{Approach 3} represents another idea: adding static assertions. Line 3 shows a static assertion that \texttt{bond} references an object in state \texttt{Sold}, which serves as documentation. Unlike traditional assertions, however, the compiler checks correctness. The intent is to make it easier for programmers to determine the types of variables.

We conducted studies with participants in the first three conditions. Inspired by observations of those participants, we invented \textit{approach 4}. This approach is like approach 3 except that it \textit{removes} state specifications from local variable declarations. The removal was not part of the original design but was inspired by early results of this study.

\begin{figure}[t]
\raggedright
\textbf{Approach 1: traditional declarations}
\begin{lstlisting}[numbers=left, framexleftmargin=0em, xleftmargin=2em]
Bond@Offered bond = new Bond();
bond.buy(...);
\end{lstlisting}

\textbf{Approach 2: types in variable names}
\begin{lstlisting}[numbers=left, framexleftmargin=0em, xleftmargin=2em]
Bond bond@Offered = new Bond();
bond@Offered.buy(...);
\end{lstlisting}

\textbf{Approach 3: static assertions}
\begin{lstlisting}[numbers=left, framexleftmargin=0em, xleftmargin=2em]
Bond@Offered bond = new Bond();
bond.buy(...);
[bond@Sold];
\end{lstlisting}

\textbf{Approach 4: no states in local variable declarations}
\begin{lstlisting}[numbers=left, framexleftmargin=0em, xleftmargin=2em]
Bond bond = new Bond();
bond.buy(...);
[bond@Sold];
\end{lstlisting}

\caption{Variable declaration approaches}
\label{decl-approaches}
\end{figure}


\subsubsection{Participants}
We required that participants be familiar with Java and we administered a simple Java pre-test. We recruited five students (P21--P25). Based on self-reports, they had an average of about four years of Java experience (ranging from one to ten years) and an average of one year of professional  (paid) software development experience (ranging from zero to three years).

\subsubsection{Procedure}
Participants spent between 1 and 1.5 hours on the study. We used a Qualtrics survey to ask participants a series of questions regarding Obsidian programs, but the study took place in a lab and an experimenter was available to answer questions. The survey both taught aspects of the language and provided an opportunity for assessment. Most of the questions were typical \textit{code understanding} questions, which gave snippets of code and asked whether the compiler would give an error or what the code meant. Rather than assigning participants to conditions randomly and ensuring equal numbers of participants in each condition, we conducted each trial according to the particular questions we wanted to obtain insight on at that time. We assigned P22 to approach 1, P21 to approach 2,  P23 and P24 to approach 3, and P25 to approach 4. 

\begin{figure}
\includegraphics[width = .7\columnwidth]{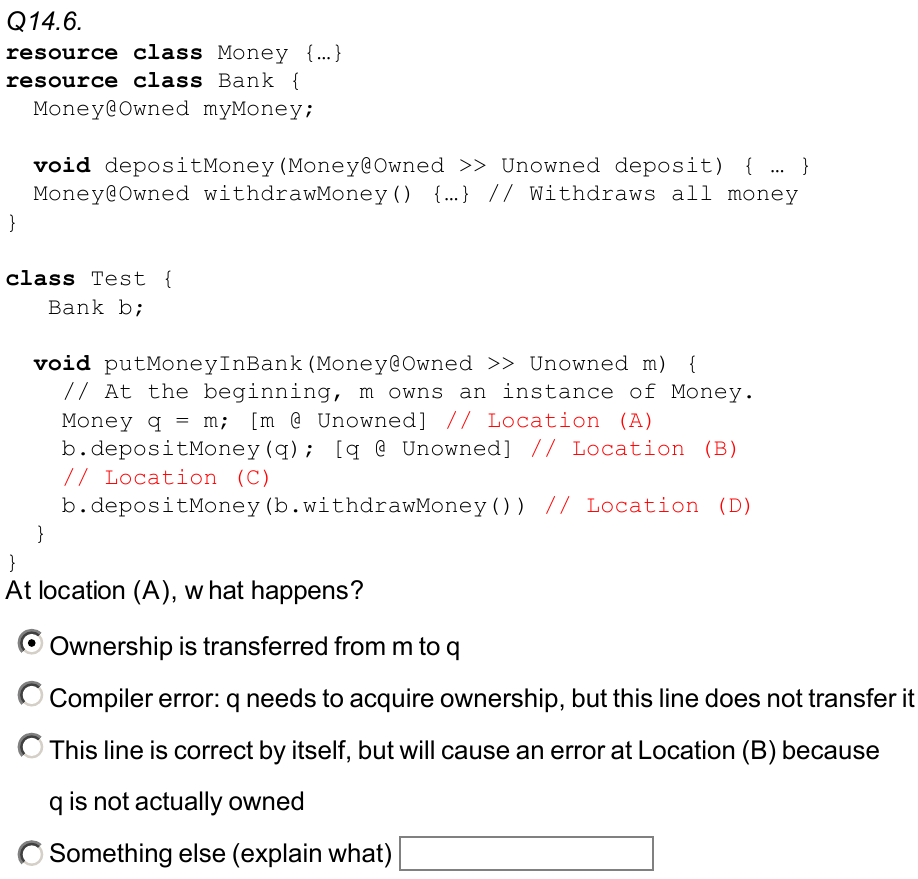}
\caption{An example question assessing understanding of ownership transfer. The correct answer is selected, since assignment transfers ownership.}
\label{P25-Q14.6}
\end{figure}

\subsubsection{Results and Discussion}
P22, who was given approach 1 (with permissions and states specified only in declarations), tried to guess the compiler's behavior, saying things like ``If the compiler was smart\ldots". For example, P22 expected that the language would infer an implicit \texttt{@Off} in the declaration \texttt{LightSwitch s1 = new LightSwitch()}.
P22 also expected that although changes of state were permitted via transactions, state-mismatching assignment to variables would be forbidden, even though approach 1 assumes that states can be inconsistent with type declarations. This approach would be inconsistent and P22's confusion suggests that the type-declaration approach is problematic.

Including types in variables names seemed to be confusing as well. P21 expected that ownership was \textit{not} passed into method calls even when an owned reference was passed. P21 was also surprised that no ownership annotation meant that there was no ownership, instead expecting this to mean that ownership was unknown. 

Participants in condition 3 seemed to do much better. For example, although the materials did not use the word \textit{assertion}, P23 observed that the annotations were assertions. P23 liked the system, commenting ``Perfect, I like this, this is very nice. I wish Java had this; it would have saved me a lot of bugs." 
As we obtained additional confidence in the value of approach \#3, we added additional material. For P24, we changed assertions to use \texttt{@} rather than the initial \texttt{>>} so that we could use \texttt{>>} to specify type changes in transaction parameters. With P25, we used \texttt{?} to indicate lack of static state knowledge. We later simplified the system because this approach was ambiguous, leaving only notations \texttt{Owned}, \texttt{Unowned}, \texttt{Shared}, and unions of specific states (separated with \texttt{|}).

P24 was confused because state specifications on local variables were redundant. For example, in \texttt{LightSwitch@Off s = new LightSwitch()}, the \texttt{@Off} portion is redundant because the compiler already knows the state of the new object due to the constructor's declaration. To resolve this, we added approach 4, removing typestate and permission annotations from local variable declarations; in contrast, permissions are always specified for fields and formal parameters. In those cases, the annotations are important because they constrain types of variables at the end and beginning of transactions. This approach eliminated those as sources of confusion for P25. However, P25 was still confused by whether ownership was transferred in one case after an owned reference was passed to a transaction because of a failure to read the type of the transaction that was being invoked, in spite of a nearby annotation indicating the final permission. We concluded that this difficulty would likely be addressable with a small amount of training, so we proceeded with approach 4 in our final design.

In summary, this study motivated the removal of state specifications from local variable declarations and provided initial evidence that static assertions are likely to be a convenient way for programmers to specify states and permissions of local variables. We also obtained evidence that with these other changes, static state assertions are understandable by current Java users with little extra training. 

\subsection{Threats to validity}
The studies share common threats to validity, many of which correspond to the \textit{external validity} challenge: our participants may not be representative of the population of blockchain programmers; we had limited numbers of participants in each trial; and our tasks may not reflect the reality of blockchain programming. We believe, however, that the population of likely language users is \textit{more} skilled than our participant population, which mostly consisted of students, so if the students are successful in completing tasks, that aspect of the result is likely to generalize. We did not seek to identify all possible usability problems, but rather to identify the most common and severe ones associated with particular design decisions so that we could try to address them. Because there were so many different design decisions, we focused on those for which we had prior evidence that there might be usability problems.

\section{Summative usability study of Obsidian}
\label{Obsidian-summative}

We finished a complete language design, including a formal proof that the design had the formal safety properties we claimed it did. We also completed implementation of the compiler and runtime environment. In order to assess whether our changes to Obsidian had resulted in a language in which programmers could be effective, we designed a summative usability study. As a usability study, this was a \textit{qualitative} study that sought to identify remaining usability barriers as well as to find out whether participants could complete relevant programming tasks. In particular, our prior studies had identified serious usability problems, so we wanted to know whether completing relevant programming tasks was feasible at all for our participants. This study preceded an RCT comparing Obsidian to Solidity. That study is described in \cref{sec:Obsidian-RCT}.

We gave six participants the complete Obsidian language, including its compiler, and asked them to complete three programming tasks. We were interested in whether the participants would experience the same usability problems as the prior participants and whether there were sufficiently serious usability problems left to prevent them from completing their tasks. All of the participants were able to complete the first programming task, but some of the participants ran out of time before completing the other tasks. The second task focused on ownership transfer, since our earlier study found significant usability problems in our earlier prototype. All of the participants who started the second task completed it, suggesting that we had successfully improved the usability of ownership. The third task, which was more open-ended, presented additional challenges for participants in part because it required reasoning about how the states of different objects related to each other.

The design of the study was informed by several of our methodological contributions, and thus also served to assess their value. Key aspects of the PLIERS process that were helpful in designing the study included integrating training into the study to allow recruiting participants who only had Java language experience, not Obsidian experience; recruiting from a population that included many people with some professional experience; and using multiple programming tasks, rather than one long one.

\subsection{Participants}
\label{sec:obsidian-summative-participants}

We solicited experienced Java programmers to take a short screening test online, which took an average of about 9 minutes to complete. We accepted into the three-hour study only those who answered at least five of six basic Java questions correctly. The six questions concerned: Java constructor syntax; the definition of encapsulation; whether changes to a list through a reference would be visible through another reference; whether methods in interfaces may include bodies; whether abstract classes may be instantiated; and whether concrete subclasses of abstract classes must implement methods that were abstract in the superclass. Of 18 completed surveys, 11 people met our screening criteria. We got six participants (P35-P40), whom we compensated with \$50 Amazon gift cards. The participants had an average of 9 years of programming experience, 2 years of professional experience, and 2 years of Java experience. One self-identified as female; the rest identified as male. \Cref{prescreening-example} shows an example question. A copy of the screening instrument is included in the supplement.

\begin{figure}[b]
\centering
\includegraphics[width = 8cm]{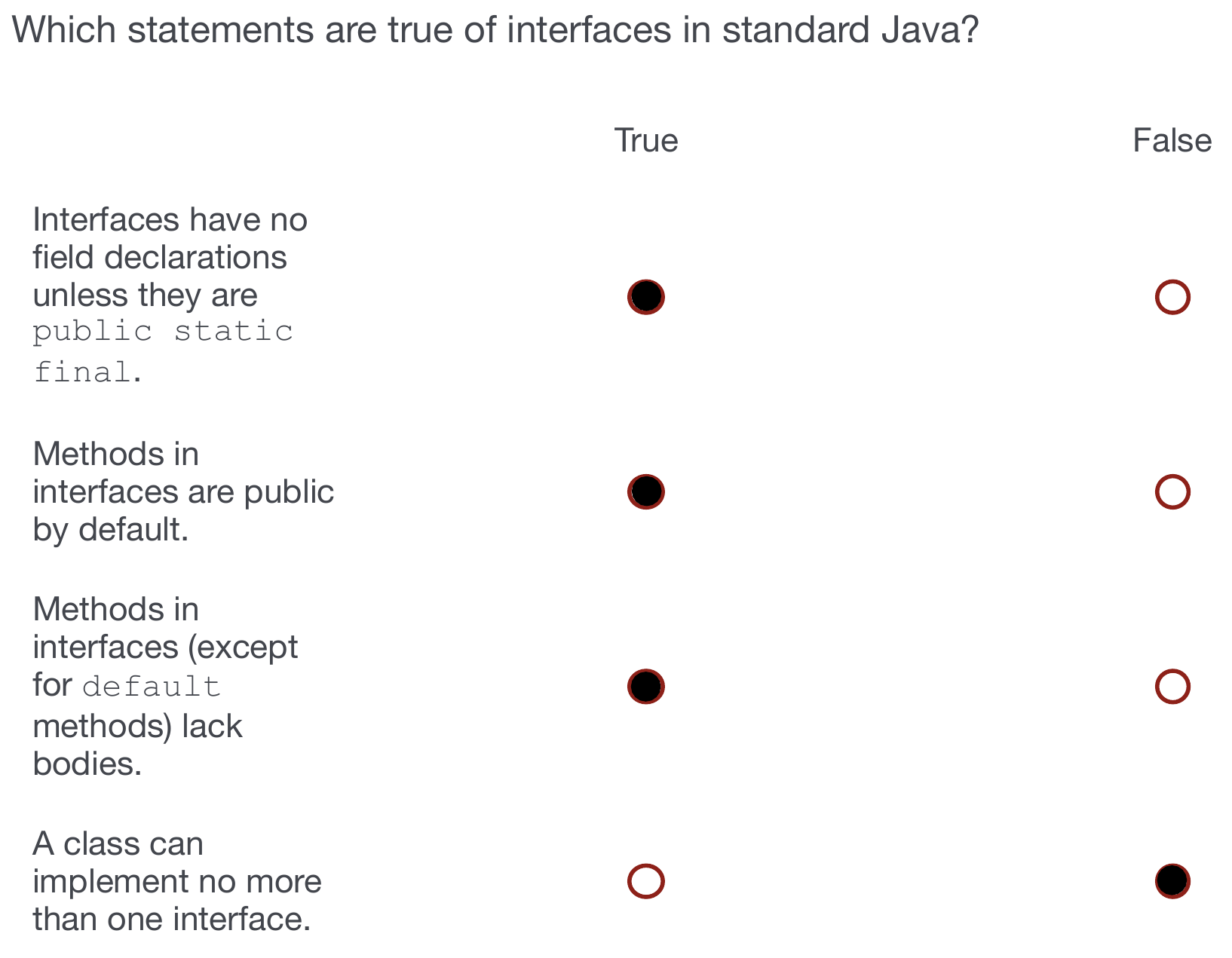}
\caption{An example question from the screening test.}
\label{prescreening-example}
\end{figure}

\subsection{Procedure}
\label{sec:obsidian-summative-procedure}
The previous studies focused on particular aspects of the design, in many cases by giving participants languages that were not precisely Obsidian. To evaluate Obsidian, we conducted a usability evaluation. Because Obsidian provides stronger safety guarantees than existing languages such as Solidity, and because of our prior experience showing that it would be very challenging to develop a linear type system that would be usable at all, we focused the study on examining whether people could effectively complete tasks. Our randomized controlled trial (RCT), described in \cref{sec:Obsidian-RCT}, focused on whether people could complete tasks faster than in existing languages.

The experimenter gave low-level guidance, such as how to invoke the compiler. Also, the experimenter provided assistance that simulated more mature tools. For example, when a participant attempted to debug an error that was reported on line 38 by examining line 33, the experimenter pointed out the discrepancy, since the IDE we provided did not highlight the appropriate line.

After completing the tutorial, which included seven programming exercises, we gave participants starter code for the three main tasks, described below.  Although participants used the compiler, they were not given tests or a runtime environment, since the focus of our usability study was the type system (recall that Obsidian is designed to detect as many bugs as possible at compile time, since runtime detection may be too late to ensure safety). Although the first two tasks were short in order to reduce variance, we allowed the third task to be more open-ended to see whether participants would be able to complete a more challenging task.

The first task, \textit{Auction}, simulated an English auction, in which bids are public, and the bidder who offers the highest price must pay that price for the item. We added the additional constraint that bids were required to come with \code{Money} so that bids could be guaranteed to be viable (a bidder could not issue a bid and then fail to pay for the item). As a starter task, we asked participants to finish implementing \code{createBid}, requiring them to invoke a constructor. They also needed to finish implementing \code{makeBid}, which records a new bid from a client. In \code{makeBid}, we were interested in whether they initially wrote code that accidentally lost the previous bid, which held the associated \code{Money} (before receiving a compiler error), indicating that Obsidian's typechecker had helped them avoid losing track of an asset. \Cref{fig:Auction-task} shows the Auction task.

\begin{figure}[tb]
\centering
\lstset{numbers=left, belowskip=0pt, basicstyle=\ttfamily\tiny, xleftmargin=12pt, linewidth=0.85\columnwidth}
\begin{lstlisting}
main asset contract Auction {
  Participant@Unowned seller;

  state Open;
  state BidsMade {
    // the bidder who made the highest bid so far
    Participant@Unowned maxBidder; 
    Money@Owned maxBid;
  }
  state Closed;

  (*@\ldots@*)

  transaction bid (Auction@Shared this, Money@Owned >> Unowned money, Participant@Unowned bidder) {
      if (this in Open) {
        // Initialize destination state, and then transition to it.
        BidsMade::maxBidder = bidder;
        BidsMade::maxBid = money;
        ->BidsMade;
      }
      else {
        if (this in BidsMade) {
          //if the new bid  > current Bid
          if (money.getAmt() > maxBid.getAmt()) { 
            //1. TODO: fill this in. 
            // Can call other transactions as needed.
  \end{lstlisting}
  \begin{lstlisting}[backgroundcolor=\color{yellow}, numbers=left, firstnumber=last, aboveskip=0pt, belowskip=0pt, basicstyle=\ttfamily\tiny]
            maxBidder.receivePayment(maxBid);
            maxBidder = bidder;
            maxBid = money;
  \end{lstlisting}
  \begin{lstlisting}[numbers=left, firstnumber=last, aboveskip=0pt, belowskip=0pt, basicstyle=\ttfamily\tiny]
          }
          else {
            //2. TODO: return money to the bidder,  since the new bid was too low.
            // Can call other transactions as needed.
  \end{lstlisting}
  \begin{lstlisting}[backgroundcolor=\color{yellow}, numbers=left, firstnumber=last, aboveskip=0pt, belowskip=0pt, basicstyle=\ttfamily\tiny]
            bidder.receivePayment(money);
  \end{lstlisting}
  \begin{lstlisting}[numbers=left, firstnumber=last, aboveskip=0pt, belowskip=0pt, basicstyle=\ttfamily\tiny]
          }
        }
        else {
          revert("Can't bid on closed auctions.");
        }
      }
    }
}
\end{lstlisting}

\caption{The Auction task. Code highlighted in yellow represents a correct solution; the rest was given to participants as starter code. Line 27 transfers ownership of the object referenced by \code{maxBid} to the \code{receivePayment} parameter. The new type of \code{maxBid} from then until line 29 is \code{Money@Unowned}. Line 29 re-establishes ownership in \code{maxBid} by transferring ownership from \code{money} to \code{maxBid}.}
\label{fig:Auction-task}
\end{figure}

The second task, \textit{Prescription}, corresponded to the medical records system in the Permissions study section (\S \ref{permissions-usability}); we were interested in whether our improvements enabled participants to reason more effectively about the code than we had observed in the previous studies. We asked participants to fill in the type signature for the \code{consumeRefill} and \code{depositPrescription} transactions, which mirrored the previous study. We also asked them to complete the implementation of \code{fillPrescription}. 

The \textit{Casino} task was more open-ended and included directions and requirements for what operations should be supported, as well as low-level starter code, such as implementations of \code{Money} and \code{Bet}. It asked participants to implement a \code{Casino} that takes bets on games. When games are complete, the casino enables winners to collect their winnings. The requirements were as follows:

\begin{enumerate}
\item If a \code{Bettor} predicts the outcome correctly, the \code{Bettor} gets twice the \code{Money} they put down. For example, if \code{Bettor} b puts down 5 tokens on the correct outcome, they should receive 10 tokens after the Game is played.
\item If the \code{Bettor} predicted incorrectly, the \code{Casino} keeps their tokens.
\item \label{bets-before-start} Bets can only be made before the \code{Game} starts.
\item \label{winnings-after-finished} Winnings can only be distributed after the \code{Game} is finished.
\item \code{Bettor}s must collect winnings themselves from the \code{Casino} after a Game by calling code, which you need to write. Until winnings are collected, the \code{Casino} keeps track of them.
\item A \code{Bettor} can have one active bet per game. If a \code{Bettor} bets more than once, their original bet should be replaced by the new one and any previous bet should be refunded.
\item A \code{Bettor} MUST put down tokens at the same time that they're making a Bet.
\item If the \code{Casino} does not have enough tokens available to pay out winnings, the invocation to collect winnings can fail.
\end{enumerate}

We also provided a sequence diagram to show participants what operations should be supported. In this way, we conveyed the requirements without also specifying the transaction signatures, since we wanted to see if the participants could infer those themselves.

We were primarily interested in participants' abilities to reason about ownership and typestate and to design architectures that could effectively use ownership.

\subsection{Results and Discussion}
Results for the tasks are summarized in Table \ref{usability-results}. All the participants completed the first task (Auction). All the participants who spent less than two hours on the tutorial completed the second task (Prescription). All the participants who started the third task (Casino), which was substantially more complex than the other two, and had at least an hour available to work on it, finished it. Note that the two successful completion times for the third task were longer than the times that the other participants had available to spend on it. With P38, to assess to what extent the tutorial materials stood alone, the experimenter declined to answer Obsidian-related and debugging-related questions. However, this made the first task perhaps unrealistically difficult and lengthy, resulting in insufficient time for the other tasks.   

\begin{table}[t]
\begin{tabular}{lllll}
\toprule
& \multicolumn{4}{c}{Task completion times (hours:minutes)}\\
\cmidrule(lr){2-5}
	& \textbf{Tutorial} & \textbf{Auction} & \textbf{Prescription} & \textbf{Casino} \\
P35 & 1:31 	& 0:13	& 0:18	& 1:01 \\
P36	& 2:12	& 0:28	& N/A	& N/A \\
P37	& 1:03	& 0:33	& 0:46	& 0:36* \\
P38	& 2:18	& 0:46	& N/A	& N/A \\
P39	& 1:14	& 0:22	& 0:27	& 0:51* \\
P40 & 1:11  & 0:12  & 0:22  & 0:58 \\
\bottomrule
\end{tabular}
\caption{Usability test results. * indicates insufficient time to finish the task. N/A indicates insufficient time to start the task.}
\label{usability-results}
\end{table}


In the \textbf{Auction} exercise, two of the six participants accidentally introduced a bug in which an asset was lost: they overwrote \code{maxBid}, which held money. The compiler gave an error message and they corrected their mistake, but if they had been using Solidity, its compiler would not have caught the bug.  After P36, we slightly simplified the Auction exercise by removing a subtask and refactoring to inline a TODO that had been put in a helper transaction. The above times are adjusted to remove the extra time P35 and P36 spent on the removed task (1 and 8 minutes, respectively).

Some participants seemed to think carefully about ownership and wrote the correct code quickly. Others seemed to focus on satisfying the compiler, and their work took longer. For example, P38 got an error message after overwriting the owned \code{maxBid} reference, and ``fixed'' it with \code{disown}. This choice may be a result of weaker programming skills and lack of help in the tutorial; P38 took the longest on the tutorial, and was surprised to not be given a design diagram for the ($< 300 \text{-line}$) Auction starter code. We changed the tutorial to emphasize that \code{disown} should be used to throw away assets.

In the \textbf{Prescription} task, as with other tasks, variance was large. For example, one reason for P38's long completion time was that P38 had used Python most recently and, despite the tutorial, sometimes wrote Python-like syntax, which did not parse (one example took four minutes to fix). At the time, we were hoping that participants would be able to complete the tasks entirely on their own, but in retrospect, we may obtain \textit{more relevant} results by carefully providing appropriate help (which we provided to all the other participants).

We were interested in participants' ability to reason effectively about ownership. All of the participants who started  Prescription were able to complete it. P37 encountered some difficulties due to shortcomings in Obsidian's support for dynamic state tests. Currently, Obsidian does not allow dynamic state tests to be used as arbitrary Boolean expressions, e.g. \code{if (x in S \&\& e)} where \code{e} is an arbitrary Boolean expression. Likewise, \code{if (x not is Owned)} is not supported (perhaps this was inspired by Python's \code{is} operator). In the latter case, P37 developed some intuition: ``Ownership doesn't feel like something I should be using in this way\ldots'' and restructured the code to check \code{if (maybeRecord in Full)}, which was correct. In another case, the compiler found a bug in which the code assumed that a collection must contain an element, a benefit of not allowing \code{null} in the language.

The \textit{Casino} task was substantially more open-ended than the other tasks, requiring substantially more time, but participants who had a full hour for the task were able to finish it. Some participants defined states in the \code{Casino} contract (P35, P39), whereas others relied only on the states in the Game contract (P37, P40). Both approaches led to a lot of dynamic state tests, since the Casino object had to check to make sure the Game object was in an appropriate state. These checks could have been avoided if the different states of Casino had different typestate specifications for their references to the Game, an idea that occurred to P40 in retrospect. This observation represents an opportunity for a future version of Obsidian in which states of owning objects are coupled to states of owned objects, reducing the need for dynamic checks.

We noticed that participants who did better on the ``advanced Java'' portion of our screening test seemed to complete tasks faster. We found that those test scores were positively correlated with completion speed in the Auction task ($r (4) = .96$, $p < .01$).

\section{Comparing Obsidian to Solidity}
\label{sec:Obsidian-RCT}

We conducted a randomized control trial (RCT) comparing Obsidian to Solidity. The results of this RCT are described in detail in another paper~\cite{Coblenz2020:Can}, but in summary, we recruited 21 Java programmers and randomly assigned them to use either Obsidian or Solidity. This study used a tutorial and tasks that were similar to those in the summative usability study (\cref{Obsidian-summative}); the Prescription task was modified to more exactly match the earlier permissions study described in \cref{permissions-usability}. 

Each study session lasted up to four hours and participants were compensated with a \$75 gift certificate. We allowed them as much time as needed to complete the tutorial. Then, we gave 30 minutes for Auction, 35 minutes for Prescription, and any remaining time for Casino. Finally, we gave them a survey regarding their opinions.

\subsection{Results and Discussion}

In the \textbf{Auction} task, seven of ten Obsidian participants completed the task successfully. Two did not finish the task, and one did so incorrectly, accidentally refunding money to the wrong bidder. Two of the seven successful Obsidian participants had received compiler errors indicating that they had lost assets, suggesting that Obsidian had helped them avoid that bug. In contrast, of the ten Solidity participants, only two completed the task correctly. Seven said they finished the task but had bugs in their code; one ran out of time. The difference in success rates was significant with $p \approx .015$. That is, participants were more likely to finish successfully if they used Obsidian (odds ratio 0.053).


The Prescription task investigated whether participants could use ownership to statically address a security problem. Six of the ten Obsidian participants did so, suggesting that ownership is learnable. Six of the ten Solidity participants attempted a dynamic solution, but only two of them were able to finish it in the time available. In both conditions, some participants made \code{Prescription} mutable, even though that was explicitly disallowed by a comment in the program. We had selected an immutable design following standard security advice, but the results suggest that a \textit{mutable} design for \code{Presciption} might have been more natural for some participants.

\begin{table}[tb]
\begin{tabular}{@{}lll@{}}
\toprule
& \textbf{Solidity} & \textbf{Obsidian} \\
\midrule
Static solutions: correct solutions / attempts & N/A / 5 & 6 / 6 \\
Dynamic solutions: correct solutions / attempts & 2 / 6 & 1 / 3 \\
Completed within time limit & 3 & 9 \\
Mean time among successful participants; [95\% CI] & 20 min.; [0, 45] & 22 min. [12, 33] \\
\bottomrule
\end{tabular}
\caption{Summary of Prescription task results. $N=10$ in each condition. Two Solidity participants tried both static and dynamic approaches, and one Solidity participant did not modify the starter code at all, resulting in 11 Solidity attempts.}
\label{table:prescription-summary}
\end{table}

The \textbf{Casino} task was substantially more open-ended and offered more opportunities for mistakes. In part due to time spent on earlier tasks, only nine Solidity participants and five Obsidian participants had enough time to arrive at a solution with which they were satisfied within their four-hour time window; we did not set a separate limit for the task. We discarded data from one Obsidian participant who encountered a compiler bug. 

Results are summarized in \cref{table:casino-successes}. Notably, the four Obsidian participants who finished the task all abused the \code{disown} keyword, despite verbiage in the training materials warning about this. The result is that they lost track of assets, since their usage suppressed errors that the compiler would have otherwise given. This warrants further investigation in how to safely provide language features that are necessary for some kinds of programs, but which nonetheless can be used unsafely. 

The Obsidian participants spent significantly longer on the task than the Solidity participants did ($p \approx 0.02$, Mann-Whitney U test, $d \approx 1.9$). This gives an approximation of the cost of the stronger type system in implementing a software prototype (but perhaps not in implementing a production-quality system). Of course, this cost may be worth bearing, since the safety guarantees may result in a more efficient and safer software creation process by reducing the testing burden. However, this benefit may require either language modifications or more training to avoid the risk of abusing \code{disown}. Future work will need to investigate mitigating this risk and whether additional training and practice mitigate the cost of the stronger type system.

\begin{table}[t]
\begin{tabular}{@{}lll@{}}
\toprule
& \textbf{Solidity ($N = 8$)} & \textbf{Obsidian ($N = 4$)} \\
\midrule
Had enough time to try Casino & 9 & 5 \\
Completed Casino with a program that compiled & 8 & 4 \\
Completed task correctly (no identified bugs) & 1 & 0 \\
Winnings collection emits error if Casino is out of tokens & 5 & 2 \\
Only used disown safely & N/A & 0\\\
Managed tokens correctly (not fabricating or losing them) & 4 & 0 \\
Mean completion time & 37 min. & 64 min.\\
\bottomrule
\end{tabular}
\caption{Summary of Casino task results among completed programs that compiled, showing correct solution rates among errors made by more than one participant.}
\label{table:casino-successes}
\end{table}

We asked participants several questions about their opinions of the languages they used in the study. Participants who were assigned to use Obsidian rated ownership as \textit{more useful} than participants who used Solidity ($p \approx 0.002$, $d \approx 2.5$). However, the Solidity participants indicated that they felt they understood states better than the Obsidian participants did ($p \approx 0.04$, $d \approx 1.3$). Results are shown in \cref{table:survey-summary}.

\begin{table}[b]
\begin{tabular}{@{}lll@{}}
\toprule
& \parbox[t]{1.1cm}{\textbf{Solidity\\(N=6)}} & \parbox[t]{1.3cm}{\textbf{Obsidian\\(N=8)}} \\
\midrule
How much did you like the language you used? & 3.7 (0.82) & 4.0 (0.53) \\
How well do you feel you understand the concept of ownership? & 3.8 (0.98) & 3.75 (0.99) \\ 
*How useful do you think ownership is? & 3.0 (1.1) & 4.88 (0.36) \\ 
*How well do you feel you understand the concept of states? & 4.8 (0.41) & 4.1 (0.64) \\ 
How useful do you think states are? & 4.3 (0.81) & 4.1 (0.64) \\
\bottomrule
\end{tabular}
\caption{Perceptions of ownership, states, and assets on a 1--5 scale (5 is best). Cells show average (standard deviation). *~indicates that a Mann-Whitney U test shows a significant difference at $p < 0.05$.}
\label{table:survey-summary}
\end{table}

The survey also asked for additional comments. Three Solidity participants wrote that they wished ownership were checked by the compiler (as is the case in Obsidian). Some participants using Solidity wished they had a notion of state. For example, one wrote:

\begin{quote}
It also seemed like there should be some syntactic sugar for writing things like:
\begin{lstlisting}[aboveskip=1pt, belowskip=1pt]
   enum State { Foo, Bar, Buzz }
   State s
\end{lstlisting}
since they are so common.
\end{quote}

Some Obsidian participants wrote that they appreciated the tutorial and exercises, and that ownership seemed natural after some practice. 


\subsection{Implications on PLIERS}

The randomized controlled trial comparing Obsidian and Solidity served in part to evaluate Obsidian and in part to evaluate PLIERS. In this RCT, we were able to show a safety benefit of Obsidian in the Auction task, and were able to show that most of the Obsidian participants were able to use ownership successfully in the Prescription task. This shows that the tutorial method was mostly successful (though more success could likely have been obtained with more practice) and that the language design was effective overall (modulo the abuse of \code{disown} that we observed). Every study design involves making tradeoffs. The results here may show a tradeoff between training time and success rates; users of PLIERS will need to decide, based on their own design and research goals, how to balance the risks when designing their studies. However, the overall PLIERS design process did result in a language that had significant benefits relative to the status quo, which we were able to measure in a relatively low-cost study.

In retrospect, since only one of 20 participants completed the Casino task successfully (across both conditions), that task was too hard for the amount of time we allowed. We recommend that users of PLIERS carefully select success criteria in pilot studies in order to set appropriate task time limits and difficulties.

\section{Study Design Challenges and Solutions}
\label{challenges}

\Cref{approach-summary-table} summarizes the challenges that PLIERS addresses. Our primary interest is in programmers' abilities to achieve their goals \textit{after} they have become proficient in the programming language, not on how easy it is for novices to \textit{learn} the language. Thus, our evaluation approach requires first teaching people a language and then observing their performance on tasks.

\begin{table}[htbp]
\begin{tabular}{l p{9.5cm}}
\toprule
\textbf{Challenge} & \textbf{Approaches} \\
\midrule
Training & \begin{itemize}[leftmargin=0pt, nosep, before=\compress]
				\item Include knowledge assessments and practice problems in tutorial
				\item Divide tutorial into small pieces
				\item Answer questions during training phase of study
				\item Automatically provide feedback for wrong answers
			\end{itemize}\\
Recruiting & \begin{itemize}[leftmargin=0pt, nosep, before=\compress]
				\item In academic settings, recruit master's students, who frequently have professional experience that may be representative of many practitioners
				\item Recruit professionals, but only when their expertise is needed
				\item Appeal to professionals' altruism for recruiting (they may not be incentivized by typical study budgets)
				\item Screen participants carefully; set a high bar for student participation
				\item Evaluate language design research questions in the context of a language with which many possible participants are familiar

  			   \end{itemize}\\
\parbox[t]{3cm}{High prototyping\\cost} & \begin{itemize}[leftmargin=0pt, nosep, before=\compress]
											\item Back-port language design questions to existing languages (also helps isolate effects of independent variables)
											\item Use Wizard of Oz to simulate tools that do not exist yet: use a plain text editor rather than a real IDE, and have an experimenter provide feedback in lieu of a real compiler or interpreter
										  \end{itemize}\\
\parbox[t]{3cm}{Interdependence\\ of features} & \begin{itemize}[leftmargin=0pt, nosep, before=\compress]
                                                \item Isolate design questions by back-porting them to a familiar language
                                                \item Mitigate non-orthogonality risk with summative studies
                                            \end{itemize}\\
\parbox[t]{3cm}{Variance and\\external validity} & 	\begin{itemize}[leftmargin=0pt, nosep, before=\compress]
														\item Triangulate with multiple study types
														\item Break tasks into subtasks
														\item Recruit from populations with sufficient programming skills and knowledge; pre-screen participants.
													\end{itemize}\\
Time management & 	\begin{itemize}[leftmargin=0pt, nosep, before=\compress]
						\item Pilot repeatedly to assess how long tasks usually take
						\item Set cutoff times so that most people will succeed at most tasks
						\item Allow participants extra time when possible, then report these successes separately from the ``within time limit'' results
					\end{itemize}\\

\parbox[t]{3cm}{Bias toward\\familiar languages} & \begin{itemize}[leftmargin=0pt, nosep, before=\compress]
														\item Staged natural programming approach: sequentially expose additional constraints to participants
														\item Request that participants do tasks using specific language designs that are being evaluated
													\end{itemize}\\
Unsound proposals & \begin{itemize}[leftmargin=0pt, nosep, before=\compress]
						\item Provide sound alternatives and ask participants to use them
						\item Provide participants with expert feedback on design ideas
					\end{itemize}\\
\bottomrule
\end{tabular}
\caption{How PLIERS addresses common challenges in running user studies on programming languages.}
\label{approach-summary-table}
\end{table}

When we initially tried to apply HCI methods in our language design work, we were thwarted by several challenges, described in the introduction: \textit{training}, \textit{recruiting}, \textit{high prototyping cost}, and \textit{variance}. We also encountered additional challenges, such as \textit{interdependence of features}, \textit{time management} in studies, \textit{participant bias toward familiar languages}, and \textit{unsound proposals by participants}. In this section, we describe techniques we used when designing user studies in order to address each challenge.

\subsection{Training}

Evaluating a programming language requires first \textit{teaching} the programming language. Many universities offer term-length courses in specific programming languages or techniques; requiring this kind of time commitment would make it extremely difficult to recruit participants. Furthermore, most courses ensure a consistent experience for all students by having all students learn the material in parallel (for example, with one session per topic, where all students participate at the same time). In contrast, our design approach was iterative, consistent with design methods used in other areas of HCI~\cite{Dumas1999:Practical}. We were interested in addressing a variant of our \textit{training} challenge that asks: what would be an effective way to teach a programming language in a consistent way to many participants in sequence?

Initially, we created a textual guide to the new programming language, and asked participants to read it before doing the tasks relevant to each study. The guide was relatively short; it could be read thoroughly in under an hour. Unfortunately, this approach had very significant limitations. Although it was effective for some participants, others only skimmed the material and were then unable to complete the programming tasks. Because the guide was not structured as reference material and it included substantial conceptual information, skimming the guide was insufficient.

\begin{figure}[htb]
\raggedright

\begin{lstlisting}[numbers=none, framexleftmargin=0em, xleftmargin=.6em, basicstyle=\LSTfont, language=obsidian]
contract Money {
  int amount;
  transaction getAmount() returns int {
    return amount;
  } 
}
contract Wallet {
  Money@Owned m;
  Wallet@Owned() {
    m = new Money();
  }
  
  transaction spendMoney() {
    ...
  }
  
  transaction receiveMoney(Money@Owned >> Unowned mon) returns Money@Owned
    Money temp = m;
    m = mon;
    return temp;
  }
    
  transaction checkMoney() returns Money@Owned {
    return m;
  } 
}
\end{lstlisting}

\includegraphics[width = 11cm]{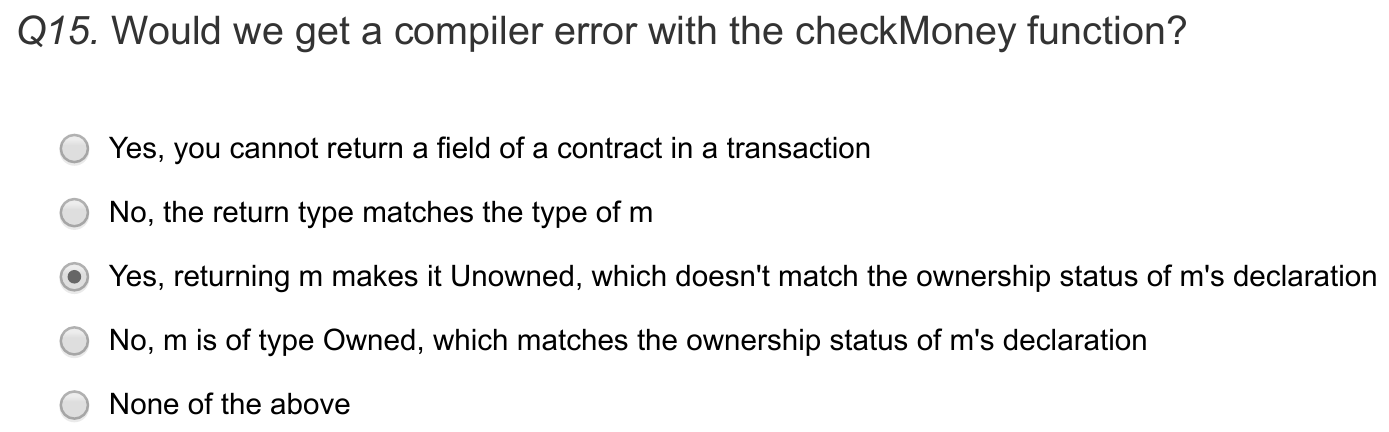}
\caption{One question from the Obsidian tutorial. The question assesses whether the participant has understood that at the ends of transactions, fields must be have types that match their declarations, and that returning a variable consumes any ownership in the variable. If a participant submits an incorrect answer, the survey tool informs them of their error so they can fix their misunderstanding.}
\label{checkMoney-question}
\end{figure}

We were able to solve the problem with two adaptations: (1) break the guide into much smaller pieces; (2) ask participants to answer questions or complete small tasks to assure they had absorbed the material of each piece. For example, we broke the Obsidian tutorial into ten parts, and still the average participant completed it in under 90 minutes. We found that we were able to design tasks that checked understanding that were brief and did not require substantial experimenter intervention (helpful for ensuring consistency). We used a web survey tool (Qualtrics~\cite{Qualtrics}) to guide participants through the tutorial and ask questions to check understanding. The tool also offered automatic feedback on participants' answers to multiple-choice questions. For example, \cref{checkMoney-question} shows a question about a code fragment with the correct answer selected. The relevant language details are explained in \S \ref{obsidian-language}.

Although we originally wanted to make the tutorial stand alone so that every participant would have the same experience, we found that to be impractical; participants inevitably had questions about the materials, and forcing them to continue without having their questions answered resulted in them being unable to complete the tasks. However, we found that if an experimenter was available to answer questions, most participants asked only a small number of questions, which could be addressed rapidly. This approach is arguably more similar to a real-world language learning experience than an approach in which no questions are answered; normally, learners can search the Internet for answers to their questions, ask friends for help, etc. 

\begin{figure}
\includegraphics[width=11cm]{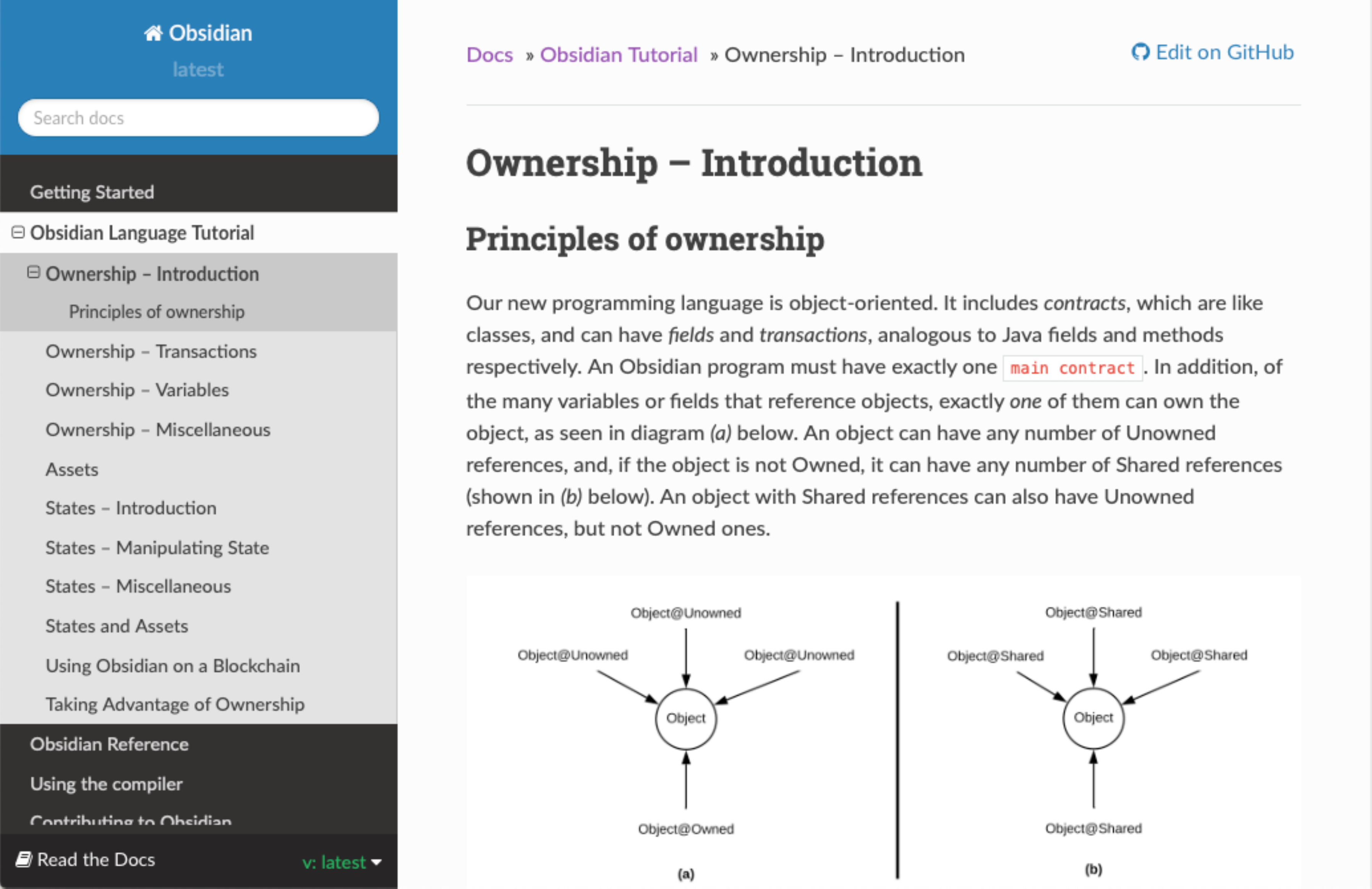}
\caption{A page from the Obsidian tutorial. The navigation bar at left shows how the tutorial was divided into 11 sections.}
\label{fig:obsidian-tutorial}
\end{figure}

In summary, although our initial tutorial was not an effective way of teaching the language, and the final tutorial was not sufficient by itself, dividing the tutorial into small pieces, providing tasks to help participants check and reinforce their understanding, and having an expert who could answer questions allowed most of our participants to learn the needed material in a short period of time. \Cref{fig:obsidian-tutorial} shows how the tutorial was broken into 11 different sections, each of which was followed by exercises for participants to complete.

\subsection{Recruiting}

Evaluation requires participants who are sufficiently skilled that they can rapidly learn a new programming language and then complete tasks using the new language. This would seem to require lengthy user studies with skilled participants, who can be challenging to recruit and retain for the required period of time. \textit{Iterative} evaluation requires a large number of participants, since participants who learned an earlier version of the language can no longer provide fresh perspectives on new ideas. Although some user interfaces for experts in other domains require recruiting members of a small population, many of those interfaces are for short-term, focused tasks rather than lengthy problem-solving tasks. Furthermore, although it is typical to conduct studies with students, this relates to our \textit{external validity} challenge: to what extent do results from students apply to the professional software engineers that are the target of our language? 

We found in our work on Glacier and Obsidian that we were able to usefully combine results from different populations. Rather than trying to exclusively obtain professional software engineers, we found that we could design studies that yield meaningful results from students; for other aspects of the research, we recruited limited numbers of professionals. For example, when we wanted to interview software engineers to find out their experiences of using immutability constructs in the Glacier work, we recruited senior-level professional software engineers. However, for the other studies, we made three observations that enabled us to do our studies with various kinds of students. 

First, about 41\% of professional developers have been programming professionally for less than five years~\cite{StackOverflow2019:Survey}. Many graduate students have some professional experience. For example, students at the Professional Master's program in Computer Science \& Engineering at the University of Washington were reported to have an average of five years of professional experience~\cite{UW2019:PMP}. Similarly, the Carnegie Mellon Master of Software Engineering program requires all students to have at least two years of experience~\cite{CMU2019:MSE}. By recruiting from graduate students, we were able to attract a population that is similar to a significant fraction of professional programmers and software engineers.


Second, in usability studies, it is typical to assume that usability problems encountered by even one user may be experienced by many others. Not every usability problem can be addressed without risking introducing new usability problems, but our experience is that many can be. For example, error messages, documentation, and keywords can be interpreted in ways that were not intended by the author; clarifying the text can prevent others from being confused in the same way. Syntax borrowed from other languages can be evocative in useful ways, but when the semantics do not match precisely, confusion can result; this can be addressed by choosing distinct syntax. On the other hand, semantic or structural changes can have consequences on users that are hard to predict, especially since one high-level change may necessitate a series of lower-level changes, which each have their own impact. For example, in Obsidian, moving transactions so that they were no longer lexically scoped in states necessitated adding special syntax for specifying the initial type of the receiver, \code{this}. We had selected that approach based on consistency with Java, which already uses that design. Unfortunately, that approach was surprising to some of our participants.

By addressing problems that student participants encounter, we prevent professionals from encountering those problems as well. Of course, some of the problems may not be ones that professionals would encounter, but nonetheless, addressing them may improve learnability, making the system better overall. When changes that would improve the system for the participants might degrade performance for experienced users, then the designer can make an informed tradeoff, potentially addressing the problem in training materials rather than in a design change.

Third, for Obsidian studies, we developed a screening instrument so that we could include only participants who had appropriate programming skills. The instrument, which is a web-based survey, takes most participants under ten minutes to complete. The instrument also included more difficult questions; because of the difficulty, we did not use this portion for screening. However, we found that performance on the more difficult portion of the instrument was positively correlated with speed in one of our programming tasks even in a small, six-participant study. At the time of the Glacier studies, we had not yet developed this instrument; the more-complex programming tasks in the Obsidian studies motivated us to screen our participants more carefully.

Designing a screening instrument (or deciding not to use one) depends on an assessment of what knowledge and skills are required of participants, and of how honest the prospective participants will be in their self-assessment. If prospective participants can reliably self-assess preparedness for the study, and they can be assumed to be honest, then screening may be unnecessary. On the other hand, even in this case, assessing programming knowledge and skills can be useful for understanding how these relate to task performance. In the Obsidian studies, we invited participants based on a ``basic Java'' portion of the screening instrument and observed that performance on one of the tasks was correlated with performance on the ``advanced Java'' portion of the instrument, suggesting that Java programming knowledge was a significant influence on task performance. This is somewhat surprising, since the screening test examined language-specific knowledge, but did not give any actual programming tasks. We would encourage others to consider using this kind of screening instrument, since it is low-cost (generally under 10 minutes per participant), resulted in participants who were generally capable Java programmers, and portions of it correlated with task performance.

%

We found that relatively small incentives were sufficient to motivate students to participate in our studies. For three-hour studies, we offered a \$50 Amazon gift card; for four-hour studies, we offered a \$75 Amazon gift card. For shorter studies, we paid \$10/hour. We recruited professionals from among our personal networks and did not offer them a specific incentive to participate.

\subsection{High prototyping cost}

Programming language designers are accustomed to creating high-cost implementations, not low-cost \textit{prototypes}, but traditional HCI methods assume that low-cost prototypes can be created. Traditional ways of evaluating programming languages typically require a compiler or interpreter as well as theoretical work to create a sound design (informally, one in which programs mean what they are supposed to mean and the safety guarantees that the type system claims to provide can actually be provided). If one insists on creating a sound, formal model of the language before evaluating it with users, iteration can require so much time that it is impractical. Furthermore, the cost is increased by the expectation of sophisticated language-dependent tooling in IDEs: syntax highlighting, autocomplete, high-quality error messages, and the like.

Instead, we do not insist on doing this work at the beginning. We outline a potentially sound underlying formalism without proving all the relevant properties. Then, we design a surface language and evaluate it with users so that we can obtain feedback early. In doing so, we accept the risk that the formal system cannot be made sound without invalidating the data we gathered from users, but in practice, we found that usually any mistakes are minor and can be corrected without having to redo the user studies.

Late in the project, we found that designing and running user studies of low-level features typically required much more time than implementing the features; for those, it make sense to implement the alternatives rather than simulating them. On other other hand, early in the project, many high-level design decisions would have required substantial design and implementation work. Among those, we carefully selected questions for which user input would be the most impactful. A key approach in minimizing cost of language changes was to re-use training materials across phases of the studies to the extent possible, allowing us to amortize the cost of their development across multiple studies. The training materials co-evolved with the implementation and represented a significant investment.

\subsection{Interdependence of features}
Suppose a comparison between two languages showed that one allowed participants to complete tasks faster or more successfully. If the two languages were very different from each other, it would be unclear which aspects of the new language were actually helpful. For example, a comparison between a particular functional language and a particular object-oriented language would not result in fine-grained, actionable design guidance for a new language. Furthermore, if the study was done in the context of a language that was new to participants, confusion might be due to unfamiliar aspects of the language that are unrelated to the design question of interest.

By using the \textit{back-porting} approach described above, we isolated particular design questions in the context of an existing language. Although this does not enable us to address very high-level design questions, such as whether the language should be object-oriented or functional, it allowed us to obtain actionable data about particular design decisions. 

Theoretical refinement is another approach that helps address feature interactions, since frequently, key theoretical issues relate to interactions between language features. Likewise, case studies, natural programming, and usability studies with appropriate tasks can lead to insight regarding cross-cutting concerns. 

Of course, it is still the case that the design choices are not orthogonal. To address this, we integrate the results into a new language and conduct summative studies on the completed language as a whole as well.

\subsection{High variance and external validity}

The nature of programming is that there is huge variance in performance on tasks among different programmers~\cite{Nichols2019:End}. When asking participants to complete programming tasks to help a designer iterate on a language design, participants frequently get stuck on problems that are not of interest to the designer. For example, in one Obsidian study, a participant spent significant time writing code to recurse through a data structure, even though code had been provided to do exactly that. Issues involving the details of the data structure were intended to be out of scope for the study. On the other hand, constraining tasks too much may result in artificial tasks that do not represent the complexity of real-world programming problems, which limits the external validity of the studies.

We use three techniques to address these problems. First, we combine the results of different kinds of studies (\textit{triangulation}~\cite{Given2008:Sage}). Qualitative studies of varied tasks with varied participants, in which timing is not an important dependent variable, can identify usability problems, and an experimenter can guide participants away from problems that are not intended to be part of the study. Quantitative studies typically involve fairly constrained tasks, but we can hope to obtain statistical significance in a comparison between two different designs. Finally, although this paper does not focus on our case study work, we also used case studies to address questions of \textit{expressiveness}: elucidating what happens when the language is used to solve a larger programming problem, which cannot be completed in a single-session user study; for more information, see \citet{Coblenz2017:Glacier} and \citet{Coblenz19:Obsidian-arxiv}.

Second, particularly in RCTs (in which the experimenter cannot provide any guidance), we give several independent tasks rather than one long task. Then, we analyze the tasks separately, although of course the performance on the tasks is not independent because the same participant completed all of the tasks. Furthermore, dividing tasks into multiple pieces enables separate analysis of complete vs. incomplete tasks. For example, in the Glacier studies, we gave both simple and complex immutability specification tasks rather than one combined task. In the Obsidian studies, we separated a complex task, Casino, from simpler tasks, Auction and Prescription, even though the research questions overlapped. This allowed participants to succeed in the simpler task even if the more complex task was too difficult for them. It also offered an opportunity for participants to apply knowledge gained in the simpler task when working on the more complex task.

Third, recruiting from a constrained population reduces the impact of uninteresting noise. The primary technique to use is a screening survey, which participants must complete before being selected to participate in the study. This allows the experimenter to ensure that programmers have sufficient programming skills and knowledge. Of course, one must be careful to avoid screening out participants that may, in fact, be representative of the population to which the results should generalize.

In qualitative studies, it is sometimes unclear how many participants to recruit. Nielsen and Landauer found that the best benefit/cost ratio occurred at 3.2 participants in a set of their usability studies, which were of a medium-large software project~\cite{Nielsen:Mathematical}. We found it was effective to consider the following factors in assessing when to stop recruiting more participants:

\begin{itemize}
    \item To what extent new data (from the most recent participants) duplicates existing data
    \item To what extent the researcher is willing to tolerate risk of missing usability problems
    \item Fidelity of the current prototype (it may not be worth exhaustively testing low-fidelity prototypes)
    \item Specific research objectives: have the primary research questions been addressed yet?
\end{itemize}

\subsection{Time management}
As a practical matter, one needs to keep each participant's commitment brief in order to be able to recruit and retain enough participants and to minimize study cost. However, the experiment designer needs to allow enough time for most participants to finish the given programming tasks (at least in some of the experimental conditions). To address this problem, we conducted enough pilot studies (when preparing for an RCT, we found typically five or so pilots sufficed) that we could estimate the range of times that most participants would spend on each task. We found that we could allow participants enough time such that when the participant did not finish in the allotted time, the experimenter usually believed that even given substantial additional time, the participant would not have completed the task. This belief was driven by observing the difficulties that participants were facing at the end of the time window. Sometimes the problem was a design choice by the participant that made the problem much more challenging than anticipated; other times we believe it was due to lack of programming skill, since we observed some participants making basic programming errors. 
Of course, it is difficult to generalize about participants who do not finish particular tasks when we expected them to have enough time, but we found that the above approach resulted in studies that were practical to run and which yielded useful results.

The choice of study pre-screening method introduces a tradeoff. A lax pre-screening procedure makes it easier to obtain enough participants from a population that generalizes to a broader community. A strict pre-screening procedure that admits only the most expert participants may reduce times as well as variance, but may make it difficult to recruit participants and harder to generalize the findings. In university settings, with many novices, we advise erring on the stricter end of the spectrum, since most real practitioners will be more skilled than most students.

Rather than giving fixed limits for each task in advance, we aimed to maximize effective use of participants' time. When participants had additional time remaining in their commitment (for example, in one study, we told participants that the study would take four hours), we could let the participants spend longer than budgeted on the later tasks if their earlier tasks took less time than expected. Then, when reporting results, we could consider what the success rate would have been if everyone had had only the time available of the participant with the minimum time window for that task. In addition, we could report which participants succeeded given the additional time. This allowed us to make the best use of our participants' time while maintaining experimental validity.

\subsection{Bias toward familiar languages}
In a user study of a new programming language in which the participants are experienced programmers, one might expect that the language that performs ``best'' might be one with which participants are already familiar. Furthermore, when asked to join in participatory design exercises, perhaps participants might be likely to guide the design toward languages with which they are already familiar. 

We used three techniques to address this problem. First, to find out what approaches might be easily learnable and would make immediate sense to participants, we adapted the \textit{natural programming} elicitation technique~\cite{naturalprogramming}. In it, participants are given blank paper or a text editor and asked to write programs without being given a specific language to use. As a form of participatory design, the goal is to elicit from participants the way they would \textit{naturally} express the ideas in question. Although traditional natural programming studies give the programmer no training at all, we took a \textit{staged} approach. First, we asked participants to write programs on a blank screen with no training. Then, we told them information about the language design, and asked them to do additional programming tasks with the new (but still underspecified) design. For example, we gave participants a state transition diagram and asked them to write a program that expresses the state transitions. By scaffolding the participants' work in stages, we were able to answer both questions about participants' initial expectations as well as identifying what approaches might be most natural given our preliminary language design assumptions.

Second, in most of the studies, we constrained the participants' work according to our design ideas. Because the languages were designed to provide particular formal safety guarantees, we were interested in the impact of the language features related to those properties. In general, providing stronger guarantees requires that programmers enable the compiler to prove safety properties, which may require additional work from programmers. We were interested, then, in whether participants could complete tasks in the language even though they were obtaining stronger safety guarantees. 

Third, we focused on observing and understanding behavior rather than preferences. By doing so, participants' prior experience was not an obstacle to overcome, but instead background we could leverage in teaching participants our language.

To encourage innovative responses (rather than ones that merely reflected prior training), we used natural programming for situations in which commonly-used languages could not directly represent the requirements we gave participants. We also used natural programming for low-level syntactic choices (e.g. keyword selection). We also instructed participants explicitly to be creative and not write in any particular existing language. Finally, we were careful to interpret the results in the context of participants' prior knowledge. For example, when participants use curly braces to denote blocks, the content of the blocks may be interesting even though the choice of curly braces is not.

\subsection{Unsound proposals by participants}
Another common limitation of natural programming is that participants lack expertise in language design, resulting in unsound proposals. This problem occurs with participatory design in other domains as well, and the usual solution is to use participant ideas as input to an expert-led design process~\cite{Nielsen:BadSuggestions}, which applies here as well. 

Our language design process typically involves writing multiple example programs, each of which assumes a particular language design and explores a particular kind of programming problem. The examples typically expose tradeoffs in language design; choosing which tradeoff to make can be informed with user input. We were able to use some of these prototypes to develop user studies, in which we presented participants with several options rather than expecting them to compose designs from scratch. In some cases, we tried to generate all feasible options in a particular design space (due to various technical constraints, this might result in three or four options), and then narrowed this down to the most promising approaches based on the tradeoffs that were apparent. We asked participants to \textit{complete tasks} using the best candidates so that we could come to an informed conclusion about which of the options were best, rather than merely asking participants for their opinions. This allowed us to focus the process on designs that would fulfill the technical requirements while still obtaining relevant design insights.

\section{Future work}
\label{sec:future-work}
In this paper, we evaluated PLIERS by applying it to two different language designs. This approach greatly facilitated developing and iterating on the PLIERS process. However, in the future, we would like to show that these methods can be used by language designers with a variety of backgrounds and goals. As a practical matter, recruiting language designers to participate in a language study is a challenging and heavyweight endeavor, but future work may identify promising contexts in which to evaluate PLIERS more broadly. We have begun by teaching PLIERS to students in an undergraduate programming language design class, and found that the students were able to use some of the methods to help them iterate on their language designs.

One particularly challenging aspect of applying PLIERS is that the theoretical aspects of the design work require substantial background. Perhaps in the future, mechanized tools could help those who are not programming language experts design safe languages in their own application domains. By using program synthesis techniques, a language synthesis tool might be able to search the space of languages that have particular formal properties to help users identify safe design candidates. Creating Obsidian required spending months developing the underlying core calculus and proving it sound; if this effort could be mechanized, language iteration would be much faster, and could be feasible for those without formal programming language training.

The Wizard of Oz approach that we proposed in this paper relies on an experimenter who can accurately simulate the kinds of error messages that a compiler might generate. We envision a utility that would help promote consistency and improve reliability. Such a tool would accept error messages entered by experimenters during experiments and deliver them to participants in a realistic way. By recording the errors that were delivered, the experimenter could re-use existing error messages as well as record participants' reactions. This approach might lead toward refined error messages that are clearer for users to understand.

Another limitation of the methods we describe in this paper is that although one can do studies that assess the usability of particular language design choices, in some cases design choices interact with each other. As a result, it is not clear that designers can combine the results of different studies and expect that the resulting language will be usable. In this work, we mitigate this threat in two ways. Summative studies address the integrated language and can reveal problems that arise from combinations of design choices. Likewise, by triangulating design through multiple kinds of studies, some of which crosscut multiple language design choices, we obtain different perspectives on various combinations of features. However, future method development work may be able to address this problem more directly.

In the future, we hope to explore how PLIERS could be used to develop tools for other problem-solving contexts beyond programming. Attributes of programming that are shared with other kinds of problem-solving activities include:

\begin{description}
\item[High variance:] Problem-solving can be unpredictable~\cite{Loksa2016:Programming}; in user studies, some participants typically complete tasks almost instantly whereas others can spend hours working and still not finish. This large variance makes running quantitative user studies very challenging. 

\item[Range of working styles:] Bergstr\"{o}m and Blackwell described a diverse collection of different approaches to programming problems~\cite{bergstrom2016practices}, such as \textit{bricolage/tinkering} and \textit{engineering}. These different styles may be used even by different people using the \textit{same language}, impeding a designer's attempts to anticipate a user's strategy or behavior.

\item[High stakes:] Errors when programming can contribute to serious real-world safety problems, e.g., in avionics or health care systems.
\end{description}

For example, CAD tools affect their users' creative processes~\cite{Robertson2009:Impact}; likewise with process engineering tools~\cite{Braunschweig2002:Software} and even drug design tools~\cite{Stewart2006:Drug}. All of these domains involve expert problem-solving by a variety of different people with high costs of failure. As such, they might be amenable to use the PLIERS process to help designers in those domains create tools that are effective for their target users.

\section{Conclusion}
\label{sec:conclusion}
PLIERS represents a new approach to designing programming languages for software engineers. PLIERS is exemplified in Glacier and Obsidian, which reflect a new way of designing programming languages that integrates user-centered techniques into many stages of the design process. By incorporating feedback from users, we obtained insights that led to two languages in which programmers can be effective at obtaining stronger safety guarantees than prior languages provided. We expect our new approach to language design is applicable to the design of other programming languages, and even to the design of a wide variety of different kinds of problem-solving tools.

\section{Acknowledgments}
  We appreciate the help of Eliezer Kanal at the Software Engineering Institute, who helped start this project, as well as Jim Laredo, Rick Hull, Petr Novotny, and Yunhui Zheng at IBM, who provided useful technical and real-world insight.
  
  This material is based upon work supported by the
  \grantsponsor{GS100000001}{National Science
    Foundation}{http://dx.doi.org/10.13039/100000001} under Grants
  \grantnum{GS100000001}{CNS-1423054} and \grantnum{GS100000001}{CCF-1814826}, by the U.S. Department of Defense, and by \grantsponsor{Ripple}{Ripple}{https://www.ripple.com}. In addition, the first author is supported by an IBM PhD Fellowship. 
  Any opinions, findings, and
  conclusions or recommendations expressed in this material are those
  of the author and do not necessarily reflect the views of the
  National Science Foundation.  
%

%
%
%
%
%

\balance{}

\bibliographystyle{ACM-Reference-Format}
\bibliography{obsidian-tochi2020}


\begin{thebibliography}{96}


\ifx \showCODEN    \undefined \def \showCODEN     #1{\unskip}     \fi
\ifx \showDOI      \undefined \def \showDOI       #1{#1}\fi
\ifx \showISBNx    \undefined \def \showISBNx     #1{\unskip}     \fi
\ifx \showISBNxiii \undefined \def \showISBNxiii  #1{\unskip}     \fi
\ifx \showISSN     \undefined \def \showISSN      #1{\unskip}     \fi
\ifx \showLCCN     \undefined \def \showLCCN      #1{\unskip}     \fi
\ifx \shownote     \undefined \def \shownote      #1{#1}          \fi
\ifx \showarticletitle \undefined \def \showarticletitle #1{#1}   \fi
\ifx \showURL      \undefined \def \showURL       {\relax}        \fi
\providecommand\bibfield[2]{#2}
\providecommand\bibinfo[2]{#2}
\providecommand\natexlab[1]{#1}
\providecommand\showeprint[2][]{arXiv:#2}

\bibitem[\protect\citeauthoryear{Ahmad, Battle, Malkani, and Kamvar}{Ahmad
  et~al\mbox{.}}{2011}]%
        {Ahmad2011:Jabberwocky}
\bibfield{author}{\bibinfo{person}{Salman Ahmad}, \bibinfo{person}{Alexis
  Battle}, \bibinfo{person}{Zahan Malkani}, {and} \bibinfo{person}{Sepander
  Kamvar}.} \bibinfo{year}{2011}\natexlab{}.
\newblock \showarticletitle{The Jabberwocky programming environment for
  structured social computing}. In \bibinfo{booktitle}{\emph{User interface
  software and technology}} \emph{(\bibinfo{series}{UIST '11})}. ACM,
  \bibinfo{pages}{53--64}.
\newblock
\urldef\tempurl%
\url{https://doi.org/10.1145/2047196.2047203}
\showDOI{\tempurl}


\bibitem[\protect\citeauthoryear{Aldrich, Sunshine, Saini, and Sparks}{Aldrich
  et~al\mbox{.}}{2009}]%
        {Aldrich09:Typestate}
\bibfield{author}{\bibinfo{person}{Jonathan Aldrich}, \bibinfo{person}{Joshua
  Sunshine}, \bibinfo{person}{Darpan Saini}, {and} \bibinfo{person}{Zachary
  Sparks}.} \bibinfo{year}{2009}\natexlab{}.
\newblock \showarticletitle{Typestate-oriented Programming}. In
  \bibinfo{booktitle}{\emph{Companion of Object Oriented Programming Systems,
  Languages, and Applications}} \emph{(\bibinfo{series}{OOPSLA '09})}.
  \bibinfo{pages}{1015–1022}.
\newblock
\urldef\tempurl%
\url{https://doi.org/10.1145/1639950.1640073}
\showDOI{\tempurl}


\bibitem[\protect\citeauthoryear{Alt and Reitwiessner}{Alt and
  Reitwiessner}{2018}]%
        {Alt2018:SMT-based}
\bibfield{author}{\bibinfo{person}{Leonardo Alt} {and}
  \bibinfo{person}{Christian Reitwiessner}.} \bibinfo{year}{2018}\natexlab{}.
\newblock \showarticletitle{SMT-Based Verification of Solidity Smart
  Contracts}. In \bibinfo{booktitle}{\emph{Leveraging Applications of Formal
  Methods, Verification and Validation. Industrial Practice}}.
\newblock


\bibitem[\protect\citeauthoryear{Barnaby, Coblenz, Etzel, Kanal, Sunshine,
  Myers, and Aldrich}{Barnaby et~al\mbox{.}}{2017}]%
        {barnaby}
\bibfield{author}{\bibinfo{person}{Celeste Barnaby}, \bibinfo{person}{Michael
  Coblenz}, \bibinfo{person}{Tyler Etzel}, \bibinfo{person}{Eliezer Kanal},
  \bibinfo{person}{Joshua Sunshine}, \bibinfo{person}{Brad Myers}, {and}
  \bibinfo{person}{Jonathan Aldrich}.} \bibinfo{year}{2017}\natexlab{}.
\newblock \showarticletitle{A User Study to Inform the Design of the Obsidian
  Blockchain DSL}. In \bibinfo{booktitle}{\emph{Workshop on Evaluation and
  Usability of Programming Languages and Tools}}
  \emph{(\bibinfo{series}{PLATEAU '17})}.
\newblock


\bibitem[\protect\citeauthoryear{Beckman, Kim, and Aldrich}{Beckman
  et~al\mbox{.}}{2011}]%
        {Beckman:2011:ESO:2032497.2032501}
\bibfield{author}{\bibinfo{person}{Nels~E. Beckman}, \bibinfo{person}{Duri
  Kim}, {and} \bibinfo{person}{Jonathan Aldrich}.}
  \bibinfo{year}{2011}\natexlab{}.
\newblock \showarticletitle{An Empirical Study of Object Protocols in the
  Wild}. In \bibinfo{booktitle}{\emph{European Conference on Object-oriented
  Programming}} \emph{(\bibinfo{series}{ECOOP '11})}. \bibinfo{pages}{2–26}.
\newblock
\urldef\tempurl%
\url{http://dl.acm.org/citation.cfm?id=2032497.2032501}
\showURL{%
\tempurl}


\bibitem[\protect\citeauthoryear{Berger, Blackburn, Hauswirth, and
  Hicks}{Berger et~al\mbox{.}}{2018}]%
        {sigplan-checklist}
\bibfield{author}{\bibinfo{person}{E.~D. Berger}, \bibinfo{person}{S.~M.
  Blackburn}, \bibinfo{person}{M. Hauswirth}, {and} \bibinfo{person}{M.
  Hicks}.} \bibinfo{year}{2018}\natexlab{}.
\newblock \bibinfo{title}{Empirical Evaluation Checklist (beta)}.
\newblock
\newblock
\urldef\tempurl%
\url{http://www.sigplan.org/Resources/EmpiricalEvaluation/}
\showURL{%
\tempurl}


\bibitem[\protect\citeauthoryear{Bergstr{\"o}m and Blackwell}{Bergstr{\"o}m and
  Blackwell}{2016}]%
        {bergstrom2016practices}
\bibfield{author}{\bibinfo{person}{Ilias Bergstr{\"o}m} {and}
  \bibinfo{person}{Alan~F Blackwell}.} \bibinfo{year}{2016}\natexlab{}.
\newblock \showarticletitle{The practices of programming}. In
  \bibinfo{booktitle}{\emph{Visual Languages and Human-Centric Computing}}
  \emph{(\bibinfo{series}{VL/HCC '16})}. IEEE, \bibinfo{publisher}{IEEE},
  \bibinfo{pages}{190--198}.
\newblock
\urldef\tempurl%
\url{https://doi.org/10.1109/VLHCC.2016.7739684}
\showDOI{\tempurl}


\bibitem[\protect\citeauthoryear{Bhargavan, Swamy, Zanella-B{\'{e}}guelin,
  Delignat-Lavaud, Fournet, Gollamudi, Gonthier, Kobeissi, Kulatova, Rastogi,
  and Sibut-Pinote}{Bhargavan et~al\mbox{.}}{2016}]%
        {Bhargavan2016}
\bibfield{author}{\bibinfo{person}{Karthikeyan Bhargavan},
  \bibinfo{person}{Nikhil Swamy}, \bibinfo{person}{Santiago
  Zanella-B{\'{e}}guelin}, \bibinfo{person}{Antoine Delignat-Lavaud},
  \bibinfo{person}{C{\'{e}}dric Fournet}, \bibinfo{person}{Anitha Gollamudi},
  \bibinfo{person}{Georges Gonthier}, \bibinfo{person}{Nadim Kobeissi},
  \bibinfo{person}{Natalia Kulatova}, \bibinfo{person}{Aseem Rastogi}, {and}
  \bibinfo{person}{Thomas Sibut-Pinote}.} \bibinfo{year}{2016}\natexlab{}.
\newblock \showarticletitle{{Formal Verification of Smart Contracts}}. In
  \bibinfo{booktitle}{\emph{ACM Workshop on Programming Languages and Analysis
  for Security}} \emph{(\bibinfo{series}{PLAS '16})}.
\newblock
\showISBNx{9781450345743}
\urldef\tempurl%
\url{https://doi.org/10.1145/2993600.2993611}
\showDOI{\tempurl}


\bibitem[\protect\citeauthoryear{Bierhoff and Aldrich}{Bierhoff and
  Aldrich}{2007}]%
        {Bierhoff:2007:MTC:1297027.1297050}
\bibfield{author}{\bibinfo{person}{Kevin Bierhoff} {and}
  \bibinfo{person}{Jonathan Aldrich}.} \bibinfo{year}{2007}\natexlab{}.
\newblock \showarticletitle{Modular Typestate Checking of Aliased Objects}. In
  \bibinfo{booktitle}{\emph{Object-oriented programming systems, languages, and
  applications}} \emph{(\bibinfo{series}{OOPSLA '07})}.
  \bibinfo{pages}{301–320}.
\newblock
\showISBNx{978-1-59593-786-5}
\urldef\tempurl%
\url{https://doi.org/10.1145/1297027.1297050}
\showDOI{\tempurl}


\bibitem[\protect\citeauthoryear{Bierhoff and Aldrich}{Bierhoff and
  Aldrich}{2008}]%
        {Bierhoff:2008:PCP:1370175.1370213}
\bibfield{author}{\bibinfo{person}{Kevin Bierhoff} {and}
  \bibinfo{person}{Jonathan Aldrich}.} \bibinfo{year}{2008}\natexlab{}.
\newblock \showarticletitle{PLURAL: Checking Protocol Compliance Under
  Aliasing}. In \bibinfo{booktitle}{\emph{Companion of International Conference
  on Software Engineering}} \emph{(\bibinfo{series}{ICSE Companion '08})}.
  \bibinfo{pages}{971–972}.
\newblock
\showISBNx{978-1-60558-079-1}
\urldef\tempurl%
\url{https://doi.org/10.1145/1370175.1370213}
\showDOI{\tempurl}


\bibitem[\protect\citeauthoryear{Blackwell and Burnett}{Blackwell and
  Burnett}{2002}]%
        {Blackwell2002:Applying}
\bibfield{author}{\bibinfo{person}{Alan Blackwell} {and}
  \bibinfo{person}{Margaret Burnett}.} \bibinfo{year}{2002}\natexlab{}.
\newblock \showarticletitle{Applying Attention Investment to End-User
  Programming}. In \bibinfo{booktitle}{\emph{Human Centric Computing Languages
  and Environments}} \emph{(\bibinfo{series}{HCC '02})}.
  \bibinfo{publisher}{IEEE Computer Society}, \bibinfo{address}{Washington, DC,
  USA}, \bibinfo{pages}{28--30}.
\newblock
\showISBNx{0-7695-1644-0}
\urldef\tempurl%
\url{https://doi.org/10.1109/HCC.2002.1046337}
\showDOI{\tempurl}


\bibitem[\protect\citeauthoryear{Blandford and Green}{Blandford and
  Green}{2008}]%
        {Blandford2008:Methodological}
\bibfield{author}{\bibinfo{person}{Ann Blandford} {and} \bibinfo{person}{Thomas
  Green}.} \bibinfo{year}{2008}\natexlab{}.
\newblock \bibinfo{booktitle}{\emph{Methodological Development}}.
\newblock \bibinfo{publisher}{Cambridge University Press},
  \bibinfo{pages}{158--174}.
\newblock
\urldef\tempurl%
\url{https://doi.org/10.1017/CBO9780511814570}
\showDOI{\tempurl}


\bibitem[\protect\citeauthoryear{Bloch}{Bloch}{2008}]%
        {bloch2008effective}
\bibfield{author}{\bibinfo{person}{Joshua Bloch}.}
  \bibinfo{year}{2008}\natexlab{}.
\newblock \bibinfo{booktitle}{\emph{Effective Java, Second Edition}}.
\newblock \bibinfo{publisher}{Addison-Wesley}.
\newblock


\bibitem[\protect\citeauthoryear{Bostock and Heer}{Bostock and Heer}{2009}]%
        {Bostock2009:Protovis}
\bibfield{author}{\bibinfo{person}{Michael Bostock} {and}
  \bibinfo{person}{Jeffrey Heer}.} \bibinfo{year}{2009}\natexlab{}.
\newblock \showarticletitle{Protovis: A graphical toolkit for visualization}.
\newblock \bibinfo{journal}{\emph{IEEE transactions on visualization and
  computer graphics}} \bibinfo{volume}{15}, \bibinfo{number}{6}
  (\bibinfo{year}{2009}), \bibinfo{pages}{1121--1128}.
\newblock
\urldef\tempurl%
\url{https://doi.org/10.1109/TVCG.2009.174}
\showDOI{\tempurl}


\bibitem[\protect\citeauthoryear{Braunschweig and Gani}{Braunschweig and
  Gani}{2002}]%
        {Braunschweig2002:Software}
\bibfield{author}{\bibinfo{person}{Bertrand Braunschweig} {and}
  \bibinfo{person}{Rafiqul Gani}.} \bibinfo{year}{2002}\natexlab{}.
\newblock \bibinfo{booktitle}{\emph{Software architectures and tools for
  computer aided process engineering}}. \bibinfo{series}{Computer Aided
  Chemical Engineering}, Vol.~\bibinfo{volume}{11}.
\newblock \bibinfo{publisher}{Elsevier}.
\newblock


\bibitem[\protect\citeauthoryear{Buse, Sadowski, and Weimer}{Buse
  et~al\mbox{.}}{2011}]%
        {Buse2011:benefits}
\bibfield{author}{\bibinfo{person}{Raymond~PL Buse}, \bibinfo{person}{Caitlin
  Sadowski}, {and} \bibinfo{person}{Westley Weimer}.}
  \bibinfo{year}{2011}\natexlab{}.
\newblock \showarticletitle{Benefits and barriers of user evaluation in
  software engineering research}. In \bibinfo{booktitle}{\emph{Object-oriented
  Programming, Systems, Languages, and Applications}}
  \emph{(\bibinfo{series}{OOPSLA '11})}. \bibinfo{pages}{643–656}.
\newblock
\urldef\tempurl%
\url{https://doi.org/10.1145/2076021.2048117}
\showDOI{\tempurl}


\bibitem[\protect\citeauthoryear{Chamberlain}{Chamberlain}{2017}]%
        {Chamberlain2017:Assessing}
\bibfield{author}{\bibinfo{person}{Roger~D. Chamberlain}.}
  \bibinfo{year}{2017}\natexlab{}.
\newblock \showarticletitle{Assessing User Preferences in Programming Language
  Design}. In \bibinfo{booktitle}{\emph{Symposium on New Ideas, New Paradigms,
  and Reflections on Programming and Software}} \emph{(\bibinfo{series}{Onward!
  2017})}. \bibinfo{publisher}{Association for Computing Machinery},
  \bibinfo{pages}{18–29}.
\newblock
\showISBNx{9781450355308}
\urldef\tempurl%
\url{https://doi.org/10.1145/3133850.3133851}
\showDOI{\tempurl}


\bibitem[\protect\citeauthoryear{Chasins}{Chasins}{2017}]%
        {Chasins2017:Helena}
\bibfield{author}{\bibinfo{person}{Sarah Chasins}.}
  \bibinfo{year}{2017}\natexlab{}.
\newblock \bibinfo{title}{Helena: Web Automation for End Users}.
\newblock
\newblock
\urldef\tempurl%
\url{http://helena-lang.org/}
\showURL{%
\tempurl}


\bibitem[\protect\citeauthoryear{Coblenz, Aldrich, Myers, and Sunshine}{Coblenz
  et~al\mbox{.}}{2020a}]%
        {Coblenz2020:Can}
\bibfield{author}{\bibinfo{person}{Michael Coblenz}, \bibinfo{person}{Jonathan
  Aldrich}, \bibinfo{person}{Brad Myers}, {and} \bibinfo{person}{Joshua
  Sunshine}.} \bibinfo{year}{2020}\natexlab{a}.
\newblock \showarticletitle{Can Advanced Type Systems Be Usable? An Empirical
  Study of Ownership, Assets, and Typestate in Obsidian}. In
  \bibinfo{booktitle}{\emph{Object-oriented programming systems, languages, and
  applications}} \emph{(\bibinfo{series}{OOPSLA '20})}.
\newblock
\newblock
\shownote{To appear.}


\bibitem[\protect\citeauthoryear{Coblenz, Aldrich, Myers, and Sunshine}{Coblenz
  et~al\mbox{.}}{2018}]%
        {Coblenz18:Interdisciplinary}
\bibfield{author}{\bibinfo{person}{Michael Coblenz}, \bibinfo{person}{Jonathan
  Aldrich}, \bibinfo{person}{Brad~A. Myers}, {and} \bibinfo{person}{Joshua
  Sunshine}.} \bibinfo{year}{2018}\natexlab{}.
\newblock \showarticletitle{Interdisciplinary Programming Language Design}. In
  \bibinfo{booktitle}{\emph{Symposium on New Ideas, New Paradigms, and
  Reflections on Programming and Software}} \emph{(\bibinfo{series}{Onward!
  '18})}. \bibinfo{pages}{133–146}.
\newblock
\urldef\tempurl%
\url{https://doi.org/10.1145/3276954.3276965}
\showDOI{\tempurl}


\bibitem[\protect\citeauthoryear{Coblenz, Kambhatla, Koronkevich, Wise,
  Barnaby, Sunshine, Aldrich, and Myers}{Coblenz et~al\mbox{.}}{2019a}]%
        {Coblenz2019:Usability}
\bibfield{author}{\bibinfo{person}{Michael Coblenz}, \bibinfo{person}{Gauri
  Kambhatla}, \bibinfo{person}{Paulette Koronkevich}, \bibinfo{person}{Jenna~L.
  Wise}, \bibinfo{person}{Celeste Barnaby}, \bibinfo{person}{Joshua Sunshine},
  \bibinfo{person}{Jonathan Aldrich}, {and} \bibinfo{person}{Brad~A. Myers}.}
  \bibinfo{year}{2019}\natexlab{a}.
\newblock \bibinfo{title}{Usability Methods for Designing Programming Languages
  for Software Engineers}.
\newblock
\newblock
\showeprint[arxiv]{1912.04719}


\bibitem[\protect\citeauthoryear{Coblenz, Nelson, Aldrich, Myers, and
  Sunshine}{Coblenz et~al\mbox{.}}{2017}]%
        {Coblenz2017:Glacier}
\bibfield{author}{\bibinfo{person}{Michael Coblenz}, \bibinfo{person}{Whitney
  Nelson}, \bibinfo{person}{Jonathan Aldrich}, \bibinfo{person}{Brad Myers},
  {and} \bibinfo{person}{Joshua Sunshine}.} \bibinfo{year}{2017}\natexlab{}.
\newblock \showarticletitle{Glacier: Transitive Class Immutability for {Java}}.
  In \bibinfo{booktitle}{\emph{International Conference on Software
  Engineering}} \emph{(\bibinfo{series}{ICSE '17})}. \bibinfo{publisher}{IEEE
  Press}, \bibinfo{pages}{496–506}.
\newblock
\urldef\tempurl%
\url{https://doi.org/10.1109/ICSE.2017.52}
\showDOI{\tempurl}


\bibitem[\protect\citeauthoryear{Coblenz, Nelson, Aldrich, Myers, and
  Sunshine}{Coblenz et~al\mbox{.}}{2020b}]%
        {Coblenz2020:Glacier}
\bibfield{author}{\bibinfo{person}{Michael Coblenz}, \bibinfo{person}{Whitney
  Nelson}, \bibinfo{person}{Jonathan Aldrich}, \bibinfo{person}{Brad Myers},
  {and} \bibinfo{person}{Joshua Sunshine}.} \bibinfo{year}{2020}\natexlab{b}.
\newblock \showarticletitle{{Glacier software and user study replication
  package}}.
\newblock  (\bibinfo{date}{7} \bibinfo{year}{2020}).
\newblock
\urldef\tempurl%
\url{https://doi.org/10.1184/R1/12108693.v1}
\showDOI{\tempurl}


\bibitem[\protect\citeauthoryear{Coblenz, Nelson, Aldrich, Myers, and
  Sunshine}{Coblenz et~al\mbox{.}}{2020c}]%
        {Coblenz2020:ObsidianReplication}
\bibfield{author}{\bibinfo{person}{Michael Coblenz}, \bibinfo{person}{Whitney
  Nelson}, \bibinfo{person}{Jonathan Aldrich}, \bibinfo{person}{Brad Myers},
  {and} \bibinfo{person}{Joshua Sunshine}.} \bibinfo{year}{2020}\natexlab{c}.
\newblock \showarticletitle{{Obsidian software and user study replication
  package}}.
\newblock  (\bibinfo{date}{8} \bibinfo{year}{2020}).
\newblock
\urldef\tempurl%
\url{https://doi.org/10.1184/R1/12771074}
\showDOI{\tempurl}


\bibitem[\protect\citeauthoryear{Coblenz, Oei, Etzel, Koronkevich, Baker,
  Bloem, Myers, Sunshine, and Aldrich}{Coblenz et~al\mbox{.}}{2019b}]%
        {Coblenz19:Obsidian-arxiv}
\bibfield{author}{\bibinfo{person}{Michael Coblenz}, \bibinfo{person}{Reed
  Oei}, \bibinfo{person}{Tyler Etzel}, \bibinfo{person}{Paulette Koronkevich},
  \bibinfo{person}{Miles Baker}, \bibinfo{person}{Yannick Bloem},
  \bibinfo{person}{Brad~A. Myers}, \bibinfo{person}{Joshua Sunshine}, {and}
  \bibinfo{person}{Jonathan Aldrich}.} \bibinfo{year}{2019}\natexlab{b}.
\newblock \bibinfo{title}{Obsidian: Typestate and Assets for Safer Blockchain
  Programming}.
\newblock
\newblock
\showeprint[arxiv]{1909.03523}


\bibitem[\protect\citeauthoryear{Coblenz, Sunshine, Aldrich, Myers, Weber, and
  Shull}{Coblenz et~al\mbox{.}}{2016a}]%
        {Coblenz2016:Exploring}
\bibfield{author}{\bibinfo{person}{Michael Coblenz}, \bibinfo{person}{Joshua
  Sunshine}, \bibinfo{person}{Jonathan Aldrich}, \bibinfo{person}{Brad Myers},
  \bibinfo{person}{Sam Weber}, {and} \bibinfo{person}{Forrest Shull}.}
  \bibinfo{year}{2016}\natexlab{a}.
\newblock \showarticletitle{Exploring Language Support for Immutability}. In
  \bibinfo{booktitle}{\emph{International Conference on Software Engineering}}
  \emph{(\bibinfo{series}{ICSE '16})}. \bibinfo{publisher}{ACM},
  \bibinfo{pages}{736–747}.
\newblock
\showISBNx{978-1-4503-3900-1}
\urldef\tempurl%
\url{https://doi.org/10.1145/2884781.2884798}
\showDOI{\tempurl}


\bibitem[\protect\citeauthoryear{Coblenz, Sunshine, Aldrich, Myers, Weber, and
  Shull}{Coblenz et~al\mbox{.}}{2016b}]%
        {Coblenz2016:Exploring-extended}
\bibfield{author}{\bibinfo{person}{Michael Coblenz}, \bibinfo{person}{Joshua
  Sunshine}, \bibinfo{person}{Jonathan Aldrich}, \bibinfo{person}{Brad Myers},
  \bibinfo{person}{Sam Weber}, {and} \bibinfo{person}{Forrest Shull}.}
  \bibinfo{year}{2016}\natexlab{b}.
\newblock \bibinfo{booktitle}{\emph{Exploring Language Support for
  Immutability}}.
\newblock \bibinfo{type}{{T}echnical {R}eport} CMU-ISR-16-106.
  \bibinfo{institution}{Carnegie Mellon University}.
\newblock


\bibitem[\protect\citeauthoryear{Dahlb{\"a}ck, J{\"o}nsson, and
  Ahrenberg}{Dahlb{\"a}ck et~al\mbox{.}}{1993}]%
        {dahlback1993wizard}
\bibfield{author}{\bibinfo{person}{Nils Dahlb{\"a}ck}, \bibinfo{person}{Arne
  J{\"o}nsson}, {and} \bibinfo{person}{Lars Ahrenberg}.}
  \bibinfo{year}{1993}\natexlab{}.
\newblock \showarticletitle{Wizard of Oz studies - why and how}.
\newblock \bibinfo{journal}{\emph{Knowledge-based systems}}
  \bibinfo{volume}{6}, \bibinfo{number}{4} (\bibinfo{year}{1993}),
  \bibinfo{pages}{258--266}.
\newblock
\urldef\tempurl%
\url{https://doi.org/10.1016/0950-7051(93)90017-N}
\showDOI{\tempurl}


\bibitem[\protect\citeauthoryear{Daian}{Daian}{2016}]%
        {DAO-details}
\bibfield{author}{\bibinfo{person}{Phil Daian}.}
  \bibinfo{year}{2016}\natexlab{}.
\newblock \bibinfo{title}{Analysis of the {DAO} exploit}.
\newblock
\newblock
\urldef\tempurl%
\url{http://hackingdistributed.com/2016/06/18/analysis-of-the-dao-exploit/}
\showURL{%
Retrieved August 21, 2018 from \tempurl}


\bibitem[\protect\citeauthoryear{Delmolino, Arnett, Kosba, Miller, and
  Shi}{Delmolino et~al\mbox{.}}{2016}]%
        {Delmolino2016:Step}
\bibfield{author}{\bibinfo{person}{Kevin Delmolino}, \bibinfo{person}{Mitchell
  Arnett}, \bibinfo{person}{Ahmed Kosba}, \bibinfo{person}{Andrew Miller},
  {and} \bibinfo{person}{Elaine Shi}.} \bibinfo{year}{2016}\natexlab{}.
\newblock \showarticletitle{Step by step towards creating a safe smart
  contract: Lessons and insights from a cryptocurrency lab}. In
  \bibinfo{booktitle}{\emph{International conference on financial cryptography
  and data security}}.
\newblock
\urldef\tempurl%
\url{https://doi.org/10.1007/978-3-662-53357-4_6}
\showDOI{\tempurl}


\bibitem[\protect\citeauthoryear{Dumas and Redish}{Dumas and Redish}{1999}]%
        {Dumas1999:Practical}
\bibfield{author}{\bibinfo{person}{Joseph~S Dumas} {and}
  \bibinfo{person}{Janice Redish}.} \bibinfo{year}{1999}\natexlab{}.
\newblock \bibinfo{booktitle}{\emph{A practical guide to usability testing}}.
\newblock \bibinfo{publisher}{Intellect books}.
\newblock


\bibitem[\protect\citeauthoryear{Elsden, Manohar, Briggs, Harding, Speed, and
  Vines}{Elsden et~al\mbox{.}}{2018}]%
        {Elsden18:Making}
\bibfield{author}{\bibinfo{person}{Chris Elsden}, \bibinfo{person}{Arthi
  Manohar}, \bibinfo{person}{Jo Briggs}, \bibinfo{person}{Mike Harding},
  \bibinfo{person}{Chris Speed}, {and} \bibinfo{person}{John Vines}.}
  \bibinfo{year}{2018}\natexlab{}.
\newblock \showarticletitle{Making Sense of Blockchain Applications: A Typology
  for HCI}. In \bibinfo{booktitle}{\emph{CHI Conference on Human Factors in
  Computing Systems}} \emph{(\bibinfo{series}{CHI '18})}.
  \bibinfo{pages}{1--14}.
\newblock
\showISBNx{978-1-4503-5620-6}
\urldef\tempurl%
\url{https://doi.org/10.1145/3173574.3174032}
\showDOI{\tempurl}


\bibitem[\protect\citeauthoryear{Endrikat, Hanenberg, Robbes, and
  Stefik}{Endrikat et~al\mbox{.}}{2014}]%
        {Endrikat:2014:ADS:2568225.2568299}
\bibfield{author}{\bibinfo{person}{Stefan Endrikat}, \bibinfo{person}{Stefan
  Hanenberg}, \bibinfo{person}{Romain Robbes}, {and} \bibinfo{person}{Andreas
  Stefik}.} \bibinfo{year}{2014}\natexlab{}.
\newblock \showarticletitle{How Do {API} Documentation and Static Typing Affect
  {API} Usability?}. In \bibinfo{booktitle}{\emph{International Conference on
  Software Engineering}} \emph{(\bibinfo{series}{ICSE '14})}.
  \bibinfo{publisher}{ACM}, \bibinfo{pages}{632--642}.
\newblock
\showISBNx{978-1-4503-2756-5}
\urldef\tempurl%
\url{https://doi.org/10.1145/2568225.2568299}
\showDOI{\tempurl}


\bibitem[\protect\citeauthoryear{Ericsson and Simon}{Ericsson and
  Simon}{1984}]%
        {Ericsson1984:Protocol}
\bibfield{author}{\bibinfo{person}{K~Anders Ericsson} {and}
  \bibinfo{person}{Herbert~A Simon}.} \bibinfo{year}{1984}\natexlab{}.
\newblock \bibinfo{booktitle}{\emph{Protocol analysis: Verbal reports as
  data.}}
\newblock \bibinfo{publisher}{the MIT Press}.
\newblock


\bibitem[\protect\citeauthoryear{{Ethereum Foundation}}{{Ethereum
  Foundation}}{2020a}]%
        {Solidity-Patterns}
\bibfield{author}{\bibinfo{person}{{Ethereum Foundation}}.}
  \bibinfo{year}{2020}\natexlab{a}.
\newblock \bibinfo{title}{Common Patterns}.
\newblock
\newblock
\urldef\tempurl%
\url{http://solidity.readthedocs.io/en/develop/common-patterns.html}
\showURL{%
Retrieved February 18, 2020 from \tempurl}


\bibitem[\protect\citeauthoryear{{Ethereum Foundation}}{{Ethereum
  Foundation}}{2020b}]%
        {Solidity}
\bibfield{author}{\bibinfo{person}{{Ethereum Foundation}}.}
  \bibinfo{year}{2020}\natexlab{b}.
\newblock \bibinfo{title}{Solidity}.
\newblock
\newblock
\urldef\tempurl%
\url{https://solidity.readthedocs.io/en/develop/}
\showURL{%
Retrieved February 18, 2020 from \tempurl}


\bibitem[\protect\citeauthoryear{Graham}{Graham}{2017}]%
        {CNBC}
\bibfield{author}{\bibinfo{person}{Luke Graham}.}
  \bibinfo{year}{2017}\natexlab{}.
\newblock \bibinfo{title}{\$32 million worth of digital currency ether stolen
  by hackers}.
\newblock
\newblock
\urldef\tempurl%
\url{https://www.cnbc.com/2017/07/20/32-million-worth-of-digital-currency-ether-stolen-by-hackers.html}
\showURL{%
Retrieved November 2, 2017 from \tempurl}


\bibitem[\protect\citeauthoryear{Graham}{Graham}{2001}]%
        {Graham}
\bibfield{author}{\bibinfo{person}{Paul Graham}.}
  \bibinfo{year}{2001}\natexlab{}.
\newblock \bibinfo{title}{Five Questions about Language Design}.
\newblock
\newblock
\urldef\tempurl%
\url{http://www.paulgraham.com/langdes.html}
\showURL{%
\tempurl}


\bibitem[\protect\citeauthoryear{Green and Petre}{Green and Petre}{1996}]%
        {green1996usability}
\bibfield{author}{\bibinfo{person}{Thomas R.~G. Green} {and}
  \bibinfo{person}{Marian Petre}.} \bibinfo{year}{1996}\natexlab{}.
\newblock \showarticletitle{Usability analysis of visual programming
  environments: a `cognitive dimensions' framework}.
\newblock \bibinfo{journal}{\emph{Journal of Visual Languages \& Computing}}
  \bibinfo{volume}{7}, \bibinfo{number}{2} (\bibinfo{year}{1996}),
  \bibinfo{pages}{131--174}.
\newblock


\bibitem[\protect\citeauthoryear{Guindon, Krasner, Curtis,
  et~al\mbox{.}}{Guindon et~al\mbox{.}}{1987}]%
        {Guindon1987:Breakdowns}
\bibfield{author}{\bibinfo{person}{Raymonde Guindon}, \bibinfo{person}{Herb
  Krasner}, \bibinfo{person}{Bill Curtis}, {et~al\mbox{.}}}
  \bibinfo{year}{1987}\natexlab{}.
\newblock \showarticletitle{Breakdowns and processes during the early
  activities of software design by professionals}. In
  \bibinfo{booktitle}{\emph{Empirical studies of programmers: Second
  Workshop}}. \bibinfo{pages}{65--82}.
\newblock


\bibitem[\protect\citeauthoryear{Gulliksen, G{\"o}ransson, Boivie, Blomkvist,
  Persson, and Cajander}{Gulliksen et~al\mbox{.}}{2003}]%
        {Gulliksen2003:Key}
\bibfield{author}{\bibinfo{person}{Jan Gulliksen}, \bibinfo{person}{Bengt
  G{\"o}ransson}, \bibinfo{person}{Inger Boivie}, \bibinfo{person}{Stefan
  Blomkvist}, \bibinfo{person}{Jenny Persson}, {and} \bibinfo{person}{{\AA}sa
  Cajander}.} \bibinfo{year}{2003}\natexlab{}.
\newblock \showarticletitle{Key principles for user-centred systems design}.
\newblock \bibinfo{journal}{\emph{Behaviour and Information Technology}}
  \bibinfo{volume}{22}, \bibinfo{number}{6} (\bibinfo{year}{2003}),
  \bibinfo{pages}{397--409}.
\newblock
\urldef\tempurl%
\url{https://doi.org/10.1080/01449290310001624329}
\showDOI{\tempurl}


\bibitem[\protect\citeauthoryear{{Harvard Business Review}}{{Harvard Business
  Review}}{2017}]%
        {HealthCare}
\bibfield{author}{\bibinfo{person}{{Harvard Business Review}}.}
  \bibinfo{year}{2017}\natexlab{}.
\newblock \bibinfo{title}{The Potential for Blockchain to Transform Electronic
  Health Records}.
\newblock
\newblock
\urldef\tempurl%
\url{https://hbr.org/2017/03/the-potential-for-blockchain-to-transform-electronic-health-records}
\showURL{%
Retrieved February 18, 2020 from \tempurl}


\bibitem[\protect\citeauthoryear{Harz and Knottenbelt}{Harz and
  Knottenbelt}{2018}]%
        {Harz2018:Towards}
\bibfield{author}{\bibinfo{person}{Dominik Harz} {and} \bibinfo{person}{William
  Knottenbelt}.} \bibinfo{year}{2018}\natexlab{}.
\newblock \bibinfo{title}{Towards Safer Smart Contracts: A Survey of Languages
  and Verification Methods}.
\newblock
\newblock
\showeprint[arxiv]{1809.09805}


\bibitem[\protect\citeauthoryear{Herlihy}{Herlihy}{2019}]%
        {herlihy2019blockchains}
\bibfield{author}{\bibinfo{person}{Maurice Herlihy}.}
  \bibinfo{year}{2019}\natexlab{}.
\newblock \showarticletitle{Blockchains from a distributed computing
  perspective}.
\newblock \bibinfo{journal}{\emph{Commun. ACM}} \bibinfo{volume}{62},
  \bibinfo{number}{2} (\bibinfo{year}{2019}), \bibinfo{pages}{78--85}.
\newblock


\bibitem[\protect\citeauthoryear{Hoare}{Hoare}{2009}]%
        {Hoare2009:Null}
\bibfield{author}{\bibinfo{person}{C.~A.~R. Hoare}.}
  \bibinfo{year}{2009}\natexlab{}.
\newblock \bibinfo{title}{Null References: The Billion Dollar Mistake}.
\newblock
\newblock
\urldef\tempurl%
\url{https://www.infoq.com/presentations/Null-References-The-Billion-Dollar-Mistake-Tony-Hoare/}
\showURL{%
Retrieved February 18, 2020 from \tempurl}


\bibitem[\protect\citeauthoryear{IBM}{IBM}{2019}]%
        {SupplyChain}
\bibfield{author}{\bibinfo{person}{IBM}.} \bibinfo{year}{2019}\natexlab{}.
\newblock \bibinfo{title}{Blockchain for supply chain}.
\newblock
\newblock
\urldef\tempurl%
\url{https://www.ibm.com/blockchain/supply-chain/}
\showURL{%
Retrieved March 31, 2019 from \tempurl}


\bibitem[\protect\citeauthoryear{Jones, Blackwell, and Burnett}{Jones
  et~al\mbox{.}}{2003}]%
        {Jones2003:User}
\bibfield{author}{\bibinfo{person}{Simon~Peyton Jones}, \bibinfo{person}{Alan
  Blackwell}, {and} \bibinfo{person}{Margaret Burnett}.}
  \bibinfo{year}{2003}\natexlab{}.
\newblock \showarticletitle{A User-centred Approach to Functions in Excel}. In
  \bibinfo{booktitle}{\emph{International Conference on Functional
  Programming}} \emph{(\bibinfo{series}{ICFP '03})}. \bibinfo{publisher}{ACM},
  \bibinfo{pages}{165--176}.
\newblock
\showISBNx{1-58113-756-7}
\urldef\tempurl%
\url{https://doi.org/10.1145/944705.944721}
\showDOI{\tempurl}


\bibitem[\protect\citeauthoryear{Kelleher and Pausch}{Kelleher and
  Pausch}{2005}]%
        {Kelleher2005:Lowering}
\bibfield{author}{\bibinfo{person}{Caitlin Kelleher} {and}
  \bibinfo{person}{Randy Pausch}.} \bibinfo{year}{2005}\natexlab{}.
\newblock \showarticletitle{Lowering the barriers to programming: A taxonomy of
  programming environments and languages for novice programmers}.
\newblock \bibinfo{journal}{\emph{ACM Computing Surveys (CSUR)}}
  \bibinfo{volume}{37}, \bibinfo{number}{2} (\bibinfo{year}{2005}),
  \bibinfo{pages}{83--137}.
\newblock


\bibitem[\protect\citeauthoryear{Ko, Abraham, Beckwith, Blackwell, Burnett,
  Erwig, Scaffidi, Lawrance, Lieberman, Myers, et~al\mbox{.}}{Ko
  et~al\mbox{.}}{2011}]%
        {Ko2011:State}
\bibfield{author}{\bibinfo{person}{Amy~J. Ko}, \bibinfo{person}{Robin Abraham},
  \bibinfo{person}{Laura Beckwith}, \bibinfo{person}{Alan Blackwell},
  \bibinfo{person}{Margaret Burnett}, \bibinfo{person}{Martin Erwig},
  \bibinfo{person}{Chris Scaffidi}, \bibinfo{person}{Joseph Lawrance},
  \bibinfo{person}{Henry Lieberman}, \bibinfo{person}{Brad Myers},
  {et~al\mbox{.}}} \bibinfo{year}{2011}\natexlab{}.
\newblock \showarticletitle{The state of the art in end-user software
  engineering}.
\newblock \bibinfo{journal}{\emph{ACM Computing Surveys (CSUR)}}
  \bibinfo{volume}{43}, \bibinfo{number}{3}, Article \bibinfo{articleno}{21}
  (\bibinfo{year}{2011}), \bibinfo{numpages}{44}~pages.
\newblock
\urldef\tempurl%
\url{https://doi.org/10.1145/1922649.1922658}
\showDOI{\tempurl}


\bibitem[\protect\citeauthoryear{Ko, LaToza, and Burnett}{Ko
  et~al\mbox{.}}{2015}]%
        {Ko2015:practical}
\bibfield{author}{\bibinfo{person}{Amy~J. Ko}, \bibinfo{person}{Thomas~D.
  LaToza}, {and} \bibinfo{person}{Margaret~M. Burnett}.}
  \bibinfo{year}{2015}\natexlab{}.
\newblock \showarticletitle{A practical guide to controlled experiments of
  software engineering tools with human participants}.
\newblock \bibinfo{journal}{\emph{Empirical Software Engineering}}
  \bibinfo{volume}{20}, \bibinfo{number}{1} (\bibinfo{year}{2015}),
  \bibinfo{pages}{110--141}.
\newblock
\urldef\tempurl%
\url{https://doi.org/10.1007/s10664-013-9279-3}
\showDOI{\tempurl}


\bibitem[\protect\citeauthoryear{Krasner, Curtis, and Iscoe}{Krasner
  et~al\mbox{.}}{1987}]%
        {Krasner1987:Communication}
\bibfield{author}{\bibinfo{person}{Herb Krasner}, \bibinfo{person}{Bill
  Curtis}, {and} \bibinfo{person}{Neil Iscoe}.}
  \bibinfo{year}{1987}\natexlab{}.
\newblock \bibinfo{booktitle}{\emph{Communication Breakdowns and Boundary
  Spanning Activities on Large Programming Projects}}.
\newblock \bibinfo{pages}{47–64}.
\newblock


\bibitem[\protect\citeauthoryear{Lazar, Feng, and Hochheiser}{Lazar
  et~al\mbox{.}}{2010}]%
        {Lazar2010:Research}
\bibfield{author}{\bibinfo{person}{Jonathan Lazar},
  \bibinfo{person}{Jinjuan~Heidi Feng}, {and} \bibinfo{person}{Harry
  Hochheiser}.} \bibinfo{year}{2010}\natexlab{}.
\newblock \bibinfo{booktitle}{\emph{Research Methods in Human-Computer
  Interaction}}.
\newblock \bibinfo{publisher}{Wiley Publishing}.
\newblock
\showISBNx{0470723378}


\bibitem[\protect\citeauthoryear{Loksa, Ko, Jernigan, Oleson, Mendez, and
  Burnett}{Loksa et~al\mbox{.}}{2016}]%
        {Loksa2016:Programming}
\bibfield{author}{\bibinfo{person}{Dastyni Loksa}, \bibinfo{person}{Amy~J. Ko},
  \bibinfo{person}{Will Jernigan}, \bibinfo{person}{Alannah Oleson},
  \bibinfo{person}{Christopher~J. Mendez}, {and} \bibinfo{person}{Margaret~M.
  Burnett}.} \bibinfo{year}{2016}\natexlab{}.
\newblock \showarticletitle{Programming, Problem Solving, and Self-Awareness:
  Effects of Explicit Guidance}. In \bibinfo{booktitle}{\emph{SIGCHI Conference
  on Human Factors in Computing Systems}} \emph{(\bibinfo{series}{CHI '16})}.
  \bibinfo{publisher}{ACM}, \bibinfo{pages}{1449--1461}.
\newblock
\showISBNx{978-1-4503-3362-7}
\urldef\tempurl%
\url{https://doi.org/10.1145/2858036.2858252}
\showDOI{\tempurl}


\bibitem[\protect\citeauthoryear{Mikhajlov and Sekerinski}{Mikhajlov and
  Sekerinski}{1998}]%
        {Mikhajlov:1998:SFB:646155.679700}
\bibfield{author}{\bibinfo{person}{Leonid Mikhajlov} {and}
  \bibinfo{person}{Emil Sekerinski}.} \bibinfo{year}{1998}\natexlab{}.
\newblock \showarticletitle{A Study of The Fragile Base Class Problem}. In
  \bibinfo{booktitle}{\emph{European Conference on Object-Oriented
  Programming}} \emph{(\bibinfo{series}{ECOOP 1998})}.
  \bibinfo{pages}{355–382}.
\newblock


\bibitem[\protect\citeauthoryear{Miller}{Miller}{1974}]%
        {Miller1974:Programming}
\bibfield{author}{\bibinfo{person}{Lance~A. Miller}.}
  \bibinfo{year}{1974}\natexlab{}.
\newblock \showarticletitle{Programming by Non-programmers}.
\newblock \bibinfo{journal}{\emph{International Journal of Man-Machine
  Studies}} (\bibinfo{year}{1974}).
\newblock
\urldef\tempurl%
\url{https://doi.org/10.1016/S0020-7373(74)80004-0}
\showDOI{\tempurl}


\bibitem[\protect\citeauthoryear{Myers, Ko, LaToza, and Yoon}{Myers
  et~al\mbox{.}}{2016}]%
        {Myers2016:Programmers}
\bibfield{author}{\bibinfo{person}{Brad~A. Myers}, \bibinfo{person}{Amy~J. Ko},
  \bibinfo{person}{Thomas~D. LaToza}, {and} \bibinfo{person}{YoungSeok Yoon}.}
  \bibinfo{year}{2016}\natexlab{}.
\newblock \showarticletitle{Programmers Are Users Too: Human-Centered Methods
  for Improving Programming Tools}.
\newblock \bibinfo{journal}{\emph{Computer}} \bibinfo{volume}{49},
  \bibinfo{number}{7} (\bibinfo{date}{July} \bibinfo{year}{2016}),
  \bibinfo{pages}{44--52}.
\newblock
\showISSN{0018-9162}
\urldef\tempurl%
\url{https://doi.org/10.1109/MC.2016.200}
\showDOI{\tempurl}


\bibitem[\protect\citeauthoryear{Myers, Pane, and Ko}{Myers
  et~al\mbox{.}}{2004}]%
        {naturalprogramming}
\bibfield{author}{\bibinfo{person}{Brad~A. Myers}, \bibinfo{person}{John~F.
  Pane}, {and} \bibinfo{person}{Amy~J. Ko}.} \bibinfo{year}{2004}\natexlab{}.
\newblock \showarticletitle{Natural Programming Languages and Environments}.
\newblock \bibinfo{journal}{\emph{Commun. ACM}}  \bibinfo{volume}{47}
  (\bibinfo{year}{2004}), \bibinfo{pages}{47--52}.
\newblock
Issue 9.
\urldef\tempurl%
\url{https://doi.org/10.1145/1015864.1015888}
\showDOI{\tempurl}


\bibitem[\protect\citeauthoryear{Newell and Card}{Newell and Card}{1985}]%
        {Newell1985:Prospects}
\bibfield{author}{\bibinfo{person}{Allen Newell} {and}
  \bibinfo{person}{Stuart~K. Card}.} \bibinfo{year}{1985}\natexlab{}.
\newblock \showarticletitle{The Prospects for Psychological Science in
  Human-computer Interaction}.
\newblock \bibinfo{journal}{\emph{Hum.-Comput. Interact.}} \bibinfo{volume}{1},
  \bibinfo{number}{3} (\bibinfo{date}{Sept.} \bibinfo{year}{1985}),
  \bibinfo{pages}{209--242}.
\newblock
\showISSN{0737-0024}
\urldef\tempurl%
\url{https://doi.org/10.1207/s15327051hci0103_1}
\showDOI{\tempurl}


\bibitem[\protect\citeauthoryear{{Nichols}}{{Nichols}}{2019}]%
        {Nichols2019:End}
\bibfield{author}{\bibinfo{person}{W.~R. {Nichols}}.}
  \bibinfo{year}{2019}\natexlab{}.
\newblock \showarticletitle{The End to the Myth of Individual Programmer
  Productivity}.
\newblock \bibinfo{journal}{\emph{IEEE Software}} \bibinfo{volume}{36},
  \bibinfo{number}{5} (\bibinfo{year}{2019}), \bibinfo{pages}{71--75}.
\newblock


\bibitem[\protect\citeauthoryear{Nielsen and Landauer}{Nielsen and
  Landauer}{1993}]%
        {Nielsen:Mathematical}
\bibfield{author}{\bibinfo{person}{Jakob Nielsen} {and}
  \bibinfo{person}{Thomas~K. Landauer}.} \bibinfo{year}{1993}\natexlab{}.
\newblock \showarticletitle{A Mathematical Model of the Finding of Usability
  Problems}. In \bibinfo{booktitle}{\emph{Proceedings of the INTERACT ’93 and
  CHI ’93 Conference on Human Factors in Computing Systems}}
  \emph{(\bibinfo{series}{CHI ’93})}. \bibinfo{publisher}{Association for
  Computing Machinery}, \bibinfo{address}{New York, NY, USA},
  \bibinfo{pages}{206–213}.
\newblock
\showISBNx{0897915755}
\urldef\tempurl%
\url{https://doi.org/10.1145/169059.169166}
\showDOI{\tempurl}


\bibitem[\protect\citeauthoryear{of~Washington}{of~Washington}{2019}]%
        {UW2019:PMP}
\bibfield{author}{\bibinfo{person}{University of Washington}.}
  \bibinfo{year}{2019}\natexlab{}.
\newblock \bibinfo{title}{Professional Master's Program}.
\newblock
\newblock
\urldef\tempurl%
\url{https://www.cs.washington.edu/academics/pmp}
\showURL{%
Retrieved February 18, 2020 from \tempurl}


\bibitem[\protect\citeauthoryear{Oney, Myers, and Brandt}{Oney
  et~al\mbox{.}}{2014}]%
        {Oney2014:Interstate}
\bibfield{author}{\bibinfo{person}{Stephen Oney}, \bibinfo{person}{Brad~A.
  Myers}, {and} \bibinfo{person}{Joel Brandt}.}
  \bibinfo{year}{2014}\natexlab{}.
\newblock \showarticletitle{InterState: a language and environment for
  expressing interface behavior}. In \bibinfo{booktitle}{\emph{User interface
  software and technology}} \emph{(\bibinfo{series}{UIST '14})}. ACM,
  \bibinfo{pages}{263--272}.
\newblock
\urldef\tempurl%
\url{https://doi.org/10.1145/2642918.2647358}
\showDOI{\tempurl}


\bibitem[\protect\citeauthoryear{Pane, Myers, and Miller}{Pane
  et~al\mbox{.}}{2002}]%
        {Pane2002:Using}
\bibfield{author}{\bibinfo{person}{John~F. Pane}, \bibinfo{person}{Brad~A.
  Myers}, {and} \bibinfo{person}{Leah~B. Miller}.}
  \bibinfo{year}{2002}\natexlab{}.
\newblock \showarticletitle{Using HCI techniques to design a more usable
  programming system}. In \bibinfo{booktitle}{\emph{Human Centric Computing
  Languages and Environments}} \emph{(\bibinfo{series}{HCC '02})}.
  \bibinfo{pages}{198--206}.
\newblock
\urldef\tempurl%
\url{https://doi.org/10.1109/HCC.2002.1046372}
\showDOI{\tempurl}


\bibitem[\protect\citeauthoryear{Pennington}{Pennington}{1987}]%
        {Pennington1987:Stimulus}
\bibfield{author}{\bibinfo{person}{Nancy Pennington}.}
  \bibinfo{year}{1987}\natexlab{}.
\newblock \showarticletitle{Stimulus structures and mental representations in
  expert comprehension of computer programs}.
\newblock \bibinfo{journal}{\emph{Cognitive Psychology}} \bibinfo{volume}{19},
  \bibinfo{number}{3} (\bibinfo{year}{1987}), \bibinfo{pages}{295 -- 341}.
\newblock
\showISSN{0010-0285}
\urldef\tempurl%
\url{https://doi.org/10.1016/0010-0285(87)90007-7}
\showDOI{\tempurl}


\bibitem[\protect\citeauthoryear{Pernice and Whintenton}{Pernice and
  Whintenton}{2017}]%
        {Nielsen:BadSuggestions}
\bibfield{author}{\bibinfo{person}{Kara Pernice} {and} \bibinfo{person}{Kathryn
  Whintenton}.} \bibinfo{year}{2017}\natexlab{}.
\newblock \bibinfo{title}{How to Deal With Bad Design Suggestions}.
\newblock
\newblock
\urldef\tempurl%
\url{https://www.nngroup.com/articles/bad-design-suggestions/}
\showURL{%
Retrieved February 18, 2020 from \tempurl}


\bibitem[\protect\citeauthoryear{Perry, Sim, and Easterbrook}{Perry
  et~al\mbox{.}}{2004}]%
        {Perry2004:Case}
\bibfield{author}{\bibinfo{person}{Dewayne~E. Perry},
  \bibinfo{person}{Susan~Elliott Sim}, {and} \bibinfo{person}{Steve~M.
  Easterbrook}.} \bibinfo{year}{2004}\natexlab{}.
\newblock \showarticletitle{Case studies for software engineers}. In
  \bibinfo{booktitle}{\emph{International Conference on Software Engineering}}
  \emph{(\bibinfo{series}{ICSE '04})}. IEEE, \bibinfo{pages}{736--738}.
\newblock


\bibitem[\protect\citeauthoryear{Pierce}{Pierce}{2002}]%
        {Pierce:TypeSystems}
\bibfield{author}{\bibinfo{person}{Benjamin~C. Pierce}.}
  \bibinfo{year}{2002}\natexlab{}.
\newblock \bibinfo{booktitle}{\emph{Types and Programming Languages}}.
\newblock \bibinfo{publisher}{MIT Press}.
\newblock


\bibitem[\protect\citeauthoryear{Pierce and Mandelbaum}{Pierce and
  Mandelbaum}{2009}]%
        {Morrisett2009:Grand}
\bibfield{author}{\bibinfo{person}{Benjamin~C. Pierce} {and}
  \bibinfo{person}{Yitzhak Mandelbaum}.} \bibinfo{year}{2009}\natexlab{}.
\newblock \bibinfo{title}{PL Grand Challenges}.
\newblock
\newblock
\urldef\tempurl%
\url{http://plgrand.blogspot.com}
\showURL{%
Retrieved February 18, 2020 from \tempurl}


\bibitem[\protect\citeauthoryear{{Qualtrics}}{{Qualtrics}}{2020}]%
        {Qualtrics}
\bibfield{author}{\bibinfo{person}{{Qualtrics}}.}
  \bibinfo{year}{2020}\natexlab{}.
\newblock \bibinfo{title}{Qualtrics Software}.
\newblock
\newblock
\urldef\tempurl%
\url{http://www.qualtrics.com/}
\showURL{%
\tempurl}


\bibitem[\protect\citeauthoryear{\relax{Microsoft Corp.}}{\relax{Microsoft
  Corp.}}{2008}]%
        {microsoft-struct}
\bibfield{author}{\bibinfo{person}{\relax{Microsoft Corp.}}}
  \bibinfo{year}{2008}\natexlab{}.
\newblock \bibinfo{title}{Framework Design Guidelines}.
\newblock
\newblock
\urldef\tempurl%
\url{https://docs.microsoft.com/en-us/dotnet/standard/design-guidelines/struct}
\showURL{%
Retrieved February 18, 2020 from \tempurl}


\bibitem[\protect\citeauthoryear{\relax{Oracle Corp.}}{\relax{Oracle
  Corp.}}{2019}]%
        {securecoding}
\bibfield{author}{\bibinfo{person}{\relax{Oracle Corp.}}}
  \bibinfo{year}{2019}\natexlab{}.
\newblock \bibinfo{title}{Secure Coding Guidelines for the {Java} {SE}, version
  4.0}.
\newblock
\newblock
\urldef\tempurl%
\url{https://www.oracle.com/technetwork/java/seccodeguide-139067.html}
\showURL{%
Retrieved February 18, 2020 from \tempurl}


\bibitem[\protect\citeauthoryear{Resnick, Maloney, Monroy-Hern{\'a}ndez, Rusk,
  Eastmond, Brennan, Millner, Rosenbaum, Silver, Silverman,
  et~al\mbox{.}}{Resnick et~al\mbox{.}}{2009}]%
        {resnick2009scratch}
\bibfield{author}{\bibinfo{person}{M. Resnick}, \bibinfo{person}{J. Maloney},
  \bibinfo{person}{A. Monroy-Hern{\'a}ndez}, \bibinfo{person}{N. Rusk},
  \bibinfo{person}{E. Eastmond}, \bibinfo{person}{K. Brennan},
  \bibinfo{person}{A. Millner}, \bibinfo{person}{E. Rosenbaum},
  \bibinfo{person}{J. Silver}, \bibinfo{person}{B. Silverman}, {et~al\mbox{.}}}
  \bibinfo{year}{2009}\natexlab{}.
\newblock \showarticletitle{Scratch: Programming for all.}
\newblock \bibinfo{journal}{\emph{Commun. Acm}} \bibinfo{volume}{52},
  \bibinfo{number}{11} (\bibinfo{year}{2009}), \bibinfo{pages}{60--67}.
\newblock
\urldef\tempurl%
\url{https://doi.org/10.1145/1592761.1592779}
\showDOI{\tempurl}


\bibitem[\protect\citeauthoryear{Robertson and Radcliffe}{Robertson and
  Radcliffe}{2009}]%
        {Robertson2009:Impact}
\bibfield{author}{\bibinfo{person}{BF Robertson} {and} \bibinfo{person}{DF
  Radcliffe}.} \bibinfo{year}{2009}\natexlab{}.
\newblock \showarticletitle{Impact of CAD tools on creative problem solving in
  engineering design}.
\newblock \bibinfo{journal}{\emph{Computer-Aided Design}} \bibinfo{volume}{41},
  \bibinfo{number}{3} (\bibinfo{year}{2009}), \bibinfo{pages}{136--146}.
\newblock
\urldef\tempurl%
\url{https://doi.org/10.1016/j.cad.2008.06.007}
\showDOI{\tempurl}


\bibitem[\protect\citeauthoryear{Rothbauer}{Rothbauer}{2008}]%
        {Given2008:Sage}
\bibfield{author}{\bibinfo{person}{Paulette~M. Rothbauer}.}
  \bibinfo{year}{2008}\natexlab{}.
\newblock \bibinfo{booktitle}{\emph{The Sage encyclopedia of qualitative
  research methods}}.
\newblock \bibinfo{publisher}{SAGE}.
\newblock
\showISBNx{9781412963909}


\bibitem[\protect\citeauthoryear{Satyanarayan, Russell, Hoffswell, and
  Heer}{Satyanarayan et~al\mbox{.}}{2015}]%
        {Satyanarayan2015:Reactive}
\bibfield{author}{\bibinfo{person}{Arvind Satyanarayan}, \bibinfo{person}{Ryan
  Russell}, \bibinfo{person}{Jane Hoffswell}, {and} \bibinfo{person}{Jeffrey
  Heer}.} \bibinfo{year}{2015}\natexlab{}.
\newblock \showarticletitle{Reactive Vega: A Streaming Dataflow Architecture
  for Declarative Interactive Visualization}.
\newblock \bibinfo{journal}{\emph{IEEE transactions on visualization and
  computer graphics}} \bibinfo{volume}{22}, \bibinfo{number}{1}
  (\bibinfo{year}{2015}), \bibinfo{pages}{659--668}.
\newblock
\urldef\tempurl%
\url{https://doi.org/10.1109/TVCG.2015.2467091}
\showDOI{\tempurl}


\bibitem[\protect\citeauthoryear{Shneiderman}{Shneiderman}{1986}]%
        {Shneiderman1986:Empirical}
\bibfield{author}{\bibinfo{person}{Ben Shneiderman}.}
  \bibinfo{year}{1986}\natexlab{}.
\newblock \showarticletitle{Empirical Studies of Programmers: The Territory,
  Paths, and Destinations}.
\newblock \bibinfo{journal}{\emph{Empirical Studies of Programmers}}
  (\bibinfo{year}{1986}), \bibinfo{pages}{1--12}.
\newblock


\bibitem[\protect\citeauthoryear{Shneiderman and Plaisant}{Shneiderman and
  Plaisant}{2006}]%
        {Shneiderman2006:Strategies}
\bibfield{author}{\bibinfo{person}{Ben Shneiderman} {and}
  \bibinfo{person}{Catherine Plaisant}.} \bibinfo{year}{2006}\natexlab{}.
\newblock \showarticletitle{Strategies for evaluating information visualization
  tools: multi-dimensional in-depth long-term case studies}. In
  \bibinfo{booktitle}{\emph{AVI workshop on BEyond time and errors: novel
  evaluation methods for information visualization}}. ACM,
  \bibinfo{publisher}{ACM}, \bibinfo{pages}{1--7}.
\newblock
\urldef\tempurl%
\url{https://doi.org/10.1145/1168149.1168158}
\showDOI{\tempurl}


\bibitem[\protect\citeauthoryear{Sime, Green, and Guest}{Sime
  et~al\mbox{.}}{1977}]%
        {Sime1977:Scope}
\bibfield{author}{\bibinfo{person}{Max~E. Sime}, \bibinfo{person}{Thomas R.~G.
  Green}, {and} \bibinfo{person}{DJ Guest}.} \bibinfo{year}{1977}\natexlab{}.
\newblock \showarticletitle{Scope marking in computer conditionals—a
  psychological evaluation}.
\newblock \bibinfo{journal}{\emph{International Journal of Man-Machine
  Studies}} \bibinfo{volume}{9}, \bibinfo{number}{1} (\bibinfo{year}{1977}),
  \bibinfo{pages}{107--118}.
\newblock


\bibitem[\protect\citeauthoryear{Sirer}{Sirer}{2016}]%
        {DAO}
\bibfield{author}{\bibinfo{person}{Emin~G{\"u}n Sirer}.}
  \bibinfo{year}{2016}\natexlab{}.
\newblock \bibinfo{title}{Thoughts on The {DAO} Hack}.
\newblock
\newblock
\urldef\tempurl%
\url{http://hackingdistributed.com/2016/06/17/thoughts-on-the-dao-hack/}
\showURL{%
Retrieved February 18, 2020 from \tempurl}


\bibitem[\protect\citeauthoryear{Soloway and Ehrlich}{Soloway and
  Ehrlich}{1984}]%
        {Soloway1984:Empirical}
\bibfield{author}{\bibinfo{person}{Elliot Soloway} {and} \bibinfo{person}{Kate
  Ehrlich}.} \bibinfo{year}{1984}\natexlab{}.
\newblock \showarticletitle{Empirical studies of programming knowledge}.
\newblock \bibinfo{journal}{\emph{IEEE Transactions on software engineering}}
  \bibinfo{number}{5} (\bibinfo{year}{1984}), \bibinfo{pages}{595--609}.
\newblock


\bibitem[\protect\citeauthoryear{{Stack Overflow}}{{Stack Overflow}}{2019}]%
        {StackOverflow2019:Survey}
\bibfield{author}{\bibinfo{person}{{Stack Overflow}}.}
  \bibinfo{year}{2019}\natexlab{}.
\newblock \bibinfo{title}{Developer Survey Results 2019}.
\newblock
\newblock
\urldef\tempurl%
\url{https://insights.stackoverflow.com/survey/2019}
\showURL{%
Retrieved February 18, 2020 from \tempurl}


\bibitem[\protect\citeauthoryear{Stefik and Hanenberg}{Stefik and
  Hanenberg}{2014}]%
        {Stefik14:Programming}
\bibfield{author}{\bibinfo{person}{Andreas Stefik} {and}
  \bibinfo{person}{Stefan Hanenberg}.} \bibinfo{year}{2014}\natexlab{}.
\newblock \showarticletitle{The Programming Language Wars: Questions and
  Responsibilities for the Programming Language Community}. In
  \bibinfo{booktitle}{\emph{Symposium on New Ideas, New Paradigms, and
  Reflections on Programming and Software}} \emph{(\bibinfo{series}{Onward!
  2014})}. \bibinfo{pages}{283--299}.
\newblock
\showISBNx{978-1-4503-3210-1}
\urldef\tempurl%
\url{https://doi.org/10.1145/2661136.2661156}
\showDOI{\tempurl}


\bibitem[\protect\citeauthoryear{Stefik and Hanenberg}{Stefik and
  Hanenberg}{2017}]%
        {stefik2017methodological}
\bibfield{author}{\bibinfo{person}{Andreas Stefik} {and}
  \bibinfo{person}{Stefan Hanenberg}.} \bibinfo{year}{2017}\natexlab{}.
\newblock \showarticletitle{Methodological irregularities in
  programming-language research}.
\newblock \bibinfo{journal}{\emph{Computer}} \bibinfo{volume}{50},
  \bibinfo{number}{8} (\bibinfo{year}{2017}), \bibinfo{pages}{60--63}.
\newblock
\urldef\tempurl%
\url{https://doi.org/10.1109/MC.2017.3001257}
\showDOI{\tempurl}


\bibitem[\protect\citeauthoryear{Stefik, Siebert, Stefik, and Slattery}{Stefik
  et~al\mbox{.}}{2011}]%
        {Stefik:2011:ECA:2089155.2089159}
\bibfield{author}{\bibinfo{person}{Andreas Stefik}, \bibinfo{person}{Susanna
  Siebert}, \bibinfo{person}{Melissa Stefik}, {and} \bibinfo{person}{Kim
  Slattery}.} \bibinfo{year}{2011}\natexlab{}.
\newblock \showarticletitle{An Empirical Comparison of the Accuracy Rates of
  Novices Using the Quorum, Perl, and Randomo Programming Languages}. In
  \bibinfo{booktitle}{\emph{Workshop on Evaluation and Usability of Programming
  Languages and Tools}} \emph{(\bibinfo{series}{PLATEAU '11})}.
  \bibinfo{publisher}{ACM}, \bibinfo{pages}{3--8}.
\newblock
\showISBNx{978-1-4503-1024-6}
\urldef\tempurl%
\url{https://doi.org/10.1145/2089155.2089159}
\showDOI{\tempurl}


\bibitem[\protect\citeauthoryear{Stewart, Shiroda, and James}{Stewart
  et~al\mbox{.}}{2006}]%
        {Stewart2006:Drug}
\bibfield{author}{\bibinfo{person}{Kent~D. Stewart}, \bibinfo{person}{Melisa
  Shiroda}, {and} \bibinfo{person}{Craig~A. James}.}
  \bibinfo{year}{2006}\natexlab{}.
\newblock \showarticletitle{Drug Guru: a computer software program for drug
  design using medicinal chemistry rules}.
\newblock \bibinfo{journal}{\emph{Bioorganic \& medicinal chemistry}}
  \bibinfo{volume}{14}, \bibinfo{number}{20} (\bibinfo{year}{2006}),
  \bibinfo{pages}{7011--7022}.
\newblock


\bibitem[\protect\citeauthoryear{Sunshine, Herbsleb, and Aldrich}{Sunshine
  et~al\mbox{.}}{2015}]%
        {Sunshine:2015:SSS:2820282.2820295}
\bibfield{author}{\bibinfo{person}{Joshua Sunshine}, \bibinfo{person}{James~D.
  Herbsleb}, {and} \bibinfo{person}{Jonathan Aldrich}.}
  \bibinfo{year}{2015}\natexlab{}.
\newblock \showarticletitle{Searching the State Space: A Qualitative Study of
  API Protocol Usability}. In \bibinfo{booktitle}{\emph{International
  Conference on Program Comprehension}} \emph{(\bibinfo{series}{ICPC '15})}.
  \bibinfo{publisher}{IEEE Press}, \bibinfo{address}{Piscataway, NJ, USA},
  \bibinfo{pages}{82--93}.
\newblock
\urldef\tempurl%
\url{http://dl.acm.org/citation.cfm?id=2820282.2820295}
\showURL{%
\tempurl}


\bibitem[\protect\citeauthoryear{Sunshine, Naden, Stork, Aldrich, and
  Tanter}{Sunshine et~al\mbox{.}}{2011}]%
        {sunshine2011first}
\bibfield{author}{\bibinfo{person}{Joshua Sunshine}, \bibinfo{person}{Karl
  Naden}, \bibinfo{person}{Sven Stork}, \bibinfo{person}{Jonathan Aldrich},
  {and} \bibinfo{person}{{\'E}ric Tanter}.} \bibinfo{year}{2011}\natexlab{}.
\newblock \showarticletitle{First-class state change in {Plaid}}. In
  \bibinfo{booktitle}{\emph{Object Oriented Programming Systems, Languages, and
  Applications}} \emph{(\bibinfo{series}{OOPSLA '11})}.
\newblock
\urldef\tempurl%
\url{https://doi.org/10.1145/2076021.2048122}
\showDOI{\tempurl}


\bibitem[\protect\citeauthoryear{Uesbeck, Stefik, Hanenberg, Pedersen, and
  Daleiden}{Uesbeck et~al\mbox{.}}{2016}]%
        {Uesbeck:2016:ESI:2884781.2884849}
\bibfield{author}{\bibinfo{person}{Phillip~Merlin Uesbeck},
  \bibinfo{person}{Andreas Stefik}, \bibinfo{person}{Stefan Hanenberg},
  \bibinfo{person}{Jan Pedersen}, {and} \bibinfo{person}{Patrick Daleiden}.}
  \bibinfo{year}{2016}\natexlab{}.
\newblock \showarticletitle{An Empirical Study on the Impact of C++ Lambdas and
  Programmer Experience}. In \bibinfo{booktitle}{\emph{International Conference
  on Software Engineering}} \emph{(\bibinfo{series}{ICSE '16})}.
  \bibinfo{publisher}{ACM}, \bibinfo{pages}{760--771}.
\newblock
\showISBNx{978-1-4503-3900-1}
\urldef\tempurl%
\url{https://doi.org/10.1145/2884781.2884849}
\showDOI{\tempurl}


\bibitem[\protect\citeauthoryear{University}{University}{2019}]%
        {CMU2019:MSE}
\bibfield{author}{\bibinfo{person}{Carnegie~Mellon University}.}
  \bibinfo{year}{2019}\natexlab{}.
\newblock \bibinfo{title}{Master's Programs}.
\newblock
\newblock
\urldef\tempurl%
\url{https://www.cs.cmu.edu/masters-programs}
\showURL{%
Retrieved February 18, 2020 from \tempurl}


\bibitem[\protect\citeauthoryear{Vans, von Mayrhauser, and Somlo}{Vans
  et~al\mbox{.}}{1999}]%
        {Vans1999:Program}
\bibfield{author}{\bibinfo{person}{A.~Marie Vans}, \bibinfo{person}{Anneliese
  von Mayrhauser}, {and} \bibinfo{person}{Gabriel Somlo}.}
  \bibinfo{year}{1999}\natexlab{}.
\newblock \showarticletitle{Program understanding behavior during corrective
  maintenance of large-scale software}.
\newblock \bibinfo{journal}{\emph{International Journal of Human-Computer
  Studies}} \bibinfo{volume}{51}, \bibinfo{number}{1} (\bibinfo{year}{1999}),
  \bibinfo{pages}{31 -- 70}.
\newblock
\showISSN{1071-5819}
\urldef\tempurl%
\url{https://doi.org/10.1006/ijhc.1999.0268}
\showDOI{\tempurl}


\bibitem[\protect\citeauthoryear{Venable, Pries-Heje, and Baskerville}{Venable
  et~al\mbox{.}}{2012}]%
        {Venable2012:Comprehensive}
\bibfield{author}{\bibinfo{person}{John Venable}, \bibinfo{person}{Jan
  Pries-Heje}, {and} \bibinfo{person}{Richard Baskerville}.}
  \bibinfo{year}{2012}\natexlab{}.
\newblock \showarticletitle{A Comprehensive Framework for Evaluation in Design
  Science Research}. In \bibinfo{booktitle}{\emph{Design Science Research in
  Information Systems. Advances in Theory and Practice}}.
  \bibinfo{pages}{423--438}.
\newblock
\urldef\tempurl%
\url{https://doi.org/10.1007/978-3-642-29863-9_31}
\showDOI{\tempurl}


\bibitem[\protect\citeauthoryear{Verner, Sampson, Tosic, Bakar, and
  Kitchenham}{Verner et~al\mbox{.}}{2009}]%
        {Verner2009:Guidelines}
\bibfield{author}{\bibinfo{person}{June~M. Verner}, \bibinfo{person}{Jennifer
  Sampson}, \bibinfo{person}{Vladimir Tosic}, \bibinfo{person}{N.A.~Abu Bakar},
  {and} \bibinfo{person}{Barbara~A. Kitchenham}.}
  \bibinfo{year}{2009}\natexlab{}.
\newblock \showarticletitle{Guidelines for industrially-based multiple case
  studies in software engineering}. In \bibinfo{booktitle}{\emph{International
  Conference on Research Challenges in Information Science}}. IEEE,
  \bibinfo{pages}{313--324}.
\newblock
\urldef\tempurl%
\url{https://doi.org/10.1109/RCIS.2009.5089295}
\showDOI{\tempurl}


\bibitem[\protect\citeauthoryear{Visser}{Visser}{1987}]%
        {Visser1987:Strategies}
\bibfield{author}{\bibinfo{person}{Willemien Visser}.}
  \bibinfo{year}{1987}\natexlab{}.
\newblock \showarticletitle{Strategies in programming programmable controllers:
  A field study on a professional programmer}. In
  \bibinfo{booktitle}{\emph{Empirical Studies of Programmers: Second workshop
  (ESP2)}}. \bibinfo{pages}{217--230}.
\newblock
\urldef\tempurl%
\url{https://hal.inria.fr/hal-00641376/document}
\showURL{%
\tempurl}


\bibitem[\protect\citeauthoryear{Walz, Elam, Krasner, and Curtis}{Walz
  et~al\mbox{.}}{1987}]%
        {Walz1987:Methodology}
\bibfield{author}{\bibinfo{person}{Diane~B. Walz}, \bibinfo{person}{Joyce~J.
  Elam}, \bibinfo{person}{Herb Krasner}, {and} \bibinfo{person}{Bill Curtis}.}
  \bibinfo{year}{1987}\natexlab{}.
\newblock \showarticletitle{A Methodology for Studying Software Design Teams:
  An Investigation of Conflict Behaviors in the Requirements Definition Phase}.
  In \bibinfo{booktitle}{\emph{Empirical Studies of Programmers: Second
  Workshop}}. \bibinfo{pages}{83–99}.
\newblock


\bibitem[\protect\citeauthoryear{Wilson, Pombrio, and Krishnamurthi}{Wilson
  et~al\mbox{.}}{2017}]%
        {Wilson2017:Crowdsource}
\bibfield{author}{\bibinfo{person}{Preston~Tunnell Wilson},
  \bibinfo{person}{Justin Pombrio}, {and} \bibinfo{person}{Shriram
  Krishnamurthi}.} \bibinfo{year}{2017}\natexlab{}.
\newblock \showarticletitle{Can We Crowdsource Language Design?}. In
  \bibinfo{booktitle}{\emph{Symposium on New Ideas in Programming and
  Reflections on Software}} \emph{(\bibinfo{series}{Onward! 2017})}.
  \bibinfo{pages}{1--17}.
\newblock
\urldef\tempurl%
\url{https://doi.org/10.1145/3133850.3133863}
\showDOI{\tempurl}


\bibitem[\protect\citeauthoryear{Zibin, Potanin, Ali, Artzi, Kielun, and
  Ernst}{Zibin et~al\mbox{.}}{2007}]%
        {IGJ}
\bibfield{author}{\bibinfo{person}{Yoav Zibin}, \bibinfo{person}{Alex Potanin},
  \bibinfo{person}{Mahmood Ali}, \bibinfo{person}{Shay Artzi},
  \bibinfo{person}{Adam Kielun}, {and} \bibinfo{person}{Michael~D. Ernst}.}
  \bibinfo{year}{2007}\natexlab{}.
\newblock \showarticletitle{Object and reference immutability using {Java}
  generics}. In \bibinfo{booktitle}{\emph{Foundations of Software Engineering}}
  \emph{(\bibinfo{series}{FSE '07})}. \bibinfo{publisher}{ACM},
  \bibinfo{pages}{75--84}.
\newblock


\end{thebibliography}

\end{document}